\documentclass[10pt,journal,compsoc]{IEEEtran}

\usepackage[utf8]{inputenc}
\usepackage{CJKutf8}

\usepackage{xcolor}
\usepackage{amssymb}
\usepackage{amsmath}

\usepackage{color}
\usepackage{tikz}
\usepackage[edges]{forest}
\definecolor{hidden-draw}{RGB}{20,68,106}
\definecolor{hidden-pink}{RGB}{255,245,247}

\usepackage{pifont}

\DeclareUnicodeCharacter{2194}{\ensuremath{\leftrightarrow}}
\usepackage{pdflscape}

 \definecolor{mylightpurple}{RGB}{130, 100, 150} 
 
\usepackage[flushleft]{threeparttable}
\usepackage{supertabular,booktabs}
\usepackage[figuresright]{rotating}

\usepackage[normalem]{ulem}
\useunder{\uline}{\ul}{}

\usepackage{bbm}
\usepackage{multirow}
\usepackage{booktabs}
\usepackage{graphicx}

\usepackage{color}
\usepackage{colortbl}
\usepackage{float}

\usepackage{caption}
\usepackage{enumerate}

\usepackage{ragged2e}
\usepackage{stfloats}
\usepackage{textcomp}
\usepackage{xcolor}
\usepackage{forest}

\usepackage{multirow}
\usepackage{adjustbox}
\usepackage{algorithm,algorithmic}
\usepackage{textcomp}
\usepackage{color,xcolor}
\usepackage{stmaryrd}
\usepackage{verbatimbox}
\usepackage{enumerate}
\usepackage[shortlabels]{enumitem}
\usepackage{bm}%

\usepackage{makecell}
\usepackage{booktabs}

\usepackage{multirow}

\usepackage{enumitem}
\usepackage{colortbl}

\usepackage{xspace}
\newcommand{\ie}{\textit{i.e.,}\xspace}
\newcommand{\eg}{\textit{e.g.,}\xspace}

\newcommand{\etal}{\textit{et al.}\xspace}

\newcommand{\figref}[1]{Figure~\ref{#1}\xspace}
\newcommand{\tabref}[1]{Table~\ref{#1}\xspace}

\newcommand{\numpaper}[1]{1009\xspace}
\newcommand{\numllm}[1]{62\xspace}
\newcommand{\numse}[1]{947\xspace}
\newcommand{\numpre}[1]{15\xspace}
\newcommand{\numdown}[1]{16\xspace}
\newcommand{\numcode}[1]{112\xspace}

\newcommand{\clm}[1]{LLMs of Code\xspace}

\newcommand{\mytime}[1]{August 2024}

\definecolor{codegreen}{rgb}{0,0.6,0}
\definecolor{codegray}{rgb}{0.5,0.5,0.5}
\definecolor{codepurple}{rgb}{0.58,0,0.82}
\definecolor{backcolour}{rgb}{0.95,0.95,0.92}
\usepackage{listings}
\lstdefinestyle{mystyle}{
  backgroundcolor=\color{backcolour},   commentstyle=\color{codegreen},
  keywordstyle=\color{magenta},
  numberstyle=\tiny\color{codegray},
  stringstyle=\color{codepurple},
  basicstyle=\ttfamily\footnotesize,
  breakatwhitespace=false,
  breaklines=true,
  captionpos=b,
  keepspaces=true,
  numbers=left,
  numbersep=5pt,
  showspaces=false,
  showstringspaces=false,
  showtabs=false,
  tabsize=2
}
\usepackage{url}
\usepackage{hyperref}
\hypersetup{
    colorlinks=true,
    linkcolor=blue,
    filecolor=blue,      
    urlcolor=blue,
    citecolor=blue,
}

\usepackage[most]{tcolorbox}
\newcommand{\finding}[2]{
\begin{center}
\begin{tcolorbox}[colback=gray!15, colframe=black, boxsep= -0.15cm, middle=-0.15cm, breakable]
\textbf{Answer to RQ{#1}:}
{#2}
\end{tcolorbox}
\end{center}
}

\newcommand{\delete}[1]{}

\AtBeginDocument{%
  \providecommand\BibTeX{{%
    \normalfont B\kern-0.5em{\scshape i\kern-0.25em b}\kern-0.8em\TeX}}}
    
\begin{document}

\title{A Survey on Large Language Models for Software Engineering}

\author{Quanjun Zhang, Chunrong Fang, Yang Xie, Yaxin Zhang, Yun Yang, Weisong Sun, Shengcheng Yu, Zhenyu Chen

\IEEEcompsocitemizethanks{
\IEEEcompsocthanksitem 
Quanjun Zhang, Chunrong Fang, Yang Xie, Yaxin Zhang, Shengcheng Yu and Zhenyu Chen
are with the State Key Laboratory for Novel Software Technology, Nanjing University, China. \protect\\
E-mail: 
quanjun.zhang@smail.nju.edu.cn,
fangchunrong@nju.edu.cn,
serialxy@outlook.com,
zhangyaxin032@gmail.com,
yusc@smail.nju.edu.cn,
zychen@nju.edu.cn

\IEEEcompsocthanksitem Weisong Sun is with the School of Computer Science and Engineering, Nanyang Technological University.\protect\\
E-mail: weisong.sun@ntu.edu.sg.

\IEEEcompsocthanksitem Yun Yang is with the Department of Computing Technologies, Swinburne University of Technology, Melbourne, VIC 3122, Australia.\protect\\
E-mail: yyang@swin.edu.au

}

}

\IEEEtitleabstractindextext{
\begin{abstract}
\justifying
Software Engineering (SE) is the systematic design, development, maintenance, and management of software applications underpinning the digital infrastructure of our modern world.
Very recently, the SE community has seen a rapidly increasing number of techniques employing Large Language Models (LLMs) to automate a broad range of SE tasks.
Nevertheless, existing information of the applications, effects, and possible limitations of LLMs within SE is still not well-studied.

In this paper, we provide a systematic survey to summarize the current state-of-the-art research in the LLM-based SE community.
We summarize \numllm{} representative \clm{} across three model architectures, \numpre{} pre-training objectives across four categories, and \numdown{} downstream tasks across five categories.
We then present a detailed summarization of the recent SE studies for which LLMs are commonly utilized, including \numse{} studies for \numcode{} specific code-related tasks across five crucial phases within the SE workflow. 
We also discuss several critical aspects during the integration of LLMs into SE, such as empirical evaluation, benchmarking, security and reliability, domain tuning, compressing and distillation.
Finally, we highlight several challenges and potential opportunities on applying LLMs for future SE studies, such as exploring domain LLMs and constructing clean evaluation datasets.
Overall, our work can help researchers gain a comprehensive understanding about the achievements of the existing LLM-based SE studies and promote the practical application of these techniques. 
Our artifacts are publicly available and will be continuously updated at the living repository: \url{https://github.com/iSEngLab/AwesomeLLM4SE}.
\end{abstract}

}

\maketitle
\IEEEdisplaynontitleabstractindextext

\IEEEpeerreviewmaketitle

\section{Introduction}
\label{sec:intro}

Software engineering (SE) stands as an essential pursuit focused on systematically and predictably designing, developing, testing, and maintaining software systems~\cite{biolchini2005systematic}.
As software increasingly becomes the infrastructure of various industries (\eg transportation, healthcare, and education) nowadays, SE plays a crucial role in modern society by ensuring software systems are built in a systematic, reliable, and efficient manner~\cite{zelkowitz1978perspectives}.
As a very active area, SE has been extensively investigated in the literature and has sustained attention from both the academic and industrial communities for several decades~\cite{yang2022survey, wang2023software}.

Very recently, one of the most transformative advancements in the realm of SE is the emergence of large language models (LLMs). 
Advanced LLMs (\eg BERT~\cite{devlin2018bert}, T5~\cite{raffel2020exploring} and GPT~\cite{radford2019language}) have significantly improved performance across a wide range of natural language processing (NLP) tasks, such as machine translation and text classification.
Typically, such models are pre-trained to derive generic language representations by self-supervised training on large-scale unlabeled data and then are transferred to benefit multiple downstream tasks by supervised fine-tuning on limited labeled data.
Inspired by the success of LLMs in NLP, many recent attempts have been adopted to boost numerous code-related tasks (\eg code summarization and code search) with LLMs (\eg CodeBERT~\cite{feng2020codebert} and CodeT5~\cite{wang2021codet5}).
The application of LLMs to SE has had a profound impact on the field, transforming how developers approach code-related tasks automatically.
For example, ChatGPT~\cite{chatgpt}, one of the most notable LLMs with billions of parameters, has demonstrated remarkable performance in a variety of tasks, showcasing the potential of LLMs to revolutionize the SE industry. 
Overall, the SE community has seen a rapidly increasing number of a broad range of SE studies equipped with LLMs, already yielding substantial benefits and further demonstrating a promising future in follow-up research.

However, the complex SE workflow (\eg software development, testing, and maintenance) and a mass of specific code-related tasks (\eg vulnerability detection, fault localization, and program repair) make it difficult for interested researchers to understand state-of-the-art LLM-based SE research and improve upon them.
Besides, the constant emergence of advanced LLMs with different architectures, training methods, sources, and a plethora of fine-tuning methods brings challenges in keeping pace with and effectively utilizing these advancements.
For example, researchers have conducted various studies to extensively investigate the effectiveness of LLMs in the field of program repair~\cite{xia2023automated,xia2023automated,zhang2023pre}.
These studies encompass different research aspects (\eg empirical and technical studies~\cite{xia2022less}), types of LLMs (\eg open-source or closed-source~\cite{xia2023automated}), mode architectures (\eg encoder-decoder or encoder-only~\cite{zhang2023pre}), model parameters (\eg CodeT5-60M and InCoder-6B~\cite{fried2022incoder}), types of bugs (\eg semantic bugs and security vulnerabilities~\cite{wu2023effective}), and utilization paradigms (\eg fine-tuning~\cite{yuan2022circle}, few-shot~\cite{nashid2023retrieval} and zero-shot~\cite{zhang2023gamma}).

In this paper, we summarize existing work and provide a retrospection of the LLM-based community field after years of rapid development.
Community researchers can have a thorough understanding of the advantages and limitations of the existing LLM-based SE techniques. 
We discuss how LLMs are integrated into specific tasks in the typical workflow of SE research. 
Based on our analysis, we point out the current challenges and suggest possible future directions for LLM-based SE research. 
Overall, our work provides a comprehensive review of the current progress of the LLM-based SE community, enabling researchers to obtain an overview of this thriving field and make progress toward advanced practices.

\textbf{{Contributions.}}
To sum up, the main contributions of this paper are as follows:
\begin{itemize}
    \item \textit{Survey Methodology.}
    We conduct a detailed analysis of \numpaper{} relevant SE studies empowered with LLMs in terms of publication trends and distribution of venues until mytime{}.
    
    \item \textit{LLM Perspective.}
    We summarize \numllm{} representative LLMs of Code for the SE community according to different aspects, such as model architectures, pre-training objectives, downstream tasks, and open science.
    
    \item \textit{SE Perspective.}
    We explore the typical application of leveraging the advance of recent LLMs to automate the SE research, involving \numse{} relevant studies for \numcode{} code-related tasks across five SE phases, \ie software requirements and design, software development, software testing, software maintenance and software management.

    \item \textit{Integration Perspective.}
    We discuss some crucial aspects when LLMs are integrated into the SE field, such as evaluation, benchmarking, security and reliability, and domain tuning.
    
    \item \textit{{Outlook and challenges.}}
    We pinpoint open research challenges and provide several practical guidelines on applying LLMs for future SE studies.
\end{itemize}

\textbf{Comparison with Existing Surveys.}
In 2022, Watson~\etal~\cite{watson2022systematic}, Wang~\etal~\cite{wang2022machine} and Yang~\etal~\cite{yang2022survey} present a systematic literature review of research at the intersection of SE and ML\&DL.
Such surveys mainly concentrate on the application of ML or DL techniques in SE rather than more powerful and rapidly emerging LLMs.
Besides, Niu~\etal~\cite{niu2022deep} and Zan~\etal~\cite{zan2023large} present a survey to summarize 20 and 27 LLMs for natural-language-to-code (NL4Code) tasks.
Such surveys are limited to a narrow research scope, \ie NL2Code, thus ignoring the more complex and challenging SE domain.
Thus, our work focuses on the foundations of recently emerged LLMs within the crucial SE research, particularly covering \numcode{} specific tasks across five crucial SE phases, \ie software requirement \& design, development, testing, maintenance and management phases, as well as corresponding integration studies.

In addition to the aforementioned published papers, we notice there exist some pre-print works that explore the integration of LLMs with SE.
While these unpublished papers are concurrent with our work, there remain some fundamental differences.
Wang~\etal~\cite{wang2023software} provide a review of LLMs in software testing, while our work targets the whole SE scope rather than a single SE phase.
Fan~\etal~\cite{fan2023large} discuss the achievements and challenges of LLMs in SE, and  Hou~\etal~\cite{hou2023large} conduct a systematic literature review on LLM4SE.
Despite the close relevance of these two works to this paper, the key distinction lies in our three-fold focus on LLMs (\ie Section~\ref{sec:rq1_llm}), SE (\ie Section~\ref{sec:rq2_se}), and the integration of both (\ie Section~\ref{sec:rq3_intergation}), Unlike the first work only taking a bird's-eye view of the LLM-based SE achievements or the second work involving complicated aspects, \eg data processing and metrics.
We also release the first public repository to track the latest progress through crowd-sourcing, and we believe it will contribute to the ongoing development of the community.
Lastly, our survey summarizes the existing studies until August 2024.

\textbf{Paper Organization.}
The remainder of this paper is organized as follows. 
Section~\ref{sec:methodology} provides a detailed exposition of three research questions and the methodology employed for conducting the survey. 
Sections~\ref{sec:rq1_llm} summarize existing LLMs of source code. 
Section~\ref{sec:rq2_se} illustrates existing SE studies empowered with LLMs.
Section~\ref{sec:rq3_intergation} summarizes the crucial aspects during the integration of LLMs into SE.
Section~\ref{sec:challenges} highlights the challenges and promising opportunities for future research. 
Section~\ref{sec:conclusion} draws the conclusions.

{\bf Availability.} All artifacts of this study are available in the following public repository.
The living repository continuously updates the latest research on LLMs, LLM4SE, and related studies.

\begin{center}
\url{https://github.com/iSEngLab/AwesomeLLM4SE}
\end{center}

\section{Survey Methodology}
\label{sec:methodology}

\subsection{Research Questions}

\delete{As we have previously discussed, with the development of Code LLMs, LLMs have found widespread application in the field of SE. 
Specifically, LLMs have demonstrated success in tasks related to SE, such as code completion, bug detection, and comment generation. 
However, there are still challenges to be overcome. For instance, making these models perform effectively across different types of scenarios, languages, and real-world applications can be challenging. 
We also face issues related to the need for appropriate pre-training and the necessity for proper fine-tuning.

To provide a comprehensive overview of this field, it is essential to gain a thorough understanding of these models, their current applications in SE, the challenges they encounter, and potential future research directions in SE. 
Therefore, our objective is to present a systematic literature review of the application of LLMs in SE.
This study aims to answer the following research questions (RQs):}

To provide a comprehensive overview of LLMs and the current achievements in SE, our work aims to address the  following research questions (RQs):

\begin{itemize}
    \item \textbf{RQ1 (LLM Perspective):} What LLMs are designed to support SE tasks?
    
    \begin{itemize}
        \item \textbf{RQ1.1:} What LLMs have been released?
        \item \textbf{RQ1.2:} What pre-training tasks have been used to train LLMs?
        \item \textbf{RQ1.3:} What downstream tasks are LLMs spread to?
        \item  \textbf{RQ1.4:} How are LLMs open-sourced to support the open science community?
    \end{itemize}

    \item \textbf{RQ2 (SE Perspective):} What SE tasks are facilitated by LLMs?

    \item \textbf{RQ3 (Integration Perspective):} What are the key factors during the integration of LLMs into SE?

\end{itemize}
To answer RQ1, we summarize LLMs in the SE literature from four aspects: LLM categories in Section~\ref{sec:rq1.1_llms}, pre-training tasks in Section~\ref{sec:rq1.2_pre_training}, downstream tasks in Section~\ref{sec:rq1.3_downstream}, and open-science in Section~\ref{sec:rq1.4_analysis}.
To answer RQ2, we investigate SE tasks that have been facilitated by LLMs from five aspects: software requirements \& design in Section~\ref{sec:rq2_se_requirement}, software development in Section~\ref{sec:rq2_se_development}, software testing in Section~\ref{sec:rq2_se_testing}, software maintenance in Section~\ref{sec:rq2_se_maintenance} and software management in Section~\ref{sec:rq2_se_management}.
To answer RQ3, we analyze challenges and achievements during the integration of LLMs into SE from four aspects: \ie evaluation and benchmarking in Section~\ref{sec:rq3.1_evaluation}, security and reliability in Section~\ref{sec:rq3.2_security}, domain tuning in Section~\ref{sec:rq3.3_tuning}, compressing and distillation in Section~\ref{sec:rq3.4_compressing}.

\begin{figure*}[htbp]
    \centering
    \graphicspath{{graphs/}}
    \includegraphics[width=0.7\linewidth]{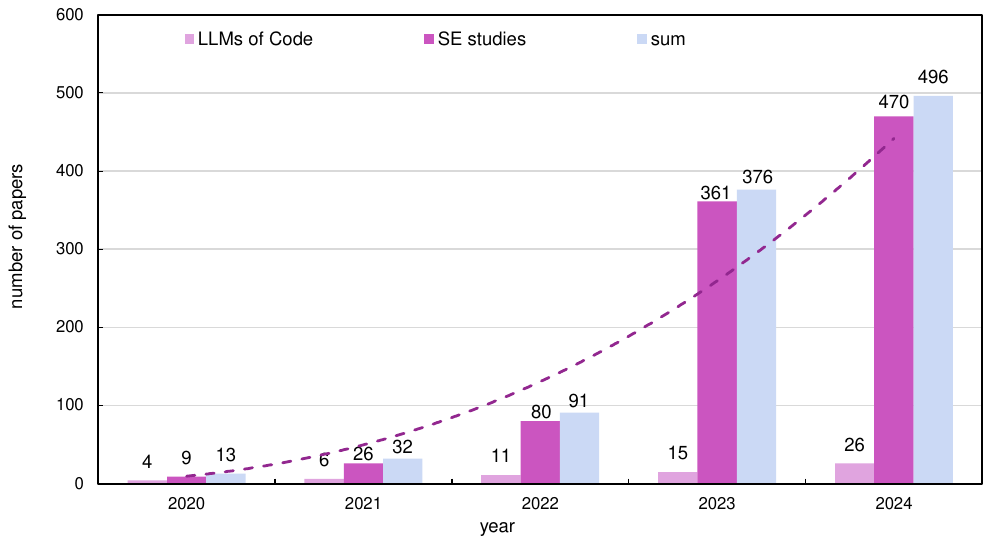}
    \caption{Number of collected papers over years}
    \label{fig:papers_year}
\end{figure*}

\subsection{Search Strategy}

Following existing DL for SE surveys~\cite{wang2022machine,yang2022survey,watson2022systematic}, we divide the search keywords used for searching papers into two groups: (1) a SE-related group containing some commonly used keywords related to SE research; and (2) an LLM-related group containing some keywords related to LLM research.
Besides, considering a significant amount of relevant papers from SE, AI, and NLP communities, following Zhang~\etal~\cite{zhang2011identifying}, we attempt to identify some preliminary search keywords from three sources: 
(1) existing LLM surveys~\cite{zan2023large} to derive LLM-related keywords;
(2) existing SE surveys~\cite{wang2022machine,watson2022systematic} to derive SE-related keywords;
(3) a limited number of LLM-based SE research papers manually collected from top-tier conferences and journals beforehand to refine LLM-related and SE-related keywords.
The search strategy can capture the most relevant studies from existing surveys while achieving better efficiency than a purely manual search.
Finally, the complete set of search keywords is as follows.

\begin{itemize}
    \item \textit{SE-related Keywords:} Software Engineering, SE, Software Requirements, Software Design, Software Development, Software Testing, Software Maintenance, Code generation, Code Search, Code Completion, Code Summarization, Fault Detection, Fault Localization, Vulnerability Prediction, Testing Minimization, Test Generation, Fuzzing, GUI testing, NLP testing, Program Repair, Code Review, Vulnerability Repair, Patch Correctness.
    \item \textit{LLM-related Keywords:} LLM, Large Language Model, Language Model, LM, PLM, Pre-trained model, Pre-training, Natural Language Processing, NLP, Machine Learning, ML, Deep Learning, DL, Artificial Intelligence, AI, Transformer, BERT, Codex, GPT, T5, ChatGPT.
\end{itemize}

\delete{It's important to note that the list of LLM-related keywords includes terms like \textit{Machine Learning} and \textit{Deep Learning}, which may not seem directly related to LLMs. This is done to avoid omitting papers relevant to our research, expanding our search scope during the automated searches.}

Our survey focuses on LLMs in the field of SE, encompassing existing LLMs and their applications in SE workflow. 
Thus, we classify papers that need to be summarized into two categories.
For LLMs research, we search for papers whose titles contain the second keyword set.
For LLM-based SE research, a paper is considered relevant only if it contains both sets of keywords.
Then we conduct an automated search on three widely used databases until \mytime{}, \ie Google Scholar repository, ACM Digital Library, and IEEE Explorer Digital Library.
Finally, we retrieve a total of 32,560 papers from three databases by automated keyword searching.

\subsection{Study Selection}
Once the potentially relevant studies based on our search strategy are collected, we conduct a three-stage paper filtering to further determine which papers are relevant to this survey. 
First, we attempt to filter out the papers before 2017, considering that the Transformer architecture~\cite{vaswani2017attention} is proposed in 2017, which is the foundation of LLMs.
Second, we automatically filter out any paper less than 7 pages and duplicated papers.
Third, we inspect the remaining papers manually to decide whether they are relevant to the LLM-based SE field according to some quality assessment criteria.
The manual inspection is conducted independently by two authors, and any paper with different decisions will be handed over to a third author to make the final decision.
As a result, we collect 59 papers related to the code LLM research and 912 papers related to the LLM-based SE research.

To mitigate potential omissions in our automated search and to ensure a thorough collection of papers, we further employed a snowballing search strategy~\cite{watson2022systematic}.
Snowballing involves meticulously reviewing the reference lists and citations of each paper to uncover additional relevant studies that our initial search may have missed. 
In particular, we look at every reference within the collected papers and determine if any of those references are relevant to our study.
Through this rigorous manual analysis, we succeed in additionally identifying three papers related to LLMs and 36 papers related to LLM-based SE, thereby enriching our survey with a diverse range of insights.

\subsection{Trend Observation}

We finally obtain \numpaper{} relevant research papers after automated searching and manual inspection.
\figref{fig:papers_year} shows the number of collected papers from 2020 to 2024, 
It can be observed that studies on proposing LLMs and their applications in SE have rapidly increased since 2020, indicating the growing recognition among researchers of LLMs as a viable and promising approach to automating SE tasks.
One possible reason is that DL technologies have already shown promising performance in various SE tasks over the past several years~\cite{watson2022systematic}. 
As a derivative of DL, LLMs bring more powerful code understanding capabilities with larger model sizes and training datasets, demonstrating the potential of being a brand-new way to address SE problems. 
The second reason is the recent flourishing of the open-source community, which provides millions or even hundreds of millions of open-source code snippets, laying the foundation for training such LLMs.

\begin{table*}[t]
\footnotesize
    \centering
    \caption{Venue Distribution of Collected Papers}
    \label{tab:venues}
    \footnotesize
    \resizebox{\linewidth}{!}{
     \begin{tabular}{lllc}
     \toprule
    \textbf{Acronym} & \textbf{Venues} & \textbf{Type} & \textbf{\# Publications} \\
     \midrule
    arXiv & N.A.  &    N.A.   & 382 \\
    ICSE  & International Conference on Software Engineering & C     & 95 \\
    FSE   & International Conference on the Foundations of Software Engineering
    & C     & 51 \\
    ACL   & Meeting of the Association for Computational Linguistics & C     & 31 \\
    ASE   & International Conference on Automated Software Engineering & C     & 30 \\
    ISSTA & International Symposium on Software Testing and Analysis & C     & 28 \\
    TSE   & Transactions on Software Engineering & J     & 26 \\
    ICLR  & International Conference on Learning Representations & C     & 20 \\
    TOSEM & Transactions on Software Engineering Methodology & C     & 20 \\
    SANER & International Conference on Software Analysis, Evolution, and Reengineering & C     & 17 \\
    EMNLP & Empirical Methods in Natural Language Processing & C     & 16 \\
    ICML  & International Conference on Machine Learning & C     & 15 \\
    ICSME & International Conference on Software Maintenance and Evolution & C     & 15 \\
    MSR   & Mining Software Repositories & C     & 15 \\
    JSS   & Journal of Systems and Software & J     & 12 \\
    NeurIPS & Neural Information Processing Systems & C     & 12 \\
    ICPC  & International Collegiate Programming Contest & C     & 11 \\
    AAAI  & Association for the advance of Artificial Intelligence & C     & 10 \\
    COMPSAC & International Computer Software and Applications Conference & C     & 9 \\
    USENIX & USENIX Security Symposium & C     & 9 \\
    ICST  & International Conference on Software Testing & C     & 8 \\
    Internetware & Asia-Pacific Symposium on Internetware & C     & 8 \\
    ISSRE & International Symposium on Software Reliability Engineering & C     & 8 \\
    IST   & Information and Software Technology & J     & 8 \\
    NAACL & North American Association of computational linguistics & C     & 8 \\
    APSEC & Asia-Pacific Software Engineering Conference & C     & 7 \\
    AUSE  & Automated software engineering & J     & 6 \\
    ESEM  & International Symposium on Empirical Software Engineering and Measurement & C     & 6 \\
    EASE  & International Conference on Evaluation and Assessment in Software Engineering & C     & 5 \\
    EMSE  & Empirical Software Engineering & J     & 5 \\
    \bottomrule
    \end{tabular}%
    }
\end{table*}

\tabref{tab:venues} further shows the number of collected papers across different venues.
The first two columns list the venue and its acronym, the third column indicates whether it is a conference or journal, and the final column shows the number of publications.
We only present the top-30 venues with the highest number of publications due to page limitations.
First, we find that these papers span multiple research fields, including SE, NLP, AI, and Security, which indicates the wide range of attention this direction has received.
Second, unlike previous work~\cite{wang2020deep,zhang2023survey1}, it can be found that a significant number of papers (382/1009) have not been peer-reviewed.
The reason behind this phenomenon lies in the rapid development in this field, especially after the release of the ChatGPT model at the end of 2022, which has stimulated a considerable amount of research in SE.
Third, the top five venues are top-tier conferences (ICSE, FSE, ACL, ASE and ISSTA), and 25 venues among the top-30 ones are conferences, indicating a current inclination towards conferences in this field due to the timeliness of conference proceedings.

\section{RQ1: LLM Perspective}
\label{sec:rq1_llm}

\tikzstyle{my-box}=[
    rectangle,
    draw=hidden-draw,
    rounded corners,
    text opacity=1,
    minimum height=1.5em,
    minimum width=5em,
    inner sep=2pt,
    align=center,
    fill opacity=.5,
    line width=0.8pt,
]
\tikzstyle{leaf}=[my-box, minimum height=1.5em,
    fill=mylightpurple!60,
    text=black,
    align=left,
    font=\normalsize,
    inner xsep=2pt,
    inner ysep=4pt,
    line width=0.8pt,
]

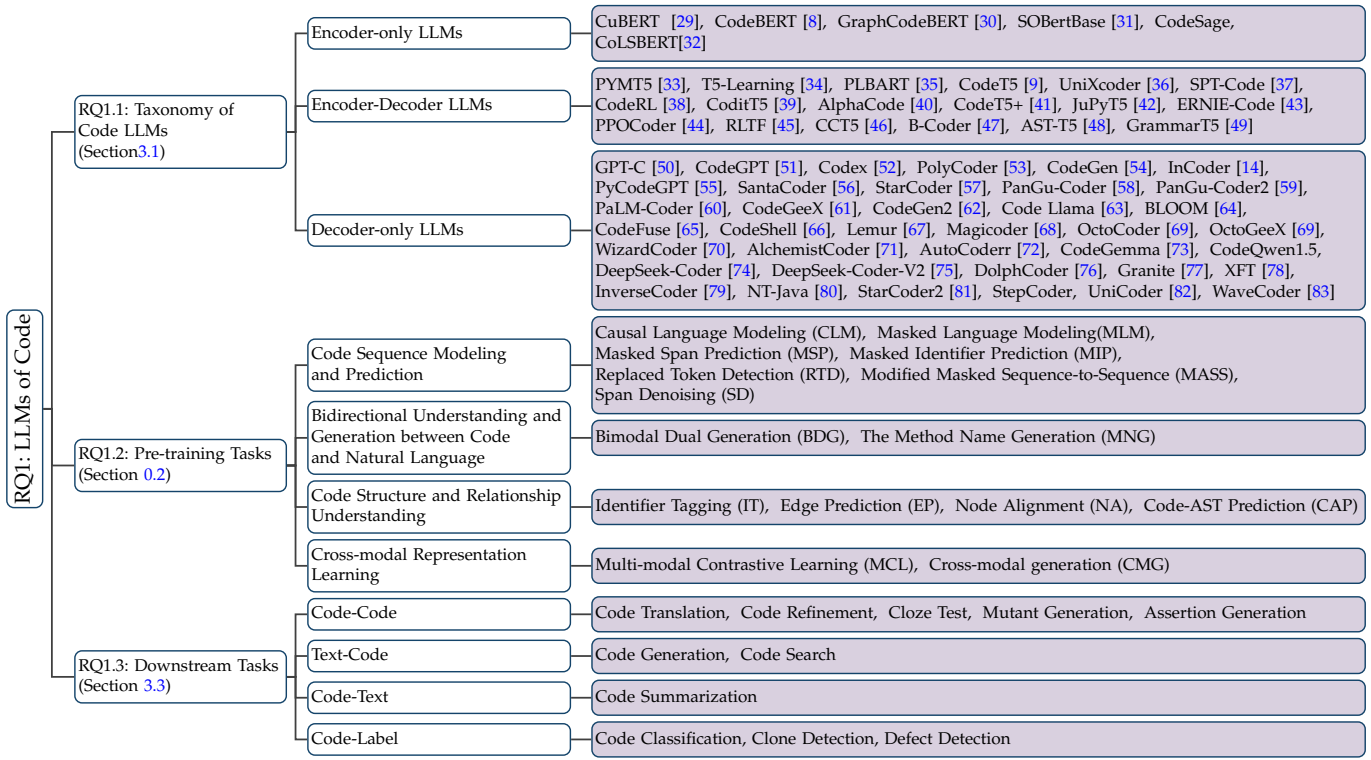
\begin{figure*}[t]
    \centering
    \resizebox{1\textwidth}{!}{
        \begin{forest}
            forked edges,
            for tree={
                grow=east,
                reversed=true,
                anchor=base west,
                parent anchor=east,
                child anchor=west,
                base=left,
                font=\large,
                rectangle,
                draw=hidden-draw,
                rounded corners,
                align=left,
                minimum width=4em,
                edge+={darkgray, line width=1pt},
                s sep=3pt,
                inner xsep=2pt,
                inner ysep=3pt,
                line width=0.8pt,
                ver/.style={rotate=90, child anchor=north, parent anchor=south, anchor=center},
            },
            where level=1{text width=12em, font=\normalsize,}{},
            where level=2{text width=15em, font=\normalsize,}{},
            where level=3{text width=6.5em, font=\normalsize,}{},
            where level=4{text width=5em, font=\normalsize,}{},
            [
                RQ1: LLMs of Code, ver
                [  
                    RQ1.1: Taxonomy of \\Code LLMs\\ (Section\ref{sec:rq1.1_llms})
                    [
                         Encoder-only LLMs
                        [
                            CuBERT~~\cite{kanade2020learning}{, }
                            CodeBERT~\cite{feng2020codebert}{, }
                            GraphCodeBERT~\cite{guo2020graphcodebert}{, }
                            SOBertBase~\cite{mukherjee2023stack}{, }
                            CodeSage{, }\\
                            CoLSBERT\cite{lin2024scaling}
                            , leaf, text width=45em
                        ]
                    ]
                    [
                        Encoder-Decoder LLMs 
                        [
                            PYMT5~\cite{clement2020pymt5}{, }
                            T5-Learning~\cite{mastropaolo2021studying}{, }
                            PLBART~\cite{ahmad2021unified}{, }
                            CodeT5~\cite{wang2021codet5}{, }
                            UniXcoder~\cite{guo2022unixcoder}{, }
                            SPT-Code~\cite{niu2022spt}{, }\\
                            CodeRL~\cite{le2022coderl}{, }
                            CoditT5~\cite{zhang2022coditt5}{, }
                            AlphaCode~\cite{li2022competition}{, }
                            CodeT5+~\cite{wang2023codet5+}{, }
                            JuPyT5~\cite{chandel2022training}{, }
                            ERNIE-Code~\cite{chai2022ernie}{, }\\
                            PPOCoder~\cite{shojaee2023execution}{, }
                            RLTF~\cite{liu2023rltf}{, }
                            CCT5~\cite{lin2023cct5}{, }
                            B-Coder~\cite{yu2023b}{, }
                            AST-T5~\cite{gong2024ast}{, }
                            GrammarT5~\cite{zhu2024grammart5},
                            leaf, text width=45em
                        ]
                    ]
                    [
                        Decoder-only LLMs 
                        [
                            GPT-C~\cite{svyatkovskiy2020intellicode}{, }
                            CodeGPT~\cite{lu2021codexglue}{, }
                            Codex~\cite{chen2021evaluating}{, }
                            PolyCoder~\cite{xu2022systematic}{, }
                            CodeGen~\cite{nijkamp2022codegen}{, }
                            InCoder~\cite{fried2022incoder}{, }\\
                            PyCodeGPT~\cite{zan2022cert}{, }
                            SantaCoder~\cite{allal2023santacoder}{, }
                            StarCoder~\cite{li2023starcoder}{, }
                            PanGu-Coder~\cite{christopoulou2022pangu}{, }
                            PanGu-Coder2~\cite{shen2023pangu}{, }\\
                            PaLM-Coder~\cite{chowdhery2022palm}{, }
                            CodeGeeX~\cite{zheng2023codegeex}{, }
                            CodeGen2~\cite{nijkamp2023codegen2}{, }
                            Code Llama~\cite{touvron2023llama}{, }
                            BLOOM~\cite{le2023bloom}{, }\\
                            CodeFuse~\cite{di2024codefuse}{, }
                            CodeShell~\cite{xie2024codeshell}{, }
                            Lemur~\cite{xu2023lemur}{, }
                            Magicoder~\cite{wei2024magicoder}{, }
                            OctoCoder~\cite{muennighoff2023octopack}{, }
                            OctoGeeX~\cite{muennighoff2023octopack}{, }\\
                            WizardCoder~\cite{luo2023wizardcoder}{, }
                            AlchemistCoder~\cite{song2024alchemistcoder}{, }
                            AutoCoderr~\cite{lei2024autocoder}{, }
                            CodeGemma~\cite{team2024codegemma}{, }
                            CodeQwen1.5{, }\\
                            DeepSeek-Coder~\cite{guo2024deepseek}{, }
                            DeepSeek-Coder-V2~\cite{zhu2024deepseek}{, }
                            DolphCoder~\cite{wang2024dolphcoder}{, }
                            Granite~\cite{mishra2024granite}{, }
                            XFT~\cite{ding2024xft}{, }\\
                            InverseCoder~\cite{wu2024inversecoder}{, }
                            NT-Java~\cite{rathinasamy2024narrow}{, }
                            StarCoder2~\cite{lozhkov2024starcoder}{, }
                            StepCoder{, }
                            UniCoder~\cite{sun2024unicoder}{, }
                            WaveCoder~\cite{yu2024wavecoder},
                            leaf, text width=45em
                        ]
                    ]
                ]
                [                
                    RQ1.2: Pre-training Tasks  \\ (Section~\ref{sec:rq1.1.2})
                    [
                         Code Sequence Modeling \\and Prediction 
                        [
                            Causal Language Modeling (CLM){, }
                            Masked Language Modeling(MLM){, }\\
                            Masked Span Prediction (MSP){, }
                            Masked Identifier Prediction (MIP){, }\\
                            Replaced Token Detection (RTD){, }
                            Modified Masked Sequence-to-Sequence (MASS){, }\\
                            Span Denoising (SD),
                            leaf, text width=45em
                        ]
                    ]
                    [
                        Bidirectional Understanding and \\Generation between Code \\and Natural Language 
                        [
                           Bimodal Dual Generation (BDG){, }
                           The Method Name Generation (MNG), 
                           leaf, text width=45em
                        ]
                    ]     
                    [
                        Code Structure and Relationship\\ Understanding 
                        [   Identifier Tagging (IT){, }
                            Edge Prediction (EP){, }
                            Node Alignment (NA){, }
                            Code-AST Prediction (CAP), 
                            leaf, text width=45em
                        ]
                    ]
                    [
                        Cross-modal Representation \\Learning 
                        [   
                            Multi-modal Contrastive Learning (MCL){, }
                            Cross-modal generation (CMG),
                            leaf, text width=45em
                        ]
                    ]
                ]
                [                
                    RQ1.3: Downstream Tasks \\(Section~\ref{sec:rq1.3_downstream})
                    [
                        Code-Code 
                        [
                           Code Translation{, } 
                           Code Refinement{, }
                           Cloze Test{, }
                           Mutant Generation{, }
                           Assertion Generation, 
                           leaf, text width=45em
                        ]
                    ]
                    [
                        Text-Code 
                        [
                            Code Generation{, }
                            Code Search, 
                            leaf, text width=45em
                        ]
                    ]
                    [               
                        Code-Text 
                        [
                            Code Summarization, 
                            leaf, text width=45em
                        ]
                    ]
                    [
                        Code-Label 
                        [
                            Code Classification{,} 
                            Clone Detection{,} 
                            Defect Detection, 
                            leaf, text width=45em
                        ]
                    ]
                ]
            ]
        \end{forest}
    }
    \caption{Taxonomy of RQ1}
    \label{fig:rq1_taxonomy}
\end{figure*}

In this section, we summarize existing representative \clm{} in Section~\ref{sec:rq1.1_llms}, pre-training tasks in Section~\ref{sec:rq1.2_pre_training}, fine-tuning tasks in Section~\ref{sec:rq1.3_downstream}, and discuss the open science issue in Section~\ref{sec:rq1.4_analysis}.
The detailed taxonomy is presented in \figref{fig:rq1_taxonomy}, including the three sub-RQs and their corresponding categorizations.

\subsection{RQ1.1: What LLMs have been released to support SE?}
\label{sec:rq1.1_llms}
\begin{table*}[!ht]
  \centering
  \renewcommand{\arraystretch}{1.}
  \caption{A Summary and Comparison of \clm{}.}
  \resizebox{0.98\linewidth}{!}{
    \begin{tabular}{cc|cccccc|c}
    \toprule
    Year  & Model & Publisher & Architecture & Size  & Tokenizer & Init  & Organization & Public \\
    \midrule
    2020  & CodeBERT & ICML  & Encoder-only & 350M  & N.A.  & BERT  & Google, IISC   & Yes \\
    2020  & CuBERT & EMNLP & Encoder-only & 125M  & WordPiece & Scratch & HIT, Microsoft   & Yes \\
    2020  & GPT-C & FSE   & Decoder-only & 366M  & BPE   & GPT-2 & Microsoft   & No \\
    2020  & PyMT5 & EMNLP & Encoder-Decoder & 374M  & BBPE  & GPT-2 & Microsoft   & No \\
    2021  & CodeGPT & NeurIPS & Decoder-only & 124M  & BBPE  & GPT-2 & PKU, Microsoft   & Yes \\
    2021  & CodeT5 & EMNLP & Encoder-Decoder & 60M,220M & BBPE  & Scratch & Salesforce   & Yes \\
    2021  & Codex & arXiv & Decoder-only & \makecell{12M,25M,42M,85M,\\300M,679M,2.5B,12B} & BBPE  & GPT-3 & OpenAI   & No \\
    2021  & GraphCodeBERT & ICLR  & Encoder-only & 125M  & WordPiece & Scratch & Microsoft, SYSU   & Yes \\
    2021  & PLBART & NAACL & Encoder-Decoder & 140M,406M & SentencePiece & Scratch & UC    & Yes \\
    2021  & T5-Learning & ICSE  & Encoder-Decoder & 60M   & SentencePiece & T5    & USI   & Yes \\
    2022  & AlphaCode & Science & Encoder-Decoder & 300M,3B,9B,41B & SentencePiece & Scratch & DeepMind   & No \\
    2022  & BLOOM & NeurIPS & Encoder-Decoder & 770M  & BBPE  & CodeT5 & Salesforce   & Yes \\
    2022  & CodeRL & ASE   & Encoder-Decoder & 220M  & BBPE  & CodeT5 & UTexas   & Yes \\
    2022  & CoditT5 & ACL   & Encoder-Decoder & 560M  & SentencePiece & Scratch & Baidu   & Yes \\
    2022  & JuPyT5 & arXiv & Encoder-Decoder & 300M  & BBPE  & PyMT5 & Microsoft   & Yes \\
    2022  & PaLM-Coder & arXiv & Decoder-only & 8B,62B,540B & SentencePiece & Scratch & Google   & No \\
    2022  & PanGu-Coder & arXiv & Decoder-only & 317M,2.6B & SentencePiece & Scratch & Huawei   & No \\
    2022  & PolyCoder & ICLR  & Decoder-only & 41B,160M,400M & BPE   & GPT-2 & CMU   & Yes \\
    2022  & PyCodeGPT & IJCAI & Decoder-only & 110M  & BPE   & GPT-Neo & Microsoft, CAS   & Yes \\
    2022  & SPT-Code & ICSE  & Encoder-Decoder & 262M  & BPE   & Scratch & Nanjing University   & Yes \\
    2022  & UnixCoder & ACL   & Encoder-Decoder & 125M  & WordPiece & Scratch & Microsoft, SYSU   & Yes \\
    2023  & CCT5  & arXiv & Decoder-only & 13B,7B,34B & SentencePiece & Llama2 & Meta   & Yes \\
    2023  & Code Llama & ICSE  & Decoder-only & 13B   & BPE   & GPT-NeoX & Ant Group   & Yes \\
    2023  & CodeFuse & KDD   & Decoder-only & 13B   & BBPE  & GPT-2 & Tsinghua, Zhipu.AI   & Yes \\
    2023  & CodeGen & ICLR  & Decoder-only & 350M,2.7B,6.1B,16.1B & BPE   & Scratch & Salesforce   & Yes \\
    2023  & CodeGen2 & ICLR  & Decoder-only & 16B,3.7B,7B & BPE   & Scratch & Salesforce   & Yes \\
    2023  & CodeShell & arXiv & Decoder-only & 7B    & BPE   & GPT-2 & PKU   & Yes \\
    2023  & CodeT5+ & EMNLP & Encoder-Decoder & 770M,2B,6B,16B & BBPE  & CodeT5 & Salesforce   & Yes \\
    2023  & ERNIE-Code & ICLR  & Decoder-only & 6.7B,1.3B & BBPE  & Scratch & CMU, Meta, UW   & Yes \\
    2023  & InCoder & ICLR  & Decoder-only & 70B   & BBPE  & Llama2 & HKU   & Yes \\
    2023  & PanGu-Coder2 & ICML  & Decoder-only & 7B    & BPE   & CodeLlama,DeepSeekCoder & UIUC   & Yes \\
    2023  & PPOCoder & ICLR  & Decoder-only & 16B   & BBPE  & StarCoder & Hugging Face   & Yes \\
    2023  & RLTF  & ICLR  & Decoder-only & 6B    & BBPE  & CodeGeeX2 & Hugging Face   & Yes \\
    2023  & SantaCoder & arXiv & Decoder-only & 15B   & SentencePiece & PanGu-Coder & Huawei   & No \\
    2023  & SOBertBase & TMLR  & Encoder-Decoder & 770M,220M & BBPE  & CodeT5 & Virginia Tech   & Yes \\
    2023  & StarCoder & TMLR  & Encoder-Decoder & 770M  & N.A.  & CodeT5 & Tencent   & Yes \\
    2024  & AlchemistCoder & ICLR  & Decoder-only & 1.1B  & BPE   & Scratch & Hugging Face   & Yes \\
    2024  & AST-T5 & TMLR  & Decoder-only & 15.5B & BBPE  & GPT-2 & Hugging Face   & Yes \\
    2024  & AutoCoder & ICLR  & Decoder-only & 15B   & BBPE  & StarCoder & Microsoft, HKBU   & Yes \\
    2024  & B-Coder & arXiv & Decoder-only & 7B,6.7B & N.A.  & \makecell{CodeLlama,Llama2,\\DeepSeekCoder} & Tongji University   & Yes \\
    2024  & CodeGeeX & ICML  & Encoder-Decoder & 277M  & BPE   & T5    & UC Berkeley, Meta   & Yes \\
    2024  & CodeGemma & arXiv & Decoder-only & 6.7B,33B & BPE   & DeepSeekCoder & UCONN & Yes \\
    2024  & CodeQwen1.5  & ICLR  & Encoder-Decoder & 770M  & BBPE  & T5    & UIC   & Yes \\
    2024  & CodeSage & arXiv & Decoder-only & 2B,7B & BPE   & Gemma & DeepMind   & Yes \\
    2024  & CoLSBERT & arXiv & Decoder-only & 7B    & BPE   & Qwen1.5 & Alibaba & Yes \\
    2024  & DeepSeek-Coder & ICLR  & Encoder-only & 130M,356M,1.3B & SentencePiece & Scratch & AWS AI Labs   & Yes \\
    2024  & DeepSeek-Coder-V2 & arXiv & Encoder-only & 354M,757M,124M,1.5B & BPE   & Scratch & IDEA   & Yes \\
    2024  & DolphCoder & arXiv & Decoder-only & 1.3B,6.7B,33B & BPE   & Scratch & PKU   & Yes \\
    2024  & GrammarT5 & arXiv & Decoder-only & 16B,236B & BPE   & DeepSeek-V2 & PKU   & Yes \\
    2024  & Granite & ACL   & Decoder-only & 7B,13B & N.A.  & CodeLlama & BUPT   & Yes \\
    2024  & InverseCoder & arXiv & Decoder-only & 3B,8B,20B,34B & BPE   & Scratch & IBM   & Yes \\
    2024  & Lemur & arXiv & Decoder-only & 6.7B,7B & N.A.  & CodeLlama,DeepSeekCoder & UCAS, CAS   & Yes \\
    2024  & Magicoder & arXiv & Decoder-only & 1.1B  & BBPE  & StarCoder & Infosys Limited   & Yes \\
    2024  & NT-Java & arXiv & Decoder-only & 3B,7B,15B & BBPE  & StarCoder & Hugging Face   & Yes \\
    2024  & OctoCoder & ACL   & Decoder-only & 6.7B  & N.A.  & DeepSeekCoder & Fudan   & Yes \\
    2024  & OctoGeeX & FSE   & Encoder-Decoder & 220M  & BBPE  & CodeT5 & NUDT & Yes \\
    2024  & StarCoder2 & arXiv & Decoder-only & 176B  & BPE   & Scratch & BigScience Workshop & Yes \\
    2024  & StepCoder & arXiv & Encoder-only & 109M,762M & BPE   & Scratch & CMU   & No \\
    2024  & UniCoder & ACL   & Decoder-only & 6.7B,7B & BPE   & DeepSeekCoder & BUAA   & Yes \\
    2024  & WaveCoder & ACL   & Decoder-only & 15B,7B,6.7B,13B & BBPE  & \makecell{StarCoder,CodeLlama,\\DeepSeekCoder} & Microsoft   & Yes \\
    2024  & WizardCoder & ACL   & Decoder-only & 1.3B  & BPE   & DeepSeekCoder & UIUC   & Yes \\
    2024  & XFT   & ICSE  & Encoder-Decoder & 60M,220M & BBPE  & CodeT5 & PKU   & Yes \\
    \bottomrule
    \end{tabular}%
    }
  \label{tab:rq1_llm}
\end{table*}%

Typically, existing LLMs can be classified into three types according to the model architecture, \ie encoder-only, decoder-only, and encoder-decoder models.
Table~\ref{tab:rq1_llm} presents the summary and comparison of these representative LLMs\footnote{We note that some general-purpose LLMs (not limited to source code) have been applied in the field of SE, such as PaLM, Qwen, LLaMA, and Gemini. However, such LLMs have already been extensively reviewed in NLP and AI communities, so they fall outside the scope of our work. For more details, please refer to the relevant works~\cite{zhao2023survey,yin2023survey}.}.
The columns summarize the year of release, model name, the publisher or the conference where the model is introduced, the architecture types, and the initialization method or the base model used for pre-training.

From Table~\ref{tab:rq1_llm}, it can be found that these LLMs are usually derived from foundational architectures in the NLP community and trained with some code-aware objectives.
A considerable number of LLMs (\eg CodeBERT and CodeT5) are introduced by leading companies (\eg Microsoft and Google).
The possible reason is that the resources to train these highly parametric models and to collect vast datasets far exceed the capabilities of the academic community.
Third, inspired by the success of foundational LLMs like ChatGPT, the size of model parameters continues to set new benchmarks, and decoder-only architectures are gaining increasing popularity. 
In the following, we summarize these LLMs according to their model architectures.
More detailed information about these specific LLMs is included in Appendix~A.

\subsubsection{Encoder-only LLMs}
\label{sec:rq1.1.1}

Encoder-only LLMs refer to a class of LLMs that utilize only the encoder stack of the Transformer architecture. 
Regarding architecture, encoder-only models use multiple layers of encoders and each encoder layer consists of a multi-head self-attention mechanism followed by feed-forward neural networks.
Regarding training, encoder-only LLMs are typically pre-trained on a massive corpus using a masked language modeling (MLM) task, which is used to learn to predict the identity of masked words based on their context. 
Regarding usage, because encoder-only LLMs generate fixed-size representations for variable-length input text, they are particularly suited for tasks that require understanding the context or meaning of a piece of text without generating new text, such as code search and vulnerability detection.

Among various encoder-only LLMs, BERT~\cite{devlin2018bert} has been acknowledged as a foundational work in the NLP field and provides crucial guidance for the conception and development of follow-up code-related LLM works.
For example, CuBERT~\cite{kanade2020learning} is the first adaption of BERT from NLP to the source code domain by replicating the training procedure of BERT on a Python code corpus.
CodeBERT~\cite{feng2020codebert} is a bimodal variant of BERT that takes into account both natural language and programming language.
GraphCodeBERT~\cite{guo2020graphcodebert} is a structure-aware extension of CodeBERT that incorporates data flow graphs to capture the structural and semantic relationships within the source code.

\subsubsection{Encoder-decoder LLMs}
\label{sec:rq1.1.2}

Encoder-decoder LLMs refer to a class of LLMs that utilize both the encoder and decoder parts of the Transformer architecture, working in tandem to transform one sequence into another.
In particular, the encoder takes the input sequence and compresses its information into a fixed-size hidden state, which can capture the essence or meaning of the input sequence,
while the decoder takes the hidden state and produces the corresponding output sequence, step by step, often using attention mechanisms to refer back to parts of the input sequence as needed.
Thus, this architecture is particularly suited for sequence-to-sequence tasks in NLP and SE, where the input and output sequences can be of different lengths and structures, such as code summarization and program repair.

Among existing encoder-decoder LLMs, T5 (Text-to-Text Transfer Transformer) is a significant development in the NLP field and serves as a catalyst for follow-up code-related works.
For example, similar to CuBERT~\cite{kanade2020learning} in the encoder-only LLM domain, PYMT5~\cite{clement2020pymt5} is the first attempt to apply T5 to source code by replicating the pre-training process of T5 on a code corpus.
In parallel with PYMT5~\cite{clement2020pymt5}, T5-learning~\cite{mastropaolo2021studying} empirically investigate how T5 performs when pre-trained with CodeSearchNet and fine-tuned to support four code-related tasks.
PLABRT~\cite{ahmad2021unified} is pre-trained with the denoising objective and built on the BART architecture.
CodeT5~\cite{feng2020codebert} represents a well-known adaption of T5 from NLP to the source code domain by leveraging the code semantics from the developer-assigned identifiers.
CoditT5~\cite{zhang2022coditt5} is a variant of CodeT5 particularly trained to tackle code editing tasks, such as code review.
Researchers also release some encoder-decoder LLM for specific scenarios, such as AlphaCode~\cite{li2022competition} is introduced by DeepMind to generate solutions for competitive programming problems and JuPyT5~\cite{chandel2022training} for Jupyter Notebook.

\subsubsection{Decoder-only LLMs}
\label{sec:rq1.1.3}

Decoder-only LLMs refer to a class of LLMs that utilize only the decoder portion of the Transformer architecture. 
Unlike encoder-decoder models, which map an input sequence to an output sequence, decoder-only models primarily focus on generating text based on a given context or prompt.
In particular, decoder-only models use multiple layers of decoders from the Transformer architecture. 
Each decoder layer consists of a multi-head self-attention mechanism followed by feed-forward neural networks.
These models are designed to generate text autoregressively, meaning they produce one token at a time and use what has been generated so far as context for subsequent tokens.

Among existing decoder-only LLMs,  GPT (Generative Pre-trained Transformer) and its subsequent versions (like GPT-2, GPT-3, and so on) are the most well-known examples of decoder-only LLMs.
Early efforts focus on adapting the GPT series models for the code domain, leading to LLMs such as GPT-C~\cite{svyatkovskiy2020intellicode}, CodeGPT~\cite{lu2021codexglue}, PolyCoder~\cite{xu2022systematic}, and PyCodeGPT~\cite{zan2022cert}.
Subsequent advancements have seen the introduction of various specialized LLMs tailored for code generation and understanding tasks with advanced training techniques and large-scale datasets.
Examples include CodeGen~\cite{nijkamp2022codegen}, InCoder~\cite{fried2022incoder}, SantaCoder~\cite{allal2023santacoder}, StarCoder~\cite{li2023starcoder}, CodeGeeX~\cite{zheng2023codegeex}, and CodeGen2~\cite{nijkamp2023codegen2}.
Recently, with the development of general-purpose LLMs, several variants have been released specifically in handling code-related tasks, such as CodeLlama~\cite{roziere2023code}, PaLM-Coder~\cite{chowdhery2022palm}, PanGu-Coder~\cite{christopoulou2022pangu}, PanGu-Coder2~\cite{shen2023pangu}.

\finding{1.1}{
Overall, existing LLMs are mainly developed along three directions, the encoder-decoder represented by Google's T5, the encoder-only represented by Microsoft's BERT, and the decoder-only represented by OpenAI's GPT.
While different model architectures excel in their own areas, it is challenging to pinpoint a single best LLM for all tasks.
For example, encoder-only models (like BERT) focus on representing input text and are typically not used for sequence generation tasks, while decoder-only models (like GPT) are primarily used for generating sequences of text without a separate encoding step.
}

\subsection{RQ1.2: How are LLMs used in pre-training tasks?}
\label{sec:rq1.2_pre_training}

\begin{table*}[t]
\footnotesize
  \centering
  \renewcommand{\arraystretch}{1.}
  \caption{A summary and comparison of pre-training objectiveness in existing \clm{}}
    \begin{tabular}{c|c}
    \toprule
    \textbf{Category} & \textbf{Task} \\
    \midrule
    \multirow{7}{*}{Code Sequence Modeling and Prediction} & Causal Language Modeling (CLM) \\
          & Masked Language Modeling (MLM) \\
          & Masked Span Prediction (MSP) \\
          & Masked Identifier Prediction (MIP) \\
          & Replaced Token Detection (RTD) \\
          & Modified Masked Sequence-to-Sequence (MASS) \\
          & Span Denoising (SD) \\
    \midrule
    \multirow{2}{*}{\makecell{Bidirectional Understanding and\\ Generation between Code and Natural Language}} & Bimodal Dual Generation (BDG) \\
          & The Method Name Generation (MNG)  \\
    \midrule
    \multirow{4}{*}{Code Structure and Relationship Understanding} & Identifier Tagging (IT) \\
          & Edge Prediction (EP) \\
          & Node Alignment (NA) \\
          & Code-AST Prediction (CAP) \\
    \midrule
    \multirow{2}{*}{                        Cross-modal Representation Learning} & Multi-modal Contrastive Learning (MCL) \\
          & Cross-modal generation (CMG) \\
    \bottomrule
    \end{tabular}%
  \label{tab:rq1_pre_training}%
\end{table*}%

In this section, we summarize some representative pre-training tasks utilized to train \clm{} in the literature. 
Table~\ref{tab:rq1_pre_training} categorizes pre-training tasks into four major classes, including code sequence modeling and prediction in Section~\ref{sec:rq1.2.1_modelling}, bidirectional understanding and generation in Section~\ref{sec:rq1.2.2_bidir}, code structure and relationship understanding in Section~\ref{sec:rq1.2.3_structure} and cross-modal representation learning in Section~\ref{sec:rq1.2.4_cross}.
Now, we list and summarize these pre-training tasks as follows.

\subsubsection{Code Sequence Modeling and Prediction}
\label{sec:rq1.2.1_modelling}

Such tasks involve predicting and completing code fragments, such as masked spans or identifiers, so as to enhance LLMS' ability to understand and fill in missing parts of code.

\textbf{Causal Language Modeling (CLM)}.
CLM\footnote{The training objective is called causal language modeling in LLMs such as CodeGen2, but also referred to as next token prediction in LLMs such as CodeGen, and unidirectional language modeling in LLMs such as UniXcoder.} attempts to predict the next most probable token in a sequence based on the context provided by the previous tokens.
Such a task has usually been utilized to train decoder-only LLMs (\eg CodeGen and CodeGPT) to generate complete programs from the beginning to the end for supporting auto-regressive tasks, such as code completion.
For a sequence $x = (x_1, \dots, x_n)$ with $n$ tokens, the task is to predict the token $x_i$ given previous tokens ($xj: j<i$).
For example, CLM predicts \texttt{x} for the given piece of incomplete code \texttt{int add(int x, int y)\{ return}.

\textbf{Masked Language Modeling (MLM)}.
MLM attempts to predict the original masked word from an artificially masked input sequence and is utilized in encoder-only LLMs such as CodeBERT.
Similar to the original BERT, 15\% of the code tokens from the input sequence are masked out.
As the prediction of masked tokens is made based on the bidirectional contextual tokens, LLMs need to take into account the tokens forward and backward from the masked token in the input sequence.  
MLM is instrumental in training the model to comprehend not merely isolated code tokens, but also the relationships between tokens within a piece of code.

\textbf{Masked Span Prediction (MSP)}.
MSP attempts to predict the masked code tokens in the input code snippet and is utilized in encoder-decoder LLMs such as CodeT5.
As mentioned in CodeT5~\cite{wang2021codet5}, MSP randomly masks spans with arbitrary lengths and then predicts these masked spans combined with some sentinel tokens at the decoder.
The input of LLMs is the original sequence, the mask sequence is processed by the noise function, and the output is the denoised sequence.

\textbf{Masked Identifier Prediction (MIP)}.
Instead of randomly masking spans like in MSP, MIP masks all identifiers in the code snippet, using a unique sentinel token for each different mask.
Inspired by the insight that changing identifier names does not impact code semantics, LLMs are tasked to predict the original identifiers from the masked input in an auto-regressive manner.
MIP is a more challenging task as it requires the model to comprehend the code semantics based on obfuscated code and link the occurrences of the same identifiers together.

\textbf{Replaced Token Detection (RTD)}.
Originally proposed by Clark~\etal~\cite{clark2020electra}, RTD attempts to predict whether a word is the original word or not, and is utilized in LLMs such as CodeBERT.
RTD replaces the original word at the location of the mask with an alternative text, and perform a binary classification problem by training a discriminator to determine if a word is the original one. 
The discriminator is trained as a binary classifier to distinguish between original and generated tokens. 
The process involves sampling alternative tokens $\hat{w}_i$ from $p_{Gw}(w_i|w_{masked})$ for positions $i$ in $m_w$, and sampling alternative tokens $\hat{c}_i$ from $p_{Gc}(c_i|c_{masked})$ for positions $i$ in $m_c$. 
Then, the corrupted input $x_{corrupt}$ is formed by replacing the masked words in $w$ and $c$ with their corresponding alternatives.
The RTD objective aims to improve the efficiency of training by replacing masked tokens with plausible alternatives, enabling the model to benefit from both bimodal and unimodal data during the learning process.

\textbf{Modified Masked Sequence-to-Sequence (MASS)}.
MASS attempts to reconstruct a sentence fragment by predicting the masked tokens in the encoder-decoder model architectures and is utilized in LLMs such as SPT-Code.
Given a code snippet $C$, the modified version $C^{u:v}_{origin}$ is obtained by masking the fragment from position $u$ to $v$.
The model is pre-trained using this modified version to predict the fragment of $C$ from $u$ to $v$
As a result, the model learns to predict masked parts of code sequences, enhancing its ability to understand and generate complex code structures accurately.

\textbf{Span Denoising (SD)}.
SD involves randomly masking a span of tokens in the input and then training the model to reconstruct the original tokens, and is utilized in LLMs such as CodeT5+.
In the SD task, 15\% of the tokens in the encoder inputs are randomly replaced with indexed sentinel tokens, and the decoder is required to recover these masked tokens by generating a combination of spans.
SD helps in learning deeper contextual representations of code snippets, enhancing LLMs' understanding of language structure and semantics.
In CodeT5+, spans are sampled for masking, where the span lengths are determined by a uniform distribution with a mean of 3, so as to avoid masking partial words and enhance the model's understanding of whole words in the code.

\subsubsection{Bidirectional Understanding and Generation between Code and Natural Language}
\label{sec:rq1.2.2_bidir}

Such tasks involve the conversion and understanding between source code and natural language, including generating method names.

\textbf{Bimodal Dual Generation (BDG)}.
BDG attempts to perform bidirectional translation between PL and NL and is utilized in LLMs such as CodeT5.
Specifically, the NL-$>$PL generation and PL-$>$NL generation are treated as dual tasks, and LLMs are optimized simultaneously on both tasks. 
For each NL-$>$PL bimodal data point, two training instances are created with reverse directions, and language identifiers (\eg \texttt{\textless java\textgreater} and \texttt{\textless en\textgreater} for Java PL and English NL, respectively) are included. 
The main objective of BDG is to enhance the alignment between the NL and PL, so as to generate syntactically correct NL descriptions for code snippets and code snippets for NL queries in downstream tasks.

\textbf{Method Name Generation (MNG)}.
MNG attempts to leverage method names to enhance LLMs' understanding of code intent and functionality, and has been utilized in LLMs such as SPT-Code.
In the MNG task, the model takes the input representation, denoted as $\text{Input} = C,[\text{SEP}],A,[\text{SEP}],N$, where $C$ is the code snippet, $A$ is the corresponding AST sequence, and $N$ is a natural language.

\subsubsection{Code Structure and Relationship Understanding}
\label{sec:rq1.2.3_structure}

Such tasks usually involve understanding the structure and relationships within source code, including the arrangement of code elements and their connections.

\textbf{Identifier Tagging (IT)}. 
IT attempts to make LLMs learn whether a code token is an identifier or not and is utilized in LLMs such as CodeT5~\cite{wang2021codet5}, which is inspired by the syntax highlighting in some coding tools.
IT can help LLMs to learn the code syntax and the data flow structures of source code.

\textbf{Edge Prediction (EP).}
EP attempts to learn representations from data flow in the context of code understanding and is utilized in LLMs such as GraphCodeBERT~\cite{guo2020graphcodebert}.
The primary motivation behind this task is to encourage the model to learn structure-aware representations that capture the relationships of "where-the-value-comes-from" in the code, thus enhancing its ability to comprehend code.
In this pre-training task, a graph representing the data flow is constructed, where nodes represent variables or data elements, and edges represent the flow of data between these nodes.
The objective is to predict the edges that are masked (hidden) in the graph. 
To do this, approximately 20\% of the nodes in the data flow graph are randomly sampled, and the direct edges connecting these sampled nodes are masked by adding an infinitely negative value in the mask matrix.

\textbf{Node Alignment (NA).} 
Similar to EP, NA attempts to align representations between source code and data flow, and is utilized in LLMs such as GraphCodeBERT~\cite{guo2020graphcodebert}.
This alignment helps the model better understand the relationships between code tokens and nodes in the data flow, leading to improved comprehension of code semantics.
In the NA pre-training task, a graph is constructed representing the data flow, and nodes in this graph represent variables or data elements. Additionally, code tokens in the source code are considered as another set of nodes. 
The objective is to predict the edges between code tokens and nodes, representing the alignment of variables in the code with their data flow counterparts.

\textbf{Code-AST Prediction (CAP).}
Inspired by the NSP task, CAP attempts to incorporate structural information of source code into the pre-training input and is utilized in LLMs such as SPT-Code~\cite{niu2022spt}.
The input of CAP includes code and its corresponding abstract syntax tree (AST) representation. 
The CAP task is formulated as a binarized task that can be easily generated from any given code.
In constructing the input representation, the format used is as [\text{Input} = \text{C}, [\text{SEP}], \text{A}, [\text{SEP}], \text{N}], where C represents the code snippet, A represents the corresponding AST sequence, and N is a natural language description.

\subsubsection{Cross-modal Representation Learning}
\label{sec:rq1.2.4_cross}

Such tasks involve understanding source code and other modalities (such as comments) together, enhancing the model's capabilities in understanding and representing code.

\textbf{Multi-modal Contrastive Learning (MCL)}.
MCL attempts to learn semantic embedding of code fragments by distinguishing between positive and negative samples, and has been utilized by LLMs such as UniXcoder~\cite{guo2022unixcoder}.
The positive sample refers to the same input but uses a different hidden dropout mask, while the negative sample refers to other representations in the batch. 
In UniXcoder~\cite{guo2022unixcoder}, MCL encodes the mapped AST sequence and then applies an average pooling layer on the hidden state of the source input to obtain semantic embedding.

\textbf{Cross-modal Generation (CMG)}.
CMG attempts to generate comments for code segments to aid LLMs in understanding the code semantics. 
The generation of the comment is conditioned on the code, integrating semantic information into the hidden states of code. 
Besides, to expose LLMs to diverse contexts, a strategy is employed where the source and target inputs are randomly swapped with a 50\% probability.

\finding{1.2}{
Overall, the recent trends of pre-training tasks for Code LLMs reflect a significant shift from early NLP-derived objectives towards more code-aware objectives.
Initially, LLMs are pre-trained with language modeling objectives from NLP tasks, including CLM for decoder-only LLMs (\eg CodeGPT), MLM for encoder-only LLMs (\eg CodeBERT), and MSP for encoder-decoder LLMs (\eg CodeT5).
The follow-up works evolve to consider code variables and structural features specifically, as well as cross-modal learning for source code and natural language.
This progression signifies a continuous advancement towards LLMs that not only process code as a sequence of tokens but deeply understand its semantic and functional aspects, bridging the gap between source code and natural language.
}

\subsection{RQ1.3: How are LLMs used in downstream tasks?}
\label{sec:rq1.3_downstream}

Once LLMs are trained on a vast corpus, it is critical to evaluate the effectiveness and applicability of LLMs on downstream tasks.
Fine-tuning is the primary method for transferring the knowledge acquired during pre-training to downstream tasks, requiring LLMs to demonstrate code understanding, reasoning, and generation capabilities.
A downstream task can be categorized by the task type (\ie code understanding and code generation) or data type (\ie code-code, code-text, text-code, and code-labels).
We summarize 15 representative downstream tasks that are evaluated by existing LLMs in their original papers according to a well-maintained repository\footnote{\url{https://microsoft.github.io/CodeXGLUE/}}, detailed as follows.

\textbf{Code-Code}.
Code-code tasks involve the process of transforming one code snippet into another.
For example, code translation attempts to convert code from one programming language into another while preserving its functionality.
This task has been adopted as a downstream task in LLMs like CodeT5~\cite{wang2021codet5}, CodeBERT~\cite{feng2020codebert}, and CodeT5~\cite{wang2021codet5}.
Code refinement (also known as program repair) attempts to refine existing code that might contain errors, and has been explored in LLMs like CodeT5~\cite{wang2021codet5}, GraphCodeBERT~\cite{guo2020graphcodebert} and SPT-Code~\cite{niu2022spt}.
Cloze test aims to predict a missing token in a code snippet and has been adopted in LLMs like CodeGPT~\cite{zheng2023codegeex}.
Mutant generation attempts to generate mutants by introducing small artificial faults, such as replacing the \texttt{+} operator with \texttt{-}, and has been adopted in LLMs like T5-learning~\cite{mastropaolo2021studying}.
Assert generation generates assert statements to verify the correctness of programs and validate certain assumptions, and has been adopted in LLMs, like T5-learning~\cite{mastropaolo2021studying}.

\textbf{Text-Code}.
Text-code tasks involve the process of transforme human language descriptions into code snippets.
For example, 
Code generation attempts to directly produce code snippets based on natural language descriptions, such as docstrings.
It has been widely adopted as a downstream task in LLMs, including PyMT5~\cite{clement2020pymt5}, CodeT5~\cite{wang2021codet5}, Codex~\cite{chen2021evaluating}.
Code search refers to the retrieval of relevant code samples from a codebase that matches a given natural language query, and haven been adopted in LLMs like CodeT5+~\cite{wang2023codet5+}, CodeGPT~\cite{lu2021codexglue}, UnixCoder~\cite{guo2022unixcoder} and SPT-Code~\cite{niu2022spt}.

\textbf{Code-Text}.
Code-text tasks involve the process of transforme code snippets into human language descriptions.
For example, code summarization is the task of automatically generating a concise and accurate natural language description, or docstring, that encapsulates the actions and purpose of a given source code snippet.
It has been explored in various LLMs, including CodeT5~\cite{wang2021codet5}, GraphCodeBERT~\cite{guo2020graphcodebert}, PLBART~\cite{ahmad2021unified}, CodeGPT~\cite{lu2021codexglue}, UnixCoder~\cite{guo2022unixcoder}, SPT-Code~\cite{niu2022spt} and ERNIE-Code~\cite{chai2022ernie}.

\textbf{Code-Label}.
Code-label tasks involve the process of performing classifications of code snippets.
For example, clone detection identifies whether two code snippets are functionally or semantically similar based on similarity analysis and has been adopted in LLMs like CodeBERT~\cite{feng2020codebert}, CodeT5~\cite{wang2021codet5} and CodeT5+~\cite{wang2023codet5+}.
Defect detection predicts whether a piece of source code contains bugs that could potentially make software systems vulnerable to attacks and has been adopted in LLMs like CodeT5~\cite{wang2021codet5}, PLBART~\cite{ahmad2021unified}, and CodeGPT~\cite{lu2021codexglue}.

\finding{1.3}{
Overall, as the direct applications of LLMs, these downstream tasks can be categorized into four classes according to input-output types, \ie code-code, code-test, test-code and code-labels, or into two classes according to task types, \ie code understanding and code generation.
We observe some trends in a majority of existing downstream tasks where LLMs can be directly applied.
First, these tasks involve only code snippets or the corresponding natural language comments.
Second, these tasks are usually evaluated automatically using well-designed metrics (\eg BLUE for generation tasks and Accuracy for classification tasks), thus supporting the large-scale evaluation benchmarks.
Third, these tasks can effectively reduce the programming efforts of developers and can be integrated into modern IDEs as plug-ins to aid programming. 
Finally, these tasks have received attention and have been investigated in both the fields of SE and artificial intelligence.
LLMs have shown preliminarily promising results on these tasks, importantly indicating their potential in a wider and more in-depth range of SE tasks, detailed in Section~\ref{sec:rq2_se}.
}

\subsection{RQ1.4: How are LLMs open-sourced to support the open science community?}
\label{sec:rq1.4_analysis}

\begin{table*}[t]
  \centering
  \footnotesize
  \renewcommand{\arraystretch}{1.}
  \caption{The Details of Code LLMs Availability}
  \resizebox{\textwidth}{!}{
    \begin{tabular}{l|lllll|l}
    \toprule
    Model & Hosting Site & CA & DA & MA & Model Site & URL only for Public \\
    \midrule
    CodeBERT & GitHub & Yes   & Yes   & Yes   & Hugging Face & \url{https://github.com/microsoft/CodeBERT} \\
    CuBERT & GitHub & Yes   & Yes   & Yes   & Google Cloud & \href{https://github.com/google-research/google-research/tree/master/cubert}{https://github.com/google-research/google-research} \\
    PyMT5 & GitHub & Yes   & Yes   & Yes   & PyPi  & \url{https://github.com/devcartel/pymt5} \\
    CodeGPT & GitHub & No    & No    & Yes   & Hugging Face & \url{https://github.com/microsoft/CodeXGLUE} \\
    CodeT5 & GitHub & Yes   & Yes   & Yes   & Hugging Face & \url{https://github.com/salesforce/CodeT5} \\
    GraphCodeBERT & GitHub & Yes   & Yes   & Yes   & Hugging Face & \url{https://github.com/microsoft/CodeBERT} \\
    PLBART & GitHub & Yes   & Yes   & Yes   & Google Drive & \url{https://github.com/wasiahmad/PLBART} \\
    T5-Learning & GitHub & Yes   & Yes   & Yes   & Google Drive & \url{https://github.com/antonio-mastropaolo/T5-learning-ICSE\_2021} \\
    CodeRL & GitHub & Yes   & Yes   & Yes   & Google Cloud & \url{https://github.com/salesforce/CodeRL} \\
    CoditT5 & GitHub & Yes   & Yes   & Yes   & Hugging Face & \url{https://github.com/engineeringsoftware/coditt5} \\
    JuPyT5 & GitHub & No    & No    & No    & N.A.    & \url{https://github.com/microsoft/DataScienceProblems} \\
    PolyCoder & GitHub & No    & Yes   & Yes   & Hugging Face & \url{https://github.com/VHellendoorn/Code-LMs} \\
    PyCodeGPT & GitHub & No    & Yes   & Yes   & Hugging Face & \url{https://github.com/microsoft/pycodegpt} \\
    SPT-Code & GitHub & Yes   & Yes   & Yes   & OneDrive & \url{https://github.com/NougatCA/SPT-Code} \\
    UnixCoder & GitHub & No    & Yes   & Yes   & Hugging Face & \url{https://github.com/microsoft/CodeBERT/tree/master/UniXcoder} \\
    Code Llama & GitHub & Yes   & No    & Yes   & Hugging Face & \url{https://github.com/facebookresearch/codellama} \\
    CodeFuse & GitHub & Yes   & No    & Yes   & Hugging Face & \url{https://github.com/codefuse-ai} \\
    CodeGen & GitHub & No    & Yes   & Yes   & Hugging Face & \url{https://github.com/salesforce/CodeGen} \\
    CodeGen2 & GitHub & No    & Yes   & Yes   & Hugging Face & \url{https://github.com/salesforce/CodeGen2} \\
    CodeT5+ & GitHub & No    & Yes   & Yes   & Hugging Face & \url{https://github.com/salesforce/CodeT5/tree/main/CodeT5\%2B} \\
    ERNIE-Code & GitHub & Yes   & No    & No    & N.A.    & \href{https://github.com/PaddlePaddle/PaddleNLP/tree/develop/model\_zoo/ernie-code}{https://github.com/PaddlePaddle/PaddleNLP/} \\
    InCoder & GitHub & No    & Yes   & Yes   & Hugging Face & \url{https://sites.google.com/view/incoder-code-models} \\
    PPOCoder & GitHub & Yes   & Yes   & Yes   & N.A.  & \url{https://github.com/reddy-lab-code-research/PPOCoder} \\
    RLTF  & GitHub & Yes   & Yes   &       & Hugging Face & \url{https://github.com/Zyq-scut/RLTF} \\
    SantaCoder & Hugging Face & No    & Yes   & Yes   & Hugging Face & \url{https://huggingface.co/bigcode/santacoder} \\
    StarCoder & GitHub & No    & No    & No    & N.A.  & \url{https://github.com/bigcode-project/starcoder} \\
    AlchemistCoder & GitHub & Yes   & Yes   & Yes   & Hugging Face & \url{https://github.com/InternLM/AlchemistCoder} \\
    AST-T5 & GitHub & Yes   & Yes   & Yes   & Hugging Face & \url{https://github.com/gonglinyuan/ast\_t5} \\
    AutoCoder & GitHub & No    & No    & Yes   & Hugging Face & \url{https://github.com/bin123apple/AutoCoder} \\
    CodeGeeX & GitHub & No    & No    & Yes   & N.A.  & \url{https://github.com/THUDM/CodeGeeX} \\
    CodeGemma & Hugging Face & No    & No    & No    & Hugging Face & \url{https://huggingface.co/blog/codegemma} \\
    CodeQwen1.5  & GitHub & No    & No    & Yes   & Hugging Face & \url{https://github.com/QwenLM/CodeQwen1.5} \\
    CodeSage & GitHub & Yes   & Yes   & Yes   & Hugging Face & \url{https://github.com/amazon-science/CodeSage} \\
    CoLSBERT & GitHub & Yes   & Yes   & Yes   & N.A.  & \url{https://github.com/stanford-futuredata/ColBERT} \\
    DeepSeek-Coder & GitHub & Yes   & Yes   & Yes   & Hugging Face & \url{https://github.com/deepseek-ai/DeepSeek-Coder} \\
    DeepSeek-Coder-V2 & GitHub & No    & Yes   & Yes   & Hugging Face & \url{https://github.com/deepseek-ai/DeepSeek-Coder-V2} \\
    DolphCoder & GitHub & No    & No    & No    & N.A.  & \url{https://github.com/pris-nlp/DolphCoder} \\
    CCT5  & GitHub & Yes   & No    & Yes   & Zenodo & \url{https://github.com/Ringbo/CCT5} \\
    BLOOM & Hugging Face & Yes   & Yes   & Yes   & Hugging Face & \url{https://huggingface.co/bigscience/bloom} \\
    CoLSBERT & GitHub & Yes   & Yes   & Yes   & N.A.  & \url{https://github.com/stanford-futuredata/ColBERT} \\
    GrammarT5 & GitHub & Yes   & Yes   & Yes   & N.A.  & \url{https://github.com/pkuzqh/GrammarT5} \\
    Granite & GitHub & No    & Yes   & Yes   & Hugging Face & \url{https://github.com/ibm-granite/granite-code-models} \\
    InverseCoder & GitHub & Yes   & Yes   & Yes   & Hugging Face & \url{https://github.com/wyt2000/InverseCoder} \\
    Lemur & GitHub & Yes   & Yes   & Yes   & Hugging Face & \url{https://github.com/OpenLemur/Lemur} \\
    Magicoder & GitHub & Yes   & Yes   & Yes   & Hugging Face & \url{https://github.com/ise-uiuc/magicoder} \\
    NT-Java & Hugging Face & No    & Yes   & Yes   & Hugging Face & \url{https://huggingface.co/infosys/NT-Java-1.1B} \\
    OctoCoder & Hugging Face & Yes   & Yes   & Yes   & Hugging Face & \url{https://huggingface.co/bigcode/octocoder} \\
    OctoGeeX & Hugging Face & Yes   & Yes   & Yes   & Hugging Face & \url{https://huggingface.co/bigcode/octocoder} \\
    StarCoder2 & GitHub & Yes   & Yes   & Yes   & Hugging Face & \url{https://github.com/bigcode-project/starcoder2} \\
    StepCoder & GitHub & No    & Yes   & No    & N.A.  & \url{https://github.com/Ablustrund/APPS\_Plus} \\
    UniCoder & GitHub & Yes   & Yes   & Yes   & Google Drive & \url{https://github.com/microsoft/Unicoder} \\
    WaveCoder & GitHub & Yes   & Yes   & Yes   & Hugging Face & \url{https://github.com/microsoft/WaveCoder} \\
    WizardCoder & GitHub & Yes   & Yes   & Yes   & Hugging Face & \url{https://github.com/nlpxucan/WizardLM} \\
    XFT   & GitHub & Yes   & Yes   & Yes   & N.A.  & \url{https://github.com/ise-uiuc/xft} \\
    \bottomrule
    \end{tabular}%
    }
  \label{tab:rq1_llm_open}%
\end{table*}%

Very recently, the literature has seen a surge in the application of LLMs for a variety of SE problems. 
LLM brings a fresh perspective on the challenges associated with code-related tasks, shifting the focus from traditional learning-based and rule-based approaches to a new pre-training-and-fine-tuning paradigm.
However, this shift also presents unique reproducibility challenges, distinct from those in traditional studies. 
For example, training complex LLMs can require substantial computational resources, often exceeding what academic institutions and most businesses can provide. 
Besides, the need for extensive data collection and hyper-parameter tuning adds complexity and feasibility issues for replication.
Given these challenges, there is a growing imperative to adhere to open science principles in the LLM-driven SE field. 
Open science encourages researchers to share their artifacts (\eg datasets, trained models, scripts) with the broader research community, fostering reproducibility and free knowledge exchange.
While numerous LLMs have been proposed for automating code-related tasks with promising results, there is a need for more support to address the critical issue of open science. 
In particular, we investigate the extent to which LLMs make their artifacts publicly available and how they provide this information.

Among \numllm{} investigated LLMs, 54 of them provide the corresponding open-source repositories, which are summarized in Table~\ref{tab:rq1_llm_open}.
For each LLM we collect, we check whether an accessible link for its model or data is provided in the main text or footnotes of the paper.
We only present the studies that provide the link of publicly available data or tools due to limited space, listed in the first column.
The second column lists which hosting site the available artifact is uploaded to for public access (\eg GitHub).
The third column lists whether the source code (\eg training scripts) is available in the artifacts.
The fourth column lists whether the dataset (\eg raw data and training data) is available in the artifacts.
The fifth column lists whether the trained model is available in the artifacts, and the sixth column lists the corresponding site.
We also list the accessible URL links in the last column.
After carefully checking the collected papers, we find that 54 of 62 LLMs have made their artifacts available to the public.
Almost all studies upload their artifacts on Github, which is the most popular platform for hosting open-source code publicly.
Similar to GitHub, nearly all checkpoints of LLMs are hosted on Hugging Face, with which developers can conveniently download these trained models and conduct training or inference on their own machines.
Meanwhile, we find that several papers fail to provide the source code, dataset, or already trained models~\cite{wang2023codet5+}, perhaps due to commercial reasons.

\finding{1.4}{
Overall, compared with traditional DL studies, the need for high-quality artifacts in LLMs is even more vital for replication and future research.
While numerous LLms have been introduced for code-related tasks, there remains a significant gap in their adherence to open science principles.
On one hand, abundant training time and expensive equipment (\eg GPUs) are required to train LLMs, making it much harder to reproduce existing works.
On the other hand, some LLMs require complex environment settings (\eg the hyperparameters and the random seed) and high-quality datasets.
Therefore, we hope that researchers can provide high-quality open-source code and detailed instructions for convenient reproduction.
}

\section{RQ2: SE Perspective}
\label{sec:rq2_se}

\begin{table*}[t!]
  \centering
  \renewcommand{\arraystretch}{1.}
  \footnotesize
  \caption{Distribution of SE tasks over five SE activities}
  \resizebox{0.8\linewidth}{!}{
    \begin{tabular}{cllc}
    \toprule
    \textbf{SE Activity} & \multicolumn{2}{l}{\textbf{SE Task}} & \textbf{Total} \\
    \midrule
    \multirow{8}{*}{\makecell{Software Requirements \\ \& Design (Section~\ref{sec:rq2_se_requirement})}} & Ambiguity Detection (5) & Requirement Prioritization (1) & \multirow{8}[2]{*}{43} \\
          & Class Diagram Derivation (2) & Requirement Summarization (1) &  \\
          & GUI Layouts (2) & Requirement Traceability (2) &  \\
          & Requirement Classification (7) & Requirements Quality Assurance (4) &  \\
          & Requirement Completeness Detection (1) & Software Modeling (5) &  \\
          & Requirement Elicitation (7) & Specification Generation (3) &  \\
          & Requirement Engineering (1) & Specifications Repair (1) &  \\
          & Use Case Generation (1) &       &  \\
    \midrule
    \multirow{14}{*}{\makecell{Software Development\\ (Section~\ref{sec:rq2_se_development})}} & API Documentation Smells (1) & Identifier Normalization (1) & \multirow{14}[2]{*}{341} \\
          & API Inference (4) & Microservice Recommendation (1) &  \\
          & API Recommendation (6) & Neural Architecture Search (1) &  \\
          & Code Comment Completion (1) & Program Synthesis (13) &  \\
          & Code Completion (34) & SO Post Title Generation (2) &  \\
          & Code Compression (2) & Type Inference (1) &  \\
          & Code Editing (5) & Unified Development (1) &  \\
          & Code Generation (155) & Code Recommendation (1) &  \\
          & Code Representation (6) & Control Flow Graph Generation (2) &  \\
          & Code Search (19) & Data Analysis (1) &  \\
          & Code Summarization (41) & Method Name Generation (1) &  \\
          & Code Translation (24) & Project Planning (1) &  \\
          & Code Understanding (12) & SO Question Answering (2) &  \\
          & Continuous Development (1) & Data Augmentation (2) &  \\
    \midrule
    \multirow{16}{*}{\makecell{Software Testing\\ (Section~\ref{sec:rq2_se_testing})}} & Invariant Prediction (1) & GUI Testing (6) & \multirow{16}[2]{*}{240} \\
          & Proof Generation (1) & Indirect Call Analysis (1) &  \\
          & Resource Leak Detection  (1) & Mutation Testing (12) &  \\
          & Taint Analysis  (1) & NLP Testing (7) &  \\
          & Vulnerability Detection (76) & Penetration Testing (4) &  \\
          & Actionable Warning Identification  (1) & Program Analysis (1) &  \\
          & Adversarial Attack (3) & Program Reduction (1) &  \\
          & API Misuse Detection (1) & Property-based Testing (1) &  \\
          & API Testing (1) & Simulation Testing (1) &  \\
          & Assertion Generation (10) & Static Analysis (6) &  \\
          & Binary Code Similarity Detection (4) & Static Warning Validating (2) &  \\
          & Code Execution (1) & Test Generation (46) &  \\
          & Decompilation (8) & Test Suite Minimization (1) &  \\
          & Failure-Inducing Testing (1) & Dependency Alert Detection (1) &  \\
          & Fault Localization (17) & Theorem Proving (1) &  \\
          & Fuzzing (18) & Formal Verification (4) &  \\
    \midrule
    \multirow{17}{*}{\makecell{Software Maintenance\\ (Section~\ref{sec:rq2_se_maintenance})}} & Android Permissions (1) & APP Review Analysis (2) & \multirow{17}[2]{*}{211} \\
          & Bug Report Detection (5) & Code Clone Detection (10) &  \\
          & Bug Reproduction (4) & Code Coverage Prediction (1) &  \\
          & Bug Triaging (3) & Code Evolution (1) &  \\
          & Code Review (24) & Code Refactoring (4) &  \\
          & Code Smells (1) & Commit Message Generation (5) &  \\
          & Compiler Optimization (8) & Debugging (1) &  \\
          & Exception Handling Recommendation (1) & Flaky Test Prediction (1) &  \\
          & Incident Management (2) & Log Analysis (20) &  \\
          & Issue Labeling (2) & Mobile App Crash Detection (1) &  \\
          & Log Anomaly Detection (13) & Outage Understanding (1) &  \\
          & Malware Tracker (1) & Sentiment Analysis (4) &  \\
          & Patch Correctness Assessment (6) & Tag Recommendation (1) &  \\
          & Privacy Policy (1) & Technical Debt Management (1) &  \\
          & Program Repair (66) & Test Update (3) &  \\
          & Report Severity Prediction (3) & Traceability Link Recovery (1) &  \\
          & Vulnerability Repair (12) &       &  \\
    \midrule
    \multirow{2}{*}{\makecell{Software Management\\ (Section~\ref{sec:rq2_se_management})}} & Developers' Behavior Analysis (1) & Software Tool Configuration (1) & \multirow{2}[2]{*}{5} \\
          & Effort Estimation (2) & Software Repository Mining (1) &  \\
    \bottomrule
    \end{tabular}%
    }
  \label{tab:rq2}%
\end{table*}%

In this section, we summarize existing SE studies empowered with LLMs, which can be categorized into five crucial phases within the SE life cycle, including software requirements and design in Section~\ref{sec:rq2_se_requirement}, software development in Section~\ref{sec:rq2_se_development}, software testing in Section~\ref{sec:rq2_se_testing}, software maintenance in Section~\ref{sec:rq2_se_maintenance}, and software management in Section~\ref{sec:rq2_se_management}.
Each SE phase contains several distinct code-related tasks, such as fault localization and program repair in the software maintenance phase.
Table~\ref{tab:rq2} presents the taxonomy of this section, categorizing \numse{} LLM-based SE studies into \numcode{} distinct SE tasks across five SE phases.
More detailed information about all specific studies is included in Appendix~B.
\delete{
We observe that LLMs have been successfully applied to a majority of existing SE tasks and have been compared to traditional SE methods.
First, these tasks employ one or more LLMs (discussed in Section~\ref{sec:RQ1}) in the overall design phase, either partially or entirely. 
For example, in Program Repair, the CodeT5-base model is utilized to generate code prompts that fill in the masked sections, resulting in candidate patches.
Second, these tasks typically evaluate the performance using multiple LLM models against traditional or enhanced metrics and compare them with baseline methods.
Finally, these tasks show
the applicability of LLMs to various SE tasks. However, researchers need to design specific collaboration frameworks to enhance the performance of base LLMs or choose suitable LLMs to embed into SE tasks to cover a broader range of task templates, detailed in Section~\ref{sec:LLMs4SE}.
}

\subsection{Software Requirements \& Design}
\label{sec:rq2_se_requirement}
Software requirements refer to specific descriptions of conditions or capabilities needed by users, systems, or system components, typically presented in document form. 
These requirements are categorized into functional and non-functional requirements. 
The purpose of software requirements is to ensure that the developed software can meet the expectations of users and relevant stakeholders, as well as the conditions and capabilities specified in contracts, standards, regulations, or other formal documents.
Software design involves the process of defining the structure, components, functionalities, interfaces, and their relationships within a software system. 
During the software design phase, software engineers need to create detailed plans and design blueprints based on software requirements and specifications to ensure that the software system can meet the users' needs and expectations.

\subsubsection{Software Specifications Generation}
\label{sec:specifications_generation}

Software specifications generation refers to the automated process of deriving formal descriptions and requirements for software systems from unstructured data sources such as comments or documentation within the software's source code. Traditional techniques for extracting software specifications usually involve rule-based or machine learning-based methods that necessitate manual effort and domain knowledge, and have limited the ability to generalize across various domains.

LLMs provide a promising avenue for automating the process of software specifications generation. 
LLMs, which have been utilized successfully in numerous software engineering tasks, offer the potential in automatically extracting software specifications from textual information.
For example, Xie~\etal~\cite{xie2023impact} conduct the first empirical study to assess the capabilities of LLMs for generating software specifications from software comments or documentation. 
They employ few-shot learning techniques to enable LLMs to generalize from a limited number of examples and explore various prompt construction strategies.
This work also conducts a comparative diagnosis of failure cases between LLMs and traditional methods to identify their respective strengths and weaknesses.
Considering traditional methods relying on pre-defined templates or grammar rules, SpecGen~\cite{ma2024specgen} leverages the code comprehension capabilities of LLMs to generate formal program specifications. 
SpecSyn~\cite{mandal2023large} is a specification synthesis approach that treats this task as a sequence-to-sequence learning problem,  directly translating natural language input into formal specifications.

\subsubsection{Software Requirements Classification}
\label{sec:requirements_classification}

Requirement classification refers to the process of categorizing software requirements into different classes or types, such as functional and non-functional requirements. 
Functional requirements outline the functionalities and behaviors the software should achieve, while non-functional requirements cover broader system attributes such as performance, security, reliability, and maintainability. 
As early as 2020, Hey~\etal~\cite{9218141} propose NoRBERT, a BERT-based requirement classification approach for both functional and non-functional classes by transfer learning.
Khan~\etal~\cite{khan2023non} discuss the performance of LLMs in identifying and categorizing non-functional requirements.
To address the issue of limited annotated data in non-functional requirements classification, Rahman~\etal~\cite{rahman2023pre} propose to classify non-functional requirements by extracting features from pre-trained word embedding models.
Considering that previous work utilizing LLMs in a black-box manner, Han~\etal~\cite{han2023improving} propose a requirement classification approach based on BERT and an explainable AI framework.
They train a concern extraction model to extract concerns from requirement texts, and utilize explainability to generate explanations for the predictions of the requirement classification model, which is then used to fine-tune BERT for requirement classification.

\subsubsection{Requirement Quality Assurance}
High-quality requirements are fundamental to the success of software development, providing a clear and detailed specification of what the software system should achieve.
To assess the quality of requirements, Lubos~\etal~\cite{lubos2024leveraging} empirically explores the effectiveness of LLMs (like Llama 2) to enhance the quality assurance process in the requirements engineering phase. 
Preda~\etal~\cite{preda2024supporting} present an initial study to automate the review of coverage between high-level and low-level software requirements by GPT-3.5 and GPT-4.
Poudel~\etal~\cite{poudel2023leveraging} utilize several BERT-style LLMs to assess whether design elements adequately satisfy software requirements.
Ronanki~\etal~\cite{ronanki2022chatgpt} utilizes ChatGPT to evaluate the quality of user stories, which are crucial to capture end-user needs and express requirements in agile software development projects.
These studies suggest that LLMs can play a crucial role in automating and enhancing the quality assurance processes across various aspects of requirements engineering.

\subsubsection{Software Specifications Repair}
\label{sec:specifications_repair}

Software Specifications Repair refers to the process of fixing errors in software specifications, which are formal declarations of software system requirements or behaviors. This repair process has become increasingly significant with the growing complexity and usage of declarative languages like Alloy. 
The recent integration of LLMs like ChatGPT in this domain aims to automate and enhance the effectiveness of repair techniques. These LLMs are evaluated for their ability to correct inaccuracies in specifications and compared against existing automated program repair methods. The process involves identifying and rectifying various types of errors in software specifications, including logical inconsistencies, type errors, and misuse of programming constructs.
In 2023, Hasan~\etal~\cite{hasan2023automated} evaluate the potential of ChatGPT for repairing software specifications in the Alloy declarative language. 
It aims to assess ChatGPT's capabilities in correcting errors and identifies challenges in making it a viable solution. 
This work demonstrates that ChatGPT successfully addresses unique errors that other tools fail to rectify, although it does not consistently surpass them in total repair count. 

\subsubsection{Requirement Ambiguity Detection}
Ambiguities refer to terms or phrases in requirements that can be interpreted in multiple ways, thus leading to misunderstandings and inconsistencies during the development process.
Ambiguity detection attempts to identify such unclear, vague, or imprecise statements early in the requirements engineering phase.
Moharil~\etal~\cite{moharil2023tabasco,moharil2022identification} introduce TABASCO to detect ambiguities by leveraging BERT to capture the different meanings a word can have depending on its context within a requirement. 
Further, Ezzini~\etal~\cite{ezzini2022automated} explore multiple approaches (such as SpanBERT) to handle anaphoric ambiguity by ambiguity detection and anaphora interpretation.
Sridhara~\cite{sridhara2023chatgpt} conduct a preliminary empirical study to explore the potential of ChatGPT in anaphora ambiguity resolution.

\subsubsection{GUI Layouts}
\label{sec:gui_layouts}

GUI layouts refer to the arrangement and organization of various elements, such as widgets, images, banners, and icons, within the design of a GUI. The purpose of GUI layouts is to provide a user-friendly interface and ensure a positive user experience. 
This encompasses how various interface components are effectively positioned and displayed to facilitate user understanding and interaction with the application. 
Designing layouts involves considerations of spatial relationships between elements, overall page structure, usability, aesthetics, and other factors to create an intuitive, user-friendly, and visually pleasing user interface.
Kolthoff~\etal~\cite{kolthoff2023data} fine-tune BERT to retrieve reusable GUIs from a large-scale GUI repository, which can be adapted to facilitate GUI prototyping.
Besides, Wu~\etal~\cite{brie2023evaluating} explore the application of LLMs in GUI layout and introduce Instigator, which utilizes LLMs to search and suggest GUI layouts based on textual instructions. 
Instigator aims to enhance creativity and efficiency in the GUI design process by providing designers with relevant and diverse layout options.
This work highlights the potential of LLMs in supporting GUI design tasks, particularly by automating parts of the creative process.

In addition to the above-mentioned tasks detailed above, researchers also integrate LLMs into software requirements and design from other aspects, including 
class diagram derivation~\cite{li2024llm}, 
requirement completeness detection~\cite{luitel2024improving},
requirement elicitation~\cite{ren2024combining}, 
requirement prioritization~\cite{sami2024prioritizing,jain2023transformer}, 
requirement traceability~\cite{guo2024natural,lin2021traceability}, 
software modeling~\cite{chaaben2023towards,ferrari2024model}
and use case generation~\cite{zhang2024experimenting}.

\subsection{Software Development}
\label{sec:rq2_se_development}
Software Development is a creative process involving the use of computer programming languages, tools, and techniques to transform user requirements, functionality, and performance requirements into computer programs. 
\delete{Software development can be divided into multiple stages, such as requirements analysis, design, coding, testing, deployment, and maintenance.}

\subsubsection{Code Generation}
\label{sec:code_generation}

Code generation plays a pivotal role during software development and has always been the primary focus in the application of LLMs, such as AlphaCode~\cite{li2022competition} and CodeGen~\cite{nijkamp2022codegen}.
In general, recent advancements in LLM-based code generation focus on requirement-guided generation, execution-guided, and empirical studies. 
Below, we will introduce these studies in detail.

\textbf{Requirement-guided code generation.}
In the early stages of development, LLMs commonly take a natural language description as the input and return the correct code snippet, which is evaluated by corresponding unit tests~\cite{yang2023syntax,jiang2023self,zhang2023planning}.
For example, ArchCode~\cite{han2024archcode} leverages in-context learning to interpret software requirements from textual descriptions for LLM-based code generation.
ClarifyGPT~\cite{mu2023clarifygpt} attempts to detect ambiguous requirements, and prompt LLMs to specific clarifying questions, which are used to refine the requirements and generate more accurate code solutions.
Besides, AceCoder~\cite{li2024acecoder} directs LLMs to analyze the given requirement and generate intermediate artifacts for better code generation, such as test cases, that help clarify the requirement.
Inspired by the way humans typically use planning to break down complex problems, Jiang~\etal~\cite{jiang2023self} introduce a self-planning code generation approach to help LLMs understand complex intents and simplify problem-solving.

\textbf{Execution-guided Code Generation.}
Inspired by the process of human programming, developers utilize the execution results to guide LLMs in refining generated code iteratively~\cite{ni2023lever,chen2023teaching,chen2023improving}.
For example, Self-Edit~\cite{zhang2023self} validates generated code with available test cases, and the results are fed into LLMs as supplementary comments for further improvements and corrections.
Besides, Dong~\etal~\cite{dong2023self} introduce a self-collaboration code generation framework, which assembles a team consisting of three ChaGPT roles (\ie, analyst, coder and tester).
Given a requirement, the analyst decomposes it into several manageable subtasks and develops a high-level plan.
The coder generates code according to the plan provided by the analyst, and refines code according to the test reports provided by the tester.
The tester receives the code generated by the coder and subsequently produces a test report.

\textbf{Empirical Evaluation of LLMs.}
A mass of empirical studies are conducted to investigate the actual code generation capabilities of LLMs from different aspects, such as robustness~\cite{mastropaolo2023robustness,zhong2023study}, efficiency~\cite{coignion2024performance}, human study~\cite{jin2024can,feng2023investigating,kou2024large}, ChatGPT~\cite{liu2024guiding,yan2023closer}, prompt engineering~\cite{liu2023improving}, benchmarks~\cite{cassano2023multipl,yu2024codereval}, domain-specific generation~\cite{chen2023effectiveness}
For example, Mastropaolo~\etal~\cite{mastropaolo2023robustness} conduct an empirical study to investigate how robust GitHub Copilot is in generating consistent code when provided with semantically equivalent natural language descriptions.
Kou~\etal~\cite{kou2023model} empirically explore the attention alignment of LLMs and human programmers during code generation.

\textbf{Others}.
Recently, In addition to the above-mentioned studies, there has been an exploration in cutting-edge areas such as repository-level generation~\cite{bairi2024codeplan,ma2024compositional,liu2023codegen4libs}, retrieval-augmented generation~\cite{li2023skcoder} and agent-based generation~\cite{lin2024llm,huang2023agentcoder,zhang2024codeagent}

\subsubsection{Code Search}
\label{sec:code_search}

Code search is an essential practice in software development where developers search for specific pieces of source code within extensive codebases.
Given a given query, this activity is undertaken to find code snippets, functions, classes, or entire files for potential reuse.
Overall, existing studies utilizing LLMs for code search can be classified into two categories.
First, researchers utilize training objectiveness to help LLMs learn more effective encoding representation, primarily focusing on contrastive learning~\cite{li2023mcodesearcher,liu2023contrabert,shi2022cross,shi2023improving}.
For example, CodeRetriever~\cite{li2022coderetriever} learns semantic representations for function-level code search with unimodal and bimodal contrastive learning.
Unimodal contrastive learning encourages similar functional code to cluster closely in the representation space, while bimodal contrastive learning assists in understanding the correlation between code and text.
Similarly, CoCoSoDa~\cite{shi2023cocosoda} helps LLMs better align code snippets and natural language queries for code search based on multimodal contrastive learning and data augmentation.
Second, some empirical studies are conducted to investigate the potential of LLMs in code search.
For example, Salza~\etal~\cite{salza2022effectiveness} explore how transfer learning can be applied to code search tasks by pre-training and fine-tuning BERT.
Chi~\etal~\cite{chi2024empirical} evaluate exiting code search LLMs in industry requirements based on their adaptability, scalability, robustness, and semantic sensitivity.

\subsubsection{Code Translation}
\label{sec:code_translation}

Code translation refers to the process of converting code from a source language into a target language while preserving the original functionality and behavior of the program. 
This task is crucial in software development, especially when developers want to migrate software systems to a different platform.
Yang~\etal~\cite{yang2024exploring} introduce UniTrans, which introduces test case generation to enhance the accuracy of code translation and provides mechanisms for iterative repair of translation errors.
TransMap~\cite{wang2023transmap} attempts to detect semantic mistakes in code translated by models like Codex and ChatGPT. 
Regarding empirical studies, Pan~\etal~\cite{pan2024lost} propose a taxonomy of translation bugs introduced by LLMs, while Jiao~\etal~\cite{jiao2023evaluation} provide a detailed analysis of code translation across four levels: token level, syntactic level, library level, and algorithm level.

\subsubsection{Code Co-Evolution}
Code co-evolution attempts to update code snippets in a target programming language by reflecting the changes made in the source programming language.
Different from code translation that directly generates code snippets from a source programming language to a target language, code co-evolution learns an edit sequence to update existing code snippets.
Zhang~\etal~\cite{zhang2023multilingual} propose Codeditor, which fine-tunes CoditT5 to align the code edits for two programming languages (Java and C\#) from eight open-source projects.

\subsubsection{Code Comment Completion}
\label{sec:comment_completion}

Code comment completion refers to the process by which a computer program or model automatically writes code comments for programmers. 
In this process, based on the code that has already been written, the program or model automatically generates corresponding comments to explain the function and purpose of the code.
It aims to provide real-time suggestions and assistance to programmers while writing code comments, similar to the concept of code auto-completion. 
This process does not create comments from scratch but rather assists programmers in completing comments more quickly based on the partial comments or code snippets they input. 
In 2021, Mastropaolo~\etal~\cite{mastropaolo2021empirical} address the issue of code comments using T5 and n-gram models.
The study compares a simple n-gram model and the T5 model in supporting code comment completion. The findings indicate that the T5 model performs better, although the n-gram model remains competitive.
The research experiments with a dataset containing a large number of Java methods and their associated comments. Results show that the T5 model outperforms the n-gram model in all the tested code comment completion scenarios.

\subsubsection{Code Summarization}
\label{sec:code_summarization}

Code summarization takes as input a code snippet provided by the developer and automatically generates a higher-level summary in natural language form.
These summaries are usually utilized to enhance the comprehension of software systems, thereby facilitating their maintenance. 
Integrating LLMs into code summarization can provide more accurate and natural summaries of source code, as LLMs can leverage their general knowledge on vast amounts of textual data to more accurately infer and generate natural language summaries of source code.

Existing LLM-based studies mainly fall into three areas: training optimizations, prompt engineering, and empirical studies.
First, for open-source LLMs, researchers utilize pre-training or fine-tuning methods to enhance LLMs' capabilities of the alignment between code and natural language~\cite{li2024cross}.
For example, ESALE~\cite{fang2024esale} introduce three three
code summarization-specific pre-training tasks (including two general tasks ULM and MLM, and one domain-specific task AWP) to help LLMs learn the code-summary alignment.
Other training strategies include adapter tuning~\cite{wang2023one} and joint training~\cite{li2024cross}.
Second, for commercial LLMs, prompt engineering is usually utilized to provide relevant information in a zero-shot or few-shot manner~\cite{geng2024large,wang2023generating}.
For example, Ahmed~\etal\cite{ahmed2024automatic} explores enhancing the code summarization performance of LLMs by explicitly adding semantic information to the prompts, such as parameter names, return types, and simple control flows.
Rukmono~\etal~\cite{rukmono2023achieving}  integrate static code analysis into the chain-of-thought prompting to generate component-level summaries of software systems. 
Third, empirical studies are conducted to explore the potential of LLMs from different aspects, such as explainability~\cite{li2024machines},
metrics~\cite{jin2024simllm}, binary code summarization~\cite{jin2023binary}
For example, Sun~\etal~\cite{sun2023automatic} evaluate the performance of ChatGPT on code summarization, and Li~\etal~\cite{li2024machines} utilize eye-tracking metrics from human participants to measure whether they concentrate the same parts of code as LLMs when generating summaries.

\subsubsection{Code Completion}
\label{sec:code_completion}

Code completion aims at speeding up code writing by predicting the next code token the developer is likely to write.
Researchers usually focus on improving the accuracy of the generated predictions.
For example, TeCo~\cite{nie2023learning} attempts to generate the next statement in a test method by fine-tuning CodeT5 with code semantic information.
CCTEST~\cite{li2023cctest} focuses on improving the performance of off-the-shelf code completion systems (such as Copilot and CodeGen) by utilizing a mutation strategy to detect erroneous outputs and a repair mechanism to fix these outputs,
Besides, repository-level code completion has always been challenging due to complicated contexts from multiple files in the repository.
LLMs demonstrate the potential in this task by integrating advanced strategies, such as information retrieval~\cite{liu2024graphcoder,cheng2024dataflow,wu2024repoformer,phan2024repohyper}, static analysis~\cite{liu2024stall+}, and reinforcement learning~\cite{wang2024rlcoder}.
There are also many empirical studies analyzing the actual performance of LLMs in code completion.
For example, Ciniselli~\etal~\cite{ciniselli2021bert,ciniselli2021transformer} investigates the effectiveness of T5 and RoBERTa for code completion at different granularity levels, from single tokens to entire code blocks.
Similarly, Van~\etal~\cite{van2023enriching}explores the impact of contextual information on three LLMs (UniXcoder, CodeGPT, and InCoder) for both token-level and line-level code completion.

\subsubsection{Program Synthesis}
\label{sec:program_synthesis}

Program synthesis refers to the automated process of generating computer programs based on high-level specifications or requirements. 
The objective of this process is to automate typically complex and time-consuming manual software development tasks by allowing machines to generate code based on specifications provided by users. 
The emphasis of program synthesis is to enhance the capabilities of LLMs, such as GPT-3 and Codex, enabling them to generate code from natural language specifications of programmer intent. 
For example, Jain~\etal~\cite{jain2022jigsaw} from Microsoft explore the integration of LLMs such as GPT-3, Codex, and Google's LLM in generating code from natural language descriptions of programmer intent. 
Liventsev~\etal~\cite{liventsev2023fully} introduce SED, a framework to enhance the program synthesis capabilities of LLMs like OpenAI Codex.
In SEIDR, a draft program is first synthesized based on a high-level description and is then executed to validate its correctness. 
If the program fails, specific instructions are generated to guide LLMs in fixing the issues until it meets the required specifications.
Besides, Vella~\etal~\cite{vella2024synergistic} explores the integration of multiple LLMs within evolutionary algorithms to enhance program synthesis tasks.

\subsubsection{Code Editing}
\label{sec:edits_prediction}

Code editing refers to the task of predicting changes or modifications that need to be made to a piece of code to transition from one version to another. 
It involves predicting the edits a developer will make to refactor code from one version (e.g., $\mathbf{v_1}$) to another version (e.g., $\mathbf{v_2}$ or $\mathbf{v_3}$). This task is essential in software development, especially during code refactoring or feature additions.

For example, Li~\etal~\cite{li2023codeeditor} design a pre-training specialized for code editing by rewriting mutated versions of code snippets into their correct form.
They then introduce CodeEditor initialized with CodeT5 and fine-tune it on two code editing scenarios, \ie code-to-code editing and comment\&code-to-code editing.
Besides, Gupta~\etal~\cite{gupta2023grace} propose Grace to address code editing by leveraging the generative capabilities of LLMs on previously related edits. 
By mirroring the behavior of developers, Li~\etal~\cite{liu2024automated} introduce a hybrid approach SARGAM for automated code editing, which consists of three steps.
SARGAM first retrieves similar code patches from a large repository and uses it to guide LLMs (such as PLBART, CoditT5, and NatGen) to generate a new patch, which is further modified to fit the exact context.
These studies demonstrate the potential of LLMs to automate complex editing tasks, offering developers an efficient way to handle repetitive and time-consuming code modifications.

In addition to the above-mentioned tasks detailed above, researchers also integrate LLMs into software development from other aspects, including API documentation smells~\cite{khan2021automatic}, API Inference~\cite{huang2023adaptive,wang2023measuring,zhuo2023pop}, API recommendation~\cite{huang2023let,wei2022clear,li2024ptm,zhang2023toolcoder,chen2024apigen,wu2024automatic}, code compression~\cite{gilbert2023semantic,von2022validity}, code representation~\cite{lin2024vargan,saberi2023model,cui2024api2vec++,liu2023contrabert,agarwal2024structured,he2024representation}, code understanding~\cite{zhao2023understanding,shen2022benchmarking,khakhar2023pac}, continuous development optimization~\cite{baral2023optimizing}, identifier normalization~\cite{zhang2022beqain}, microservice recommendation~\cite{alsayed2024microrec}, neural architecture search~\cite{nasir2024llmatic}, performance data synthesize~\cite{banday2024perfgen}, SO post title generation~\cite{le2024good}, type inference~\cite{jesse2022learning}, and unified development~\cite{qian2024chatdev}.

\subsection{Software Testing}
\label{sec:rq2_se_testing}
Software testing is the process of executing a program or system with the intent of finding errors or any activity aimed at evaluating an attribute or capability of a program or system to ensure it meets its required results.

\subsubsection{Fault Localization}
\label{sec:fault_localization}

Fault localization is a critical process in software engineering that involves identifying the specific locations or elements in a software system where faults or bugs are present.
By pinpointing the exact location of a fault, developers can more quickly perform software debugging and bug-fixing, leading to more efficient software development and maintenance.
In the literature, researchers have conducted several studies to explore the performance of LLMs in fault localization, where two types of LLMs are involved, \ie encoder-like and GPT-like models.

Regarding encoder-based localization, as early as 2021, Zhu~\etal~\cite{zhu2021trobo} propose TroBo, a CodeBERT-based cross-project bug localization approach by leveraging both bug reports and source code.
In 2022, Ciborowska~\etal~\cite{ciborowska2022fast} discuss how to optimize BERT for changeset-based bug localization. 
They explore various design choices for applying BERT, including how to encode code changes and match error reports to specific code changes to enhance accuracy. 
To support cross-language cross-project bug localization, Chandramohan~\etal~\cite{chandramohan2024supporting} fine-tune UniXcoder with a contrastive learning object to enhance the representation of both bug reports and source code. 
Regarding GPT-based localization, 
Wu~\etal~\cite{wu2023large} conduct a comprehensive empirical study on the large-scale open-source program Defects4J, evaluating the potential of OpenAI GPT LLMs (\ie ChatGPT-3.5 and ChatGPT-4) in fault localization research. 
Besides, to handle large codebases, Kang~\etal~\cite{kang2024quantitative} introduce AutoFL,  to select parts of the project and apply a post-processing step to match ChatGPT's answers with actual code elements. 
Yang~\etal~\cite{yang2024large} introduce LLMAO, the first CodeGen-based approach that locates buggy lines without relying on traditional test coverage information.
FuseFL~\cite{widyasari2024demystifying} leverages ChatGPT to provide explainable fault localization by integrating multiple sources of information, including test case outcomes and code descriptions.

\subsubsection{Code Decompilation}
\label{sec:code_decompilation}

Decompilation is the reverse engineering process of extracting a binary executable's code into a form that closely resembles its original source code and is understandable to humans.  
This task has important applications in security fields, such as malware analysis and vulnerability detection, as well as in software fields, such as code reuse and software supply chain analysis.

There mainly exist two key challenges during the decompilation process, \ie recovering variable names within binary executable files and producing human-readable code.
To address the first issue, Xu~\etal~\cite{xu2023lmpa} introduce LmPa to improve the recovery of variable names and other high-level information by combining LLMs with program analysis.
LmPa utilizes LLMs to provide meaningful names for variables, which are then refined and validated with program analysis techniques, ensuring that the recovered names are contextually appropriate and semantically accurate.
To address the second issue, Wong~\etal~\cite{wong2023refining} automatically refine the accuracy and quality of decompiled C code by combining LLMs with traditional decompilation techniques.
LLMs are utilized to fix syntax, inference, and memory errors in decompiled output, making it compatible with standard C/C++ compilation. 
A similar work is DeGPT\cite{hu2024degpt}, which designs a three-role mechanism to maximize the optimization capabilities of LLMs on decompiler output.
Researchers also make efforts to integrate LLMs into code decompilation from different perspectives, including domain LLMs~\cite{tan2024llm4decompile,jiang2023nova,armengol2024slade}, and WebAssembly~\cite{she2024wadec}, binary code understanding~\cite{shang2024far}.

\subsubsection{Vulnerability Detection}
\label{sec:vulunerability_detection}

Vulnerability detection, also known as vulnerability prediction, aims to identify potential security bugs in software systems.
Vulnerability detection is critical for protecting security-critical software systems from malicious attacks, providing the foundation for timely patching reported security vulnerabilities before they may be exploited (discussed in Section~\ref{sec:vulnerability_repair}).
In the literature, learning-based vulnerability detection approaches~\cite{chakraborty2021deep,li2021vulnerability} have been proposed to detect security vulnerabilities by extracting meaningful features and performing predictions automatically.
For example, Li~\etal~\cite{li2021vulnerability} propose IVDetect, to perform fine-grained vulnerability prediction based on a FA-GCN model and GNNExplainer.
However, such learning-based approaches are limited by the amount of training data, resulting in capturing a suboptimal vector representation of source code.

Thus, a mass of vulnerability detection approaches empowered with LLMs have been proposed.
For example, LineVul attempts to detect vulnerabilities at the line level by fine-tuning CodeBERT and utilizing its attention mechanism to pinpoint vulnerable lines.
In parallel to LineVul, VulBERTa is an encoder-only Transformer-based vulnerability detection approach by pre-training and fine-tuning RoBERTa.
Besides, VulLLM~\cite{du2024generalization} utilizes multi-task instruction tuning to adapt LLMs in vulnerability detection, which includes two auxiliary tasks: vulnerability localization to pinpoint the specific vulnerable parts of the code, and vulnerability interpretation to understand the underlying issues.
GPTScan~\cite{sun2024gptscan} attempts to detect smart contract logic vulnerabilities by combining GPT and static analysis.
In the literature, researchers also conduct some empirical studies.
For example, Steenhoek~\etal~\cite{steenhoek2023empirical} conduct an empirical study to investigate the performance of DL models in detecting software vulnerabilities from three aspects, \,  i.e., model capabilities, training data, and model interpretation.
This study includes nine learning-based AVD approaches, including the above-mentioned two LLM-based approaches, \ie LineVul and VulBERTa, and two off-the-shelf LLMs, \ie CodeBERT and PLBART.
In 2023, Zhang~\etal~\cite{zhang2023prompt} explore the performance of ChatGPT in software vulnerability detection with different prompts.
Meanwhile, Noever~\etal~\cite{noever2023can} evaluate the capabilities of LLMs, particularly GPT-4, in detecting software vulnerabilities, comparing their performance against traditional static code analysis tools.

\subsubsection{Unit Test Generation}
\label{sec:test_generation}

Test generation is the process of creating a set of test data or test cases for testing the adequacy of new or revised software programs. 
Unit test generation is a specialized domain within test generation, focusing on creating test cases for individual code units. 
We summarize these advanced studies that employ LLMs as follows.

\textbf{Fine-tuning LLM-based Generation.}
AthenaTest represents the first work to apply LLMs in the field of test generation.
AthenaTest first pre-trains BART on a large corpus of English and Java datasets and fine-tunes it on a labeled Java dataset.
A3Test~\cite{alagarsamy2023a3test} first pre-trains PLBART to learn assertion knowledge in a self-supervision manner and fine-tune it to support the test case generation task.
Rao~\etal~\cite{rao2023cat} introduce CAT-LM, a GPT-style LLM that is specifically trained to learn the mapping between code and its corresponding test cases.

\textbf{Prompt-based Generation.}
For example, Schafer~\etal~\cite{schafer2024empirical} introduce TestPilot, an LLM-based end-to-end adaptive test generation technique for JavaScript.
TestPilot first constructs a prompt with the information of the function under test and its usage examples. 
TestPilot then queries LLMs to generate test cases, which are validated through dynamic execution, and refined with new prompts containing failure information.
MuTAP~\cite{dakhel2023effective} utilizes mutation testing to augment the prompts, guiding LLMs to generate test cases that can detect software bugs.
SymPrompt~\cite{ryan2024code} enhances the performance of LLMs in generating high-coverage test cases by utilizing code-aware prompts, which are dynamically constructed with the execution paths of the method under test. 
Pizzorno~\etal~\cite{pizzorno2024coverup} introduce CoverUp, which integrates coverage analysis into prompts to iteratively generate Python regression tests.

\textbf{Conversation-driven Generation with ChatGPT.}
To address issues of invalid test cases and low coverage, Gu~\etal~\cite{gu2024testart} introduce ChatGPT-based TestART, which integrates test generation with an iterative template-based repair process.
To mitigate common issues such as invalid or incomplete tests generated by LLMs, Karmarkar~\etal~\cite{karmarkar2024navigating} introduce TestRefineGen to generate tests based on textual descriptions while ensuring confidentiality is maintained.
Xie~\etal~\cite{xie2023chatunitest} propose ChatUniTest, an automated unit test generation tool based on ChatGPT under the generation-validation-repair framework. 
Ni~\etal~\cite{ni2024casmodatest} introduce CasModaTest, an end-to-end LLM-based framework, by dividing the unit test generation task into two cascaded ones: test prefix generation and test oracle generation.
To enhance test coverage and improve the efficiency and effectiveness of automated testing, Lemieux~\etal~\cite{lemieux2023codamosa} introduce CODAMOSA, which integrates Codex with SBST to generate Python test cases.

\textbf{Empirical Evaluations of LLM-based Test Generation.}
Xiao~\etal~\cite{xiao2024optimizing} empirically explore how LLMs can be integrated into traditional search-based unit test generation workflows, including the initial phase, the test generation period, and the test coverage plateaus.
Tang~\etal~\cite{tang2023chatgpt} provide a comprehensive comparison between ChatGPT and EvoSuite based on several factors: correctness, readability, code coverage, and bug detection capabilities.
Similarly, Yang~\etal~\cite{yang2024empirical} explore the impact of prompt designs, comparison of open-source and commercial LLMs, and in-context learning.
Researchers also conduct more studies from different aspects, such as the investigation of ChatGPT~\cite{guilherme2023initial,yuan2024evaluating}, prompt engineering~\cite{ouedraogo2024large}, human study~\cite{deljouyi2024leveraging}, GUI text Input~\cite{cui2024large}, security tests~\cite{zhang2023well}

\subsubsection{Assertion Generation}
\label{sec:assertion_generation}

Assertion Generation refers to the process of automatically inferring or creating expected outcomes or conditions against which the output of a software system or program is validated. 
These oracles serve as a standard or benchmark against which the system's behavior is tested, allowing for the automatic detection of bugs, discrepancies, or unexpected behaviors within the software. 
It involves using various methods, ranging from hard-coded patterns, natural language processing, neural networks, and machine learning models to derive or predict the expected outcomes, assertions, or exceptional behaviors of a given code snippet, function, or method. The goal is to generate these oracles accurately to enable automated testing and the identification of potential flaws or issues within the software.

As early as in 2022, Tufano~\etal~\cite{tufano2022generating} propose to leverage the BART model to generate accurate assert statements in unit test cases.
They first perform semi-supervised pre-training on a large corpus of English text to help BART learn the semantic and statistical properties of natural language.
They then pre-train BART on abundant Java source code crawled from GitHub with a similar pre-training strategy to English pre-training.
Finally, they fine-tune it on a dataset mined from more than 9 thousand open-source GitHub projects containing unit test cases defined with JUnit.
Despite promising, the previous approach~\cite{tufano2022generating} struggles to find real-world bugs.
In 2022, Dinella~\etal~\cite{dinella2022toga} propose a transformer-based approach TOGA to infer both exceptional and assertion test oracles based on the context of the focal method.
TOGA fine-tunes CodeBERT to classify exceptional oracles and rank assertion oracles.
Unlike previous studies fine-tuning LLMs for assertion generation, Nashid~\etal~\cite{nashid2023retrieval} introduce Cedar, a few-shot learning method that utilizes a prompt-based approach for both test assertion generation and program repair. 
Cedar retrieves relevant demonstrations and conducts prompts to query Codex to generate assertions in a few-shot manner.

Regarding empirical studies and benchmarking, He~\etal~\cite{he2024empirical} investigate the effectiveness of seven test-to-code traceability techniques in assertion generation and enhance existing datasets for training and evaluating assertion generation models, such as T5.
besides, Pulavarthi~\etal~\cite{pulavarthi2024assertionbench} introduce AssertionBench, a benchmark to assess the effectiveness of LLMs in generating assertions for hardware verification.
AssertionBench focuses on generating high-quality assertions, which are crucial for detecting and diagnosing design bugs, particularly in complex hardware systems.
Endres~\etal~\cite{endres2024can} explore the potential of LLMs, such as GPT-4, to convert informal natural language descriptions of program behavior into formal postconditions that can be used for software verification.
Hossain~\etal~\cite{hossain2024togll} conduct a large-scale investigation of the ability of LLMs to automatically generate test oracles by fine-tuning LLMs (including  CodeGPT-110M, CodeParrot-110M, CodeGen-350M, PolyCoder-4B, Phi-1-1.3B, CodeGen-2B and PolyCoder-2.7B) with six different prompts.

\subsubsection{Test Suite Minimization}
\label{sec:test_minization}

Test suite minimization aims at improving the efficiency of software testing by removing redundant test cases, thus reducing testing time and resources while maintaining the effectiveness of the test suite.
Since previous test suite minimization approaches that rely on test code (black-box) are rather time-consuming, Pan~\etal~\cite{pan2023ltm} propose LTM, a black-box similarity-based approach that leverages LLM to address the scalability problem.
They explore three off-the-shelf pre-trained models for similarity measurements: CodeBERT, GraphCodeBERT and UniXcoder.
These models take the source code of test cases as input to generate numeric vectors.
Then, they employ two similarity measures for calculating the similarity between test method embeddings: Cosine Similarity and Euclidean Distance.
Cosine similarity measures the angle between two vectors, whereas Euclidean distance calculates the straight-line distance between them.
Results show that UniXcoder/Cosine is the best LTM configuration when considering both effectiveness and efficiency.
Besides, LTM outperforms prior works by achieving a significantly higher fault detection rate and faster minimization time.

\subsubsection{Fuzzing}
\label{sec:fuzzing}

Fuzzing is an automated software testing method that injects invalid, malformed, or unexpected inputs into a system to reveal software defects and vulnerabilities.
A fuzzing tool injects these inputs into the system and then monitors for exceptions such as crashes or information leakage.
In the following, we summarize and categorize existing fuzzing studies empowered with LLMs.

\textbf{DL Library Fuzzing}.
DL libraries, such as TensorFlow and PyTorch, are foundational to the burgeoning field of DL, play a pivotal role in our daily lives due to the widespread adoption of DL systems. 
Traditional fuzzing techniques often struggle to satisfy both the input language semantics and the DL API input constraints for tensor computations. 
To address this, Deng~\etal~\cite{deng2023large1} introduce TitanFuzz in 2023, the first approach to leverage LLMs directly for generating input programs for fuzzing DL libraries. 
For any given target API, TitanFuzz initially uses an LLM to generate a list of high-quality seed programs for fuzzing by querying the Codex model with a step-by-step prompt and sampling multiple completions.
However, due to the nature of LLMs, TitanFuzz tends to generate ordinary human-like DL programs, which can only cover a limited range of program patterns.
Deng~\etal~\cite{deng2023large2} propose FuzzGPT, which utilizes LLMs to generate unusual programs based on historically bug-triggering programs.
FuzzGPT features three variants: few-shot learning, zero-shot learning, and fine-tuning, leveraging different LLMs, including Codex and CodeGen. 
Additionally, FuzzGPT can utilize the directive-following capabilities of ChatGPT to generate atypical programs.

\textbf{Compiler Fuzzing}
Compilers are the foundation of modern software systems by translating high-level source code written by programmers into machine code, a lower-level language that a computer can execute.
Thus, the correctness of a compiler is crucial as it ensures the accurate and efficient execution of the intended functionality of the software.
In 2023, Yang~\etal~\cite{yang2023white} introduce WhiteFox, the first white-box compiler fuzzing tool using LLMs in conjunction with source code information to test compiler optimizations. 
WhiteFox utilizes a dual-model framework, where one analysis LLM examines low-level optimization source code and generates high-level test program requirements that can trigger optimizations, while another generation LLM produces test programs based on the summarized requirements.
Eom~\etal~\cite{eom2024covrl} introduce CovRL, a novel approach for fuzzing JavaScript engines by integrating coverage-guided reinforcement learning with LLMs.
CovRL leverages coverage feedback to guide the mutation process, aiming to uncover vulnerabilities more efficiently while minimizing syntax and semantic errors.

\textbf{Protocol Fuzzing}.
Protocol implementations are the practical realizations of communication protocols in software or hardware, where correctness is crucial to ensure reliable and secure data transmission across different systems and networks.
In 2023, Meng~\etal~\cite{meng2024large} explore the opportunity of interacting with LLMs, which have ingested millions of pages of human-readable protocol specifications, to extract machine-readable information about the protocol for use in protocol fuzz testing.
They develop CHATAFL, an LLM-guided protocol implementation fuzzing engine, to achieve structure-aware mutations concerning the state machine and input structure of the protocol. 
To fuzz Internet of Things (IoT) devices, LLMIF~~\cite{wang2024llmif} utilizes an LLM to analyze protocol specifications and generate relevant test cases.

\textbf{General-purpose Fuzzing.}
Different from previous studies targeting specific scenarios, general-purpose fuzzing (\eg AFL ) is unaware of the programs and focuses on byte-level transformations.
In 2023, Xia~\etal~\cite{xia2024fuzz4all} introduce Fuzz4All, the first universal fuzzer to support various software systems based on the multi-lingual capabilities of LLMs.
It utilizes auto-prompting techniques to generate effective LLM prompts for fuzzing and iteratively updates prompts to generate diversified fuzzy inputs.
Fuzz4All has been evaluated on nine systems across six different languages (\ie C, C++, SMT, Go, Java, and Python), demonstrating a significant improvement in code coverage compared to previous fuzzers. 
Besides, considering that traditional fuzzers (\eg AFL) struggle to generate structured test inputs efficiently and at scale, in 2023, Hu~\etal~\cite{hu2023augmenting} introduce CHATFUZZ, a grey-box fuzzing tool leveraging ChatGPT to enhance test input quality and effectiveness.
In parallel to CHATFUZZ, Dakhama~\etal~\cite{dakhama2023searchgem5} introduce an innovative approach that combines LLM and search-based fuzzing, specifically targeting the gem5 system. 
The technique leverages ChatGPT to parameterize C programs, compiles the resulting code snippets, and feeds them to the SearchGEM5 extension of the AFL++ fuzzer, utilizing custom mutation operators. 
Similarly, LLAMAFUZZ~\cite{zhang2024llamafuzz} attempts to enhance traditional greybox fuzzing by leveraging LLMs to generate structured data, which is often a challenge for traditional fuzzing methods that rely on random mutations.

\textbf{Fuzz Driver Generation}.
A fuzz driver is a piece of code written to accept inputs from fuzzers and execute the program accordingly.
It is labor-intensive and time-consuming for human experts to manually write high-quality fuzz drivers.
In 2023, Zhang~\etal~\cite{zhang2023understanding} conduct an empirical study to explore the fundamental issues of effective fuzz driver generation using LLMs.
This framework includes a quiz with 86 driver generation questions collected from 30 popular C projects and a set of criteria for precise driver effectiveness validation. In total, 189,628 fuzz drivers using 0.22 billion tokens are generated and evaluated. 
The research results indicate that enhanced query strategies and iterative methods can significantly improve the accuracy and efficiency of generating fuzz drivers.

\textbf{Others}.
Researchers also integrate LLMs into other fuzzing scenarios, including smart contracts~\cite{shou2024llm4fuzz}, kernel fuzzing~\cite{yang2023kernelgpt}, and BusyBox~\cite{oliinyk2024fuzzing}.
For example, Oliinyk~\etal~\cite{oliinyk2024fuzzing} utilize LLMs to generate initial test seeds for fuzzing BusyBox, a widely-used open-source software suite for Linux-based embedded devices, aligning these seeds more closely with the expected inputs of BusyBox's various utilities.

\subsubsection{Penetration Testing}
\label{sec:penetration_testing}

Penetration Testing refers to the systematic process of actively assessing an organization's, company's, or system's security defenses by simulating real-world cyberattacks. This method involves identifying potential vulnerabilities and testing the system's resilience against exploitation, thereby providing insights into the system's security gaps and weaknesses.
PentestGPT~\cite{deng2023pentestgpt} is an LLM-empowered automatic penetration testing tool that leverages the abundant domain knowledge inherent in LLMs.
PentestGPT consists of three core modules: 
(1) the reasoning module maintains a high-level overview of the testing process;
(2) the generation module translates specific sub-tasks from the reasoning module into concrete instructions; 
and (3)the parsing module operates as a supportive interface between the user and the other two core modules.
PTGroup~\cite{wu2024ptgroup} automates complex penetration testing scenarios by utilizing LLMs to interpret multiple testing strategies simultaneously and designing multiple prompt chains for different penetration testing tasks.
Meanwhile, Happe~\etal~\cite{happe2023getting} explores the use of LLMs like GPT-3.5 to enhance penetration testing, focusing on both high-level task planning and low-level attack execution.

\subsubsection{Property-based Testing}
\label{sec:property_testing}

Property-based testing (PBT) aims to verify whether the program properties are satisfied by generating a large number of random input data. In comparison to traditional unit testing, PBT emphasizes the program's properties and behaviors rather than singular predefined test cases. This testing method was initially popularized by the QuickCheck library in the Haskell language. The main steps of PBT include property definition, random data generation, and property validation.

Traditional PBT methods are not widely applied in actual software development because crafting diverse random input generators and meaningful test properties poses a challenge. However, developers tend to be more inclined towards documentation writing, and library API documentation contains valuable natural language specifications for PBT.
Vikram~\etal~\cite{vikram2023can} propose PBT-GPT, utilizing LLM to generate random inputs and test properties from API documentation. This study explores three different LLM prompting strategies, revealing various failure modes in PBT-GPT and outlining an evaluation methodology for generator and property quality. Preliminary research reports the results of using PBT-GPT on three Python library APIs.
The experimental findings demonstrate the design and evaluation framework of PBT-GPT. In the design phase, researchers introduce three distinct LLM prompting strategies to generate critical components of property-based tests. The evaluation section thoroughly analyzes the quality of the generated generators and properties, encompassing metrics such as validity, diversity, and strength. Additionally, strategies to address potential issues are presented, providing effective pathways for enhancing test quality. Overall, the experimental results offer valuable insights into synthesizing property-based tests using LLM, despite some quality issues. The proposed mitigation strategies and evaluation framework pave the way for subsequent enhancements and improvements.

\subsubsection{Failure-Inducing Testing}
\label{sec:failure_inducing}

Failure-inducing testing (FIT) is a software testing approach aimed at identifying test cases that can trigger software errors or faults. This method employs test inputs to provoke specific behaviors or anomalies within a program, thereby detecting and addressing errors in software.

For example, Li~\etal~\cite{li2023finding} propose Differential Prompting, a methodology that utilizes ChatGPT to infer program intentions, generate program versions, and conduct differential testing, effectively identifying test cases that trigger software errors. The research finds that ChatGPT's ability to infer program intentions enables it to bypass subtle differences in code, thus identifying the correct program intention. By leveraging this characteristic, Differential Prompting successfully identifies test cases that trigger software errors. Differential Prompting comprises three main steps: program intention inference, program generation, and differential testing. In experiments conducted on different program sets such as QuixBugs and Codeforces, Differential Prompting demonstrates significant advantages in identifying failure-inducing test cases, far surpassing existing baseline methods.

\subsubsection{Mutation Testing}
\label{sec:mutation_testing}

Mutation testing represents an established fault-based testing technique. It introduces faults into the programs under examination and requires developers to write tests that uncover these faults. These tests possess the potential to reveal numerous faults, particularly those associated with the introduced faults.

In particular, $\mu$BERT~\cite{khanfir2023efficient} represents the first attempt to integrate LLMs into mutation testing by using CodeBERT to generate natural mutants.
Unlike traditional mutation testing methods, $\mu$BERT does not rely on predefined syntactic transformation rules. Instead, it masks tokens in the input expression and utilizes CodeBERT to predict and generate mutants.
Inspired by $\mu$BERT, Ibrahimzada~\etal~\cite{ibrahimzada2023automated} propose BugFarm to transform arbitrary code into complex bugs by querying LLMs to mutate code in multiple locations.
VULGEN~\cite{nong2023vulgen} focuses on generating realistic software vulnerabilities by combining pattern mining with LLMs.
It first collects vulnerability-fixing examples to mine the patterns for vulnerability injection, and then leverages CodeT5 to learn the localization of injection.
Similarly, Garg\cite{garg2024coupling} investigates the relationship between software vulnerabilities and mutants generated by CodeBERT.

Regarding empirical studies, Wan~\etal~\cite{wang2024exploratory} investigate the performance of LLMs in generating mutations based on various criteria, including usability, fault detection potential, and their relationship with real bugs.
Geng~\etal~\cite{geng2024large} empirically investigates the feasibility of utilizing LLMs to generate comments that can reflect multiple developer intents by in-context learning.
Besides, Tian~\etal~\cite{tian2024large} explore the the effectiveness and efficiency of LLMs in detecting equivalent mutants.

\subsubsection{GUI Testing}
\label{sec:gui_testing}

Graphical user interface (GUI) testing ensures the accuracy and reliability of a mobile application's visual interface. It involves verifying that interactions with components like buttons and text boxes yield expected outcomes. 
The aim is to confirm the correct information display and intended user interactions. 
GUI testing can be manual or automated, often using metrics like error detection and code coverage for assessing performance. 
Its primary goal is to guarantee the stability and reliability of an application's visual and interactive elements.

In 2022, to recognize the diverse and semantic requirements for valid inputs in GUI testing, Liu~\etal~\cite{liu2022fill} propose an approach named QTypist based on LLMs for intelligently generating semantic input text according to the GUI context.
To boost the performance of LLMs on text input in mobile GUI, they develop a prompt-based data construction and tuning method to automatically extract the prompts and answers for model tuning.
They leverage the pre-trained GPT-3 model and fine-tune it with the tuning method.
For GUI testing, a context-aware input generation method generates the prompt and feeds it into the GPT-3 model.
Unlike QTypist~\cite{liu2022fill} focusing on text input generation, in 2023, Liu~\etal~\cite{liu2023make} propose GPTDroid, framing the mobile GUI testing problem as a question-answering task, utilizing LLM as a human tester. 
GPTDroid facilitates the passing of the application's GUI information to LLM to initiate testing scripts and receive and iterate execution results. This framework incorporates a functionality-aware memory prompting mechanism, equipping the LLM to retain testing knowledge and engage in long-term, functionality-based reasoning to guide exploration.
Besides, VisionDroid~\cite{liu2024vision} is a vision-driven automated GUI testing approach that leverages multimodal LLMs that can understand both visual and textual information to detect non-crash functional bugs in mobile apps.
DroidAgent~\cite{yoon2024intent} leverages LLMs to simulate user interactions by setting and executing tasks that mirror realistic user behaviors and intents.

\subsubsection{NLP Testing}
\label{sec:nlp_testing}

NLP testing refers to the process of assessing and evaluating the performance, accuracy, and robustness of natural language processing systems, including but not limited to machine translation, text generation, language understanding, and other NLP-related tasks.
As early as 2020, He~\etal~\cite{he2020structure} introduce SIT, a metamorphic testing approach that leverages BERT to validate machine translation software. 
SIT uses BERT to generate a set of similar sentences by altering a single word in the source sentence, helping to assess the structural consistency of translations.
Similarly, motivated by the idea that sentences with distinct meanings should yield different translations, Gupta~\etal~\cite{gupta2020machine} propose PatInv, which generates syntactically similar but semantically different sentences using BERT.
In 2022, Sun \etal~\cite{sun2022improving} introduce CAT, a BERT-driven word-replacement method for enhancing machine translation. 
Beyond these studies on machine translation, researchers have also explored using LLMs, particularly BERT, in other NLP testing scenarios, including named entity recognition software~\cite{yu2023automated}, textual content moderation software~\cite{wang2023mttm}, dialogue systems~\cite{liu2021dialtest}, and question answering systems~\cite{liu2022qatest}.

In addition to the above-mentioned tasks detailed above, researchers also integrate LLMs into software testing from other aspects, including API testing~\cite{le2024kat}, code execution~\cite{xue2024selfpico}, static analysis~\cite{li2023assisting,hao2023v,mohajer2023skipanalyzer,li2024enhancing}, vulnerable dependency alert detection~\cite{sun2023silent}
theorem proving~\cite{liu2024llm}, DL adversarial attack~\cite{zhang2024codebert} and program reduction~\cite{zhang2024lpr}.

\subsection{Software Maintenance}
\label{sec:rq2_se_maintenance}
Software maintenance is one of the foundational aspects of software engineering and encompasses the ongoing process of post-delivery software modification, aiming to rectify errors and meet emerging requirements.

\subsubsection{Program Repair}
\label{sec:repair}

Automated program repair aims to generate correct patches for a detected buggy code snippet automatically and plays a crucial role during software maintenance~\cite{li2024hybrid,xu2024aligning,hossain2024deep}.
A typical repair technique usually contains three steps: (1) applying off-the-shelf fault localization techniques to recognize the suspicious code elements; (2) modifying these elements based on a set of transformation rules to generate candidate patches; (3) adopting test suites to verify all candidate patches.

Overall, we summarize the existing program repair work involving LLMs into five stages. 
Initially, program repair is directly used as a downstream task to evaluate the capability of LLMs when such LLMs are designed and proposed in their original paper, such as CodeT5 and CodeBERT.
Subsequently, there exist explorations in the SE field using LLMs as a component in existing repair workflow, such as CURE.
Then comes the fine-tuning of LLMs as repair models on historical bug-fixing datasets, which is also the most widely researched topic in the literature. 
Later, zero-shot learning is utilized to better leverage LLMs, which also indicates a shift in the repair paradigm, \ie from an NMT task to a close test task in a fill-in-the-blank format. 
Recently, there have been attempts to combine LLMs with traditional repair techniques to address inherent problems that are difficult for traditional techniques to solve.
At the same time, there is a substantial amount of empirical research in the field exploring the actual performances of LLMs in program repair.
Now, we list and summarize the existing LLM-based repair techniques as follows.

\textbf{Repair as a downstream task of LLMs.}
Since the inception of LLMs, some researchers usually employ program repair as a downstream task (also referred to as code refinement) to evaluate the models' capabilities.
In this scenario, program repair is viewed as a general code-code generation task, which translates the buggy code snippet to a correct code snippet on top of natural machine translation.
The preliminary evaluations demonstrate the remarkable performance of LLMs in understanding code semantics and learning bug-fixing code changes, as well as fostering subsequent, deeper work introducing such LLMs into the field of program repair.

\begin{table}[htbp]
\footnotesize
  \centering
  \caption{The comparison results of representative LLMs on program repair}
    \begin{tabular}{l|cc|cc}
    \toprule
          & \multicolumn{2}{c|}{BFP-Small} & \multicolumn{2}{c}{BFP-Medium} \\
    Model & BLEU  & Accuracy & BLEU  & Accuracy \\
    \midrule
    CodeBERTER & 78.26\% & 17.75\% & N.A.  & N.A. \\
    CodeT5 & 77.43\% & 21.62\% & 87.64\% & 13.96\% \\
    CodeBERT & 77.42\% & 16.40\% & 91.07\% & 5.16\% \\
    GraphCodeBERT & 80.02\% & 17.30\% & 91.31\% & 0.09\% \\
    PLBART & 77.02\% & 19.21\% & 88.50\% & 8.98\% \\
    RoBERTa & 77.30\% & 15.90\% & 90.07\% & 4.10\% \\
    \bottomrule
    \end{tabular}%
  \label{tab:llm2repair}%
\end{table}%

\textbf{LLM as a Repair Component.}
In the SE domain, the earliest exploration is to use LLMs to enhance certain aspects of existing traditional repair techniques.
For example, in 2021, on top of CoCoNut, Jiang~\etal~\cite{jiang2021cure} propose a new NMT-based APR technique CURE empowered with GPT.
First, CURE extracts millions of methods from open-source Java projects and use subword tokenization to tokenize these methods.
Second, CURE pre-trains GPT on the extracted dataset and fine-tunes it on CoCoNuT's training dataset.
Third, CURE applies a new code-aware beam-search strategy to improve patch ranking and generate more correct patches.
Finally, CURE combines the fine-tuned GPT with CoCoNuT as the full APR pipeline and trains it for the patch generation task.
Importantly, it demonstrates the unique capabilities of combining a GPT PL model and an NMT model to learn both developer-like code and fix patterns to fix more bugs.
Besides, in 2022, Li~\etal~\cite{li2022dear} propose DEAR, a learning-based APR for multi-hunk, multi-statement bugs empowered with BERT.
DEAR fine-tunes BERT to learn the fixing-together relationships among statements, \ie whether two statements are needed to be fixed together.

\textbf{Fine-tuning LLMs as Repairers.}
Inspired by the successful application of program repair as a downstream task for LLMs, more research has delved into exploring the performance of fine-tuning such models in the repair domain~\cite{yang2024multi}.
For example, DeepDebug~\cite{drain2021generating} is an early-stage work that views program repair as a Seq2Seq learning task by fine-tuning BART on Java datasets. 
At the same time, Berabi~\etal~\cite{berabi2021tfix} present TFix, which fine-tunes T5 to fix JavaScript code errors.
To support multilingual repair, in 2022, Yuan~\etal\cite{yuan2022circle} propose CIRCLE, a T5-based APR framework equipped with continual learning ability across multiple programming languages.
Recently, some studies utilize parameter-efficient fine-tuning to reduce training costs~\cite{dehghan2024mergerepair,silva2023repairllama}.

\textbf{Zero-shot LLM-based Repair.}
Despite promising, the repair performance of fine-tuned LLMs is usually constrained by the quality and quantity of labeled bug-fixing pairs, similar to previous learning-based repair techniques.
Therefore, some researchers attempt to transform the repair problem into a cloze test task under a zero-shot setting, where LLMs are queried to directly predict the correct code tokens based on context information (\ie buggy methods) without any fine-tuning on historical datasets.
For example, Xia~\etal~\cite{xia2022less} propose AlphaRepair as the first cloze-style APR approach that leverages LLMs without any fine-tuning.
AlphaRepair replaces the entire buggy line with a line containing only mask tokens and queries CodeBERT to generate candidate patches.
Repilot~\cite{wei2023copiloting} utilizes a Java completion engine to offer real-time feedback during the auto-regressive token generation process, thus assisting LLMs in producing better patches.
FitRepair seeks to enhance the performance of cloze-style APR by combining LLMs and domain repair knowledge.

\textbf{Combination of LLMs and traditional APR.}
Most LLM-based repair techniques utilize LLMs as an end-to-end learning-based patch generator and are developed separately from mature traditional techniques.
Inspired by the fact that learning-based APR is complementary to traditional repair, Zhang~\etal~\cite{zhang2023gamma} propose \textsc{Gamma} to combine the advance of LLMs and well-known template-based repair.
\textsc{Gamma} summarizes a variety of fix templates, transforms them into mask patterns, and adopts LLMs to predict the correct code for masked code as a fill-in-the-blank task.
Ruiz~\etal~\cite{ruiz2024novel} explore the potential of LLMs to repair bugs with a round-trip translation strategy.
At the same time, Peng~\etal~\cite{peng2024domain} propose a domain-aware prompt-based approach TypeFix to repair Python type errors with fix templates.

\textbf{Empirical Evaluation of LLMs.}
In conjunction with the aforementioned repair strategies tackling certain technical obstacles, there has been a concurrent surge in empirical studies examining the development and nuances of these methodologies.
These empirical studies systematically explore the actual performance of LLMs during the repair workflow, with the sim to furnish insights for forthcoming program repair endeavors~\cite{xiang2024far,huang2023empirical}.
For example, in 2022, Xia~\etal~\cite{xia2023automated} conduct an extensive study on the application of LLMs in real-world bug-fixing, with nine LLMs of varying sizes. 
In 2023, Jiang~\etal~\cite{jiang2023impact} evaluate ten code LLMs on four APR benchmarks, to discuss the impact of fine-tuning, model size, and repair costs.
Unlike previous studies focusing on software bugs~\cite{xia2023automated,jiang2023impact}, Fan~\etal~\cite{fan2022automated} conduct a systematic study to explore whether APR techniques can fix the incorrect solutions produced by language models in LeetCode contests
Besides, with the rise of ChatGPT, a mass of research efforts have been made to explore the potential of ChatGPT in repair scenarios, such as QuixBugs~\cite{sobania2023analysis} and DL prorgam~\cite{cao2023study}.

\textbf{Domain Repair.}
In the LLM-based APR field, researchers have paid considerable attention to semantic and syntax bugs, which represent the most common application of the repair techniques discussed above.
Unlike traditional APR, LLMs are pre-trained from a wide range of datasets to learn general language knowledge and can be applied to a wider range of repair scenarios.
For example, in 2023, Jin~\etal~\cite{jin2023inferfix} propose InferFix, a Transformer-based approach based on Codex, to automatically fix both critical security and performance bugs detected by the static analysis tool Infer.
So far, researchers have utilized different strategies (\eg zero-shot learning, fine-tuning, and few-shot learning) to integrate LLMs into various repair domains, including real-world issues~\cite{zhao2024enhancing}, static warning~\cite{jin2023inferfix,wadhwa2024core}, syntax error~\cite{joshi2023repair}, benchmark~\cite{yang2024cref}, neural network implementation~\cite{charalambous2024automated}, integrated development environment~\cite{xin2024towards}, human study~\cite{yang2024revisiting}, programming problems~\cite{zhang2024pydex}, repair agent~\cite{bouzenia2024repairagent,lee2024unified}, interactive repair\cite{yin2024thinkrepair,xia2023keep}, multi-location repair~\cite{xin2024detecting,wang2024revisiting}, repair costs~\cite{hidvegi2024cigar}, and formal proof~\cite{first2023baldur}.

\subsubsection{Security Vulnerability Repair}
\label{sec:vulnerability_repair}

Software vulnerability predominantly pertains to the weaknesses or flaws found within the software's code or design, which can potentially be exploited to compromise the security or functionality of the hardware or network it operates on. 
Unlike common software bugs focused on by most repair work, securities are more damaging and require more urgent fixes, making it critical to automate vulnerability fixes.
Security researchers have to spend a huge amount of effort to manually fix such vulnerable functions, resulting in delays in vulnerability remediation and providing opportunities for attacks.
With the successful application of LLMs in program repair, researchers have begun to apply LLMs to help under-resourced security researchers fix software vulnerabilities automatically.
We summarize existing vulnerability repair studies empowered with LLMs as follows.

\textbf{Fine-tuning LLM-based Vulnerability Repair.}
VulRepair~\cite{fu2022vulrepair}  is an early-stage LLM-based vulnerability repair approach by directly fine-tuning CodeT5.
VQM~\cite{fu2024vision} is a successor of VulRepair by leveraging a cross-attention mechanism to locate vulnerable code elements during patch generation.
VulMaster~\cite{zhou2024out} fine-tunes CodeT5 with information from diverse sources, including the structure of vulnerable code and expert knowledge.

\textbf{Zero-shot LLM-based Vulnerability Repair.}
In 2023, Pearce~\etal~\cite{pearce2023examining} examine a zero-shot vulnerability repair approach, assessing the potential of large language models such as OpenAI's Codex and AI21's Jurassic J-1. The primary research challenge lies in designing prompts that prompt LLMs to generate corrected versions of insecure code. The study emphasizes the use of commercially available black-box LLMs, as well as open-source models and locally trained models, for extensive research experiments on synthetic, handcrafted, and real-world security vulnerability scenarios.

\textbf{Empirical Study of LLM-based Vulnerability Repair.}
In 2023, Zhang~\etal~\cite{zhang2023pre} conduct the first extensive empirical study to investigate the actual performance of various LLMs on vulnerability repair, involving more than 100 fine-tuned LLMs.
First, they demonstrate that through simple fine-tuning, LLMs are able to outperform state-of-the-art vulnerability repair techniques.
Second, they delved into studying the impact of LLMs on the repair workflow, including data pre-processing, model training, and repair inference phrases.
Third, they develop a straightforward vulnerability repair strategy, leveraging transfer learning from bug-fixing, and demonstrate that such a simplified approach further enhances the prediction accuracy of LLMs.
Furthermore, they offer additional insights by discussing various aspects, such as code representation and a preliminary study with ChatGPT, to illuminate the capabilities and limitations of LLM-based vulnerability repair approaches.
Finally, they precisely identify several practical guidelines, such as enhancing fine-tuning, to advance LLM-based vulnerability repair in the imminent future.
Besides, Wu~\etal~\cite{wu2023effective} compare the performance of LLMs with existing learning-based vulnerability repair techniques, including Codex, CodeGen, CodeT5, PLBART and InCoder.
A similar study is conducted by Huang~\etal~\cite{huang2023empirical}.
In 2023, Tol~\etal~\cite{tol2023zeroleak} empirically explore the potential of leveraging LLMs to automatically generate patches for side-channel vulnerabilities.

\subsubsection{Patch Correctness Assessment}
\label{sec:patch_correctness}
It is a common practice for the majority of extant program repair methodologies to predominantly utilize developer-constructed test suites as the program specification, serving to evaluate the accuracy of the patches produced.
Nevertheless, such an existing test suite represents an inherently partial specification, delineating only a segment of the program’s behavioral domain.
Thus, repair approaches may suffer from the patch overfitting issue (\ie patches passing the available test suites fail to generalize to other potential test suites), limiting the value and deployment of such repair approaches in real-world scenarios.
Patch correctness is a crucial phase for developers to further filter out overfitting patches after patch generation (detailed in Section~\ref{sec:repair}), so as to improve the quality of returned patches.
The patch overfitting issue is a long-standing challenge in the program repair community and some independent studies have been conducted to address it.
We summarize these works from three aspects, \ie LLMs as feature extractor, fine-tuning-based APCA and zero-shot-based APCA.
First, as early as 2020, Tian~\etal~\cite{tian2020evaluating,tian2023best} investigate the effectiveness of representation learning for for patch correctness assessment.
They select four embedding models, including re-trained (\ie Doc2vec, code2vec and CC2vec) and pre-trained models (\ie BERT), marking the first application of BERT in this field.
In 2022, Tian\etal~\cite{tian2022change} propose Quatrain, which utilizes CodeTrans to generate patch descriptions, and  BERT to embed bug reports and patch descriptions.
Recently, Invalidator~\cite{le2023invalidator} identifies the correctness of patches via semantic and syntactic reasoning.
Second, APPT~\cite{zhang2023boosting} is the first approach that fine-tunes LLMs to predict patch correctness.
APPT consists of three components:
(1) BERT to extract features from source code tokens;
(2) LSTM to capture dependency information between source and repaired code snippets;
and (3) a DL classifier to predict whether a patch is overfitting or not.
Third, PatchZero~\cite{zhou2023patchzero} attempts to predict patch correctness in a zero-shot manner.
PatchZero reformats the format of the patch correctness assessment task to match the original pre-training objective of the LLMs.
Molina~\etal~\cite{molina2024improving} introduce FixCheck to filter out incorrect patches by combining random testing and LLMs to generate fault-revealing test cases.

\subsubsection{Commit Message Generation}
\label{sec:commit_generation}

Commit message generation attempts to create natural language descriptions for code commits. 
It begins with modified code snippets and employs template-based, retrieval-based, or learning-based models to generate commit messages. 
The key lies in understanding the context of code modifications and accurately describing these changes in the commit messages.
For example, Jung~\etal~\cite{jung2021commitbert} introduce CommitBERT to generate messages by using CodeBERT as the initial weight during training.
Wang~\etal~\cite{wang2023delving} introduce ExGroFi to fine-tune LLMs by incorporating the correlation between commits and issues during the training process.
Li~\etal~\cite{li2024only} utilize the reasoning capabilities of LLMs to generate commit messages by framing it as a knowledge-intensive reasoning task.
Regarding empirical studies, Xue~\etal~\cite{xue2024automated} explore the effectiveness of different LLMs in generating commit messages, and Lopes~\etal~\cite{lopes2024commit} focus on the ChatGPT.

\subsubsection{Code Review}
\label{sec:code_review}

Code review is a critical process in the software development lifecycle, withe the aim of enhancing code quality by identifying and resolving issues before the code is merged into the main codebase. 
In code review practices, developers are tasked with thoroughly examining, understanding, and executing the code under review. 
This process involves evaluating various aspects of the code, such as its logic, functionality, performance, and style.

To reduce human review efforts, early work in this community primarily focuses on training domain LLMs based on T5 to automate the code review process~\cite{tufano2022using,li2022auger,li2022automating}.
For example, Li~\etal~\cite{li2022auger} introduce AUGER to generate code review comments by pre-training T5 with a masked language modeling object. 
Tufano~\etal~\cite{tufano2022using} pre-train and fine-tune T5 to support different tasks, such as comment generation and code refinement.
Initialized with CodeT5, CodeReviewer is pre-trained with four pre-training tasks specifically designed for the code review scenario: diff tag prediction, denoising code diff, denoising review comment, and review comment generation.
Researchers also conduct some empirical studies from different aspects~\cite{widyasari2023explaining,tufano2024code}
For example, Guo~\etal~\cite{guo2023exploring} explore the potential of ChatGPT in code review tasks, with a specific focus on code refinement based on code reviews.
This study involves various research aspects, including the impact of prompt and temperature settings, the comparison with CodeReviewer, the qualitative analysis, and the analysis of root causes.

\subsubsection{Bug Report Detection}
\label{sec:report_detection}

Duplicate bug report detection attempts to automatically identify and label duplicate bug reports in issue tracking systems, enabling developers to avoid redundantly dealing with the same issues.
In the report detection process, researchers utilize LLMs to compare and analyze the similarity between different bug reports to determine if they describe the same defect or problem. 
For example, in 2023, Zhang~\etal~\cite{zhang2023cupid} introduce Cupid, which integrates the traditional detection approach REP with ChatGPT. 
Cupid utilizes ChatGPT in a zero-shot setting to extract key information from bug reports, which is then used as input for REP to detect duplicate bug reports. 
Meanwhile, Plein~\etal~\cite{plein2023can} conduct empirical research, as outlined in \cite{plein2023can}, on how to utilize ChatGPT to transform user-provided software defect reports into formal test case specifications. 
They employ ChatGPT to generate test cases and evaluate the executability and validity of these generated test cases. Experimental results demonstrate the significant potential of ChatGPT in converting informal defect reports into formal test cases, holding crucial implications for automated software testing and defect resolution tasks.

\subsubsection{Bug Reproduction}
\label{sec:bug_reproduction}
Bug reproduction, also known as bug replay,  refers to the process of reproducing or recreating software defects or issues based on the information provided in a bug report. 
The process is crucial for software maintenance as it enables developers to understand, replicate, and fix the defects reported.

In 2023, Kang~\etal~\cite{kang2023large} introduce a framework named LIBRO, which utilizes LLMs to generate potential test cases from bug reports and subsequently ranks and suggests these generated solutions through post-processing steps. 
Focusing on mobile application crashes, Huang~\etal~\cite{huang2024crashtranslator} presents  CrashTranslator, which leverages LLMs to predict the necessary exploration steps required to reproduce a crash and uses reinforcement learning to improve the search process and mitigate inaccurate predictions.
Furthermore, Feng~\etal~\cite{feng2023prompting} propose a lightweight approach AdbGPT to automatically reproduce Andriod bugs from bug reports without any training.

\subsubsection{Test Update}
\label{sec:test_update}

Test update refers to the process of modifying existing test cases to align them with recent changes in the production code.
This is a crucial aspect of maintaining the quality and relevance of software tests throughout the software lifecycle.
CEPROT~\cite{hu2023identify} represents the first attempt to fine-tune CodeT5 to automatically identify and update obsolete method-level test cases in response to changes in production code.
SYNBCIATR~\cite{liu2024augmenting} designs three target-oriented contexts (\ie class contexts, usage contexts, and environment contexts), which are used to construct prompts to query GPT-4 to generate repaired test cases.
TaRGet~\cite{yaraghi2024automated} treats the test update process as a language translation task by fine-tuning LLMs with crucial context information.

\subsubsection{Log Analysis}
Logs record events that occur within a system, including user actions, system activities, error messages, and security alerts. Log analysis involves automatically examining and interpreting this log data, which is essential for developers to understand system status and diagnose potential problems.
Existing approaches can be categorized into fine-tuning~\cite{tao2022logstamp,mehrabieffectiveness}, few-shot~\cite{le2023log,ma2024llmparser,jiang2023lilac,liu2024interpretable,xu2024unilog}, and zero-shot learning~\cite{liu2024logprompt,xiao2024stronger,huang2024ulog}.
For example, Le~\etal~\cite{le2023log} explore using prompt-based few-shot learning for log parsing, leveraging the flexibility of prompts to adapt to various log formats with minimal training examples.
Researchers also empirically explore the potential of LLMs in log analysis, including the impact of data resampling~\cite{ma2024influence}, structuring logs~\cite{le2023log2}, detecting unusual behaviors~\cite{le2023log2,hadadi2024anomaly}, log-based question-answering~\cite{huang2024gloss}, interpreting and categorizing logs\cite{mudgal2023assessment}.

\subsubsection{Code Clone Detection}
Code Clone Detection attempts to identify duplicate or similar code segments within a software codebase based on code similarity analysis.
Zhang~\etal~\cite{zhang2024assessing} conduct the first empirical study to explore the potential of ChatGPT and GPT-4 in the task of code clone detection.
Moumoula~\etal~\cite{moumoula2024large} investigate the capabilities of four LLMs for detecting code clones across different programming languages with eight different prompt configurations.
Dou~\etal~\cite{dou2023towards} explores different aspects of LLMs' capabilities by examining five different perspectives: simple prompts, one-step chain-of-thought prompts, multi-step chain-of-thought prompts, code embeddings, and multiple programming languages.

In addition to the aforementioned studies, researchers also leverage LLMs to automate a variety of software maintenance tasks, including bug triaging~\cite{dipongkor2023comparative,lee2022light}, code smell~\cite{ma2023pre}, emotion-cause extraction~\cite{imran2024uncovering}, exception handling recommendation~\cite{cai2024programming}, incident management~\cite{jiang2024xpert,ahmed2023recommending}, issue labeling~\cite{colavito2024leveraging}, privacy policy~\cite{morales2024large}, code porting~\cite{mohammed2024enabling}, compiler optimization~\cite{gao2024vic,tu2023isolating}, and Android permission-related problems resolution~\cite{oishwee2024large}.

\subsection{Software Management}
\label{sec:rq2_se_management}
Software management refers to the practice of overseeing the entire software lifecycle, ensuring that software products are delivered on time, within budget, and meet quality standards. 
To date, researchers mainly utilize LLMs to perform effort estimation, software tool configuration, developers' behavior analysis, and software repository mining.
First, effort estimation attempts to predict the amount of time, resources, and effort required to complete a software project.
Alhamed~\etal~\cite{alhamed2022evaluation} explores the potential of BERT with expert features to estimate the effort required for software maintenance tasks.
Fine-SE~\cite{li2024fine} combines both semantic and expert features to develop an automatic integrated BERT-based approach for effort estimation.
Second, software tool configuration attempts to ensure that the appropriate tools are selected and configured.
Kannan~\etal~\cite{kannan2023can} discusses various prompting strategies for interacting with GPT-4 in configuring tools like machine learning frameworks and complex software systems.
Third, analyzing developers' behavior is crucial for project managers to understand team dynamics, productivity, and collaboration patterns.
Cai~\etal~\cite{cai2024programming} conduct the first attempt to utilize LLMs to detect the causes of emotions within developer communications.
They focus on zero-shot LLMs, such as ChatGPT and GPT-4, to automatically recognize and extract the causes behind emotional expressions such as frustration, happiness, or anger in platforms such as GitHub and Stack Overflow.
Imran~\etal\cite{imran2024emotion} conduct an empirical study for emotion classification by fine-tuning LLMs, such as BERT, RoBERTa, CodeBERT and GraphCodeBERT.
Fourth, software repository mining involves analyzing and extracting useful information from software repositories, thus helping project managers gain insights into development trends, team productivity, code quality, and project health.
Abedu~\etal~\cite{abedu2024llm} explore the performance of LLMs answering questions related to software repositories to lower the barrier for stakeholders by automating the extraction and analysis of repository data.

\finding{2}{
Overall, LLMs have been employed across various stages of SE research, tackling a diverse array of \numcode{} tasks. 
On the one hand, these tasks align with previous downstream ones (detailed in Section~\ref{sec:rq1.3_downstream}), yet they are explored more thoroughly, such as program repair. 
On the other hand, researchers have turned their attention to a greater number of more complex SE tasks.
These tasks may (1) include complex processes that LLMs cannot fully handle, such as fuzzing; (2) include other inputs, such as test reports and GUI testing; and (3) include some unique areas of SE, such as fuzzing.
Researchers need to devote more effort to addressing these more domain-specific issues, such as designing specific LLMs or embedding them into existing research workflow.
Notably, software testing and development have seen more extensive applications of LLMs. 
This trend may stem from the fact that these areas often serve as foundational downstream tasks for LLMs, where they have shown considerable promise. 
Besides, these tasks can be addressed naturally by existing LLMs in the form of sequence-to-sequence code generation.
However, the application of LLMs in software requirements \& design and software management is still relatively unexplored, suggesting a potential area of focus for future research in this field.
}

\section{RQ3: Integration Perspective}
\label{sec:rq3_intergation}

In the rapidly evolving field of SE, LLMs have emerged as pivotal roles, offering unprecedented opportunities for various code-related tasks.
However, the integration of LLMs in SE is not without challenges due to the inherent characteristics of LLMs, such as the record-breaking parameters making it difficult to deploy in practical scenarios.
This section delves into four crucial aspects of integrating LLMs for SE, focusing on evaluation and benchmarking in Section~\ref{sec:rq3.1_evaluation}, security and reliability in Section~\ref{sec:rq3.2_security}, domain tuning in Section~\ref{sec:rq3.3_tuning}, and compressing and distillation in Section~\ref{sec:rq3.4_compressing}.

\subsection{Evaluation and Benchmarking}
\label{sec:rq3.1_evaluation}

Despite an emerging research area, a variety of LLMs have been proposed and have continuously achieved promising results across a variety of SE tasks.
In addition to developing new techniques that address technical challenges, the LLM-based SE research field is benefiting from empirical studies and benchmarks.

\subsubsection{Empirical Study}
\label{sec:rq3.1.1_empirical}

In Section~\ref{sec:rq2_se}, when introducing specific SE tasks, we have discussed corresponding empirical studies, such as program repair~\cite{jiang2023impact} and vulnerability repair~\cite{zhang2023pre}. 
We also notice there exist some empirical studies that systematically explore the capabilities of LLMs from a more comprehensive perspective, such as human study and educational scenarios.

\textbf{Exploration for Multiple Tasks}.
These studies explore LLMs across multiple LLMs or tasks, providing macroscopic insights into future LLM-based SE work.
In 2022, Mastropaolo~\etal~\cite{mastropaolo2022using} empirically evaluate the performance of transfer learning on four code-related tasks by pre-training and fine-tuning T5, including (1) automatic bug-fixing, (2) injection of code mutants, (3) generation of assert statements, and (4) code summarization.
Lu~\etal~\cite{lu2021codexglue} conduct an empirical study to investigate the performance of three common LLMs in the CodeXGLUE benchmark (discussed in Section~\ref{sec:rq3.1.1_benchmark}), including the BERT-style, GPT-style, and Encoder-Decoder models.
As a pioneering work, this study only reports some preliminary results on a limited set of models.
In 2022, Zeng~\etal~\cite{zeng2022extensive} conduct a comprehensive study of eight LLMs across seven code-related tasks in the CodeXGLUE dataset.
The study covers three types of LLMs, including three encoder-based models (\ie CodeBERT, GraphCodeBERT, ContraCode), one decoder-only model (\ie CodeGPT), and four encoder-decoder LLMs (\ie CodeT5, CodeTrans, CoTexT and PLBART).
The selected LLMs are evaluated on three code understanding tasks (\ie defect detection, clone detection, and code search) and four code generation tasks (\ie code summarization, code repair, code translation, and code generation).
In 2023, Niu~\etal~\cite{niu2023empirical} perform a large systematic study of 19 LLMs on 13 SE tasks.
Besides, researchers explore the performance of LLMs in SE tasks regarding different aspects, including in-context learning~\cite{gao2023makes}, prompt design~\cite{white2024chatgpt}, code distributions~\cite{zhou2023devil}.

\textbf{Human Study and Empirical Replication}.
Liang~\etal~\cite{liang2024exploratory} explore how LLMs can be utilized to support human participation tasks in empirical software engineering.
This study involves four LLMs (including ChatGPT, ERNIE Bot, Gemini, and ChatGLM) and three types of prompts (including zero-shot, one-shot, and optimized one-shot prompts).
Endres~\etal~\cite{endres2024can} investigate the potential of GPT-4 to replicate empirical software engineering research. 
They query GPT-4 to generate assumptions, analysis plans, and code modules based on the methodologies described in seven papers and conduct a user study involving 14 participants to evaluate the quality of the outputs generated by GPT-4.

\textbf{Empirical Exploration of SE Education}.
In the SE community, the current research on LLMs has primarily focused on various code-related tasks and has achieved notable results.
Given the powerful NL capabilities (\eg ChatGPT) and extensive programming knowledge inherent in LLMs, their interaction with students to complete programming tasks is becoming increasingly promising.
The community has engaged in empirical discussions about the potential of LLMs in the SE education scenario from various dimensions. 
On the one hand, it is exciting for LLMs to introduce innovative methods for learning and teaching, facilitating a more interactive and dynamic educational environment. 
On the other hand, there exists a growing concern regarding the potential misuse of these LLMs by students, such as relying excessively on automated solutions without developing critical problem-solving skills.

Thus, researchers explore the potential and concerns of LLMs in the SE education scenario.
For example, to explore ChatGPT's effectiveness in answering software testing questions from a textbook, in 2023,
Jalil~\etal~\cite{jalil2023chatgpt} evaluate ChatGPT with 31 questions from five topics like software faults, test driven development, and coverage criteria. 
The study focuses on the accuracy of ChatGPT's answers and explanations under different prompting strategies. 
At the same time, to explore how well ChatGPT can perform in an introductory-level functional language programming course, Geng~\etal~\cite{geng2023can} select a second-year undergraduate computer science course to test ChatGPT.
When LLMs (\eg ChatGPT) are used by students during their learning processes, despite their potential, there may arise some concerns.
For example, students might directly employ LLMs to complete assignments without self-thinking, leading to ineffective learning and even plagiarism concerns.
To address such concerns, in 2023, Nguyen~\etal~\cite{nguyen2023snippet} present an empirical study to investigate the feasibility of automated identification of AI-generated code snippets.
They propose a CodeBERT-based classifier called GPTSniffer to detect source code written by LLMs.
Tian~\etal~\cite{tian2023chatgpt} conduct an empirical analysis of ChatGPT’s potential as a fully automated programming assistant.
Xue~\etal~\cite{xue2024does} investigate the performance of ChatGPT in assisting students in an introductory computer science course. 
This study is conducted in a classroom setting and includes both quantitative and qualitative analyses to understand the impact of using ChatGPT on students' learning outcomes and behaviors during programming assignments.

\subsubsection{Benchmarking}
\label{sec:rq3.1.1_benchmark}
Benchmarking plays a crucial role in LLM-based SE research, helping to drive the advancement of this rapidly evolving field.
Existing benchmarks in the community can be broadly classified into two categories. 
The first category consists of existing datasets created by traditional SE research, which are generally well-established and have been validated by the community over time.
Details can be found in prior SE survey papers~\cite{zhang2020machine,yang2022survey}.
The second category includes newly developed datasets tailored specifically for LLM-based SE research.
These new datasets primarily focus on two aspects: addressing challenges unique to LLMs and exploring various SE domains, summarized as follows.

First, different from traditional SE studies, the pipeline of LLM-based SE techniques is three-fold, \ie (1) a pre-training process with unsupervised learning on large datasets; (2) a fine-tuning process with supervised learning with labeled datasets; and (3) an evaluation process with limited datasets.
As a result, such LLMs are usually trained from all possible open-source projects in the wild, and it is difficult to ensure there exist no samples in the evaluation benchmark that appear in the pre-training dataset, \ie the data leakage problem.
To address this issue, researchers conduct new benchmarks from artifacts that are released after the training cutoff date of ChatGPT.
For example, Zhang~\etal~\cite{zhang2023critical} extensively explore the data leakage issue of ChatGPT in the program domain and introduce EvalGPTFix, a new benchmark based on competitive programming problems collected after 2021.
Several other benchmarks includes HumanEval-Java~\cite{jiang2023impact}, ConDefects~\cite{wu2023condefects} and HITS~\cite{wang2024hits}

Second, some benchmarks are tailored for different SE scenarios.
HumanEval~\cite{chen2021evaluating} is initially released by OpenAI to evaluate the code generation capabilities of Codex and has been the most popular benchmark in LLM-based code generation.
HumanEval-X~\cite{zheng2023codegeex} is an extended version of HumanEval, specifically designed to evaluate the multilingual code generation capabilities of LLMs.
Unlike HumanEval only for Python code generation, HumanEval-X supports both code generation and code translation tasks, spanning Python, C++, Java, JavaScript, and Go.
EvalPlus~\cite{liu2023your} enhances HumanEval by generating additional high-quality test inputs. 
CodeXGlUE~\cite{lu2021codexglue} is constructed by Microsoft to support a collection of ten code understanding and generation tasks and six programming languages.
Compared with CodeXGLUE, CrossCodeBench~\cite{niu2023crosscodebench} provides a larger scale and more diverse code-related tasks, including 216 tasks and more than 54M data instances.
Recently, LLMs have been applied in som SE scenarios that have not been previously investigated, resulting in a gap in relevant benchmarks.
As a result, with the advent of LLM-based SE techniques, researchers have also developed corresponding new datasets, such as class-level code generation~\cite{du2023classeval}, project-level code generation~\cite{yu2024codereval}, code decompilation~\cite{tan2024llm4decompile}, code retrieval~\cite{li2024coir}.

\subsection{Security and Reliability}
\label{sec:rq3.2_security}

As LLMs have been widely utilized and achieved state-of-the-art performances in various code-related tasks, the security of these models deserves an increasing number of attention.
Similar to conventional deep learning models, LLMs have been shown to be vulnerable to adversarial attacks~\cite{yang2022natural}, \ie generating totally different results given two semantically-identical source code snippets.
This is particularly alarming given that LLMs are deployed in some mission-critical applications, such as vulnerability detection and code search.
For example, an attacker can manipulate LLMs to, (1) in vulnerability detection, output a non-vulnerable label for a piece of code that actually contains vulnerabilities, with the intention of preserving the vulnerabilities in a software system; and (2) in code search,  rank the malicious code snippet high in the search results such that it can be adopted in real-world deployed software, such as autonomous driving systems.
Such attacks on LLMs may cause serious incidents and have a negative societal impact.
In the field of SE, there has been some preliminary exploration of the attacks on LLMs for different SE tasks.

In the following, we summarize existing studies on attacking LLMs for SE, which mainly fall into four categories according to attack strategies.

\textbf{Adversarial Attack for Code LLMs.}
An adversarial attack refers to an attempt to deceive models into making incorrect decisions or predictions by inserting subtle, imperceptible alterations to input data.
In 2022, Yang~\etal~\cite{yang2022natural} propose ALERT, an adversarial attack for Code LLMs to ensure that the generated adversarial examples must maintain naturalness while preserving operational semantics to cater to human reviewers' needs.
ALERT employs both Greedy-Attack and GA-Attack to search for adversarial examples, followed by conducting a user study to assess whether the substitutes generated by ALERT can produce adversarial examples that appear natural to human evaluators. 
ALERT conducts adversarial attacks on CodeBERT and GraphCodeBERT across three downstream tasks,\ie vulnerability prediction, clone detection, and authorship attribution. 
Unlike ALERT focuses on code understanding tasks, like vulnerability detection and clone detection,
Jha~\etal~\cite{jha2023codeattack} propose CodeAttack, the first work to perform adversarial attacks on different code generation tasks.
CodeAttack attempts to generate imperceptible, effective, and minimally perturbed adversarial code samples based on code structure.
CodeAttack selects representative LLMs from different categories as victim models to attack, including CodeT5, CodeBERT, GraphCodeBERT, RoBERTa, and generates adversarial samples for different tasks, including code translation, code repair, and code summarization.

\textbf{Backdoor Attack for Code LLMs.}
A backdoor attack involves secretly making poison samples embedded with triggers (\eg a specific word) during training, so that the target model normally performs on inputs without triggers (\ie clean inputs) from ordinary users, but yields targeted erroneous behaviors on inputs with triggers (\ie poison inputs) from attackers.
By using triggers to activate backdoors, attackers can manipulate the output of poisoned models and lead to severe consequences.
Such attacks enable perpetrators to manipulate the output of compromised models, potentially leading to severe consequences.
For example, attackers can attack vulnerability detection models to mislabel a vulnerable piece of code as non-vulnerable.
As early in 2021, Schuster~\etal~\cite{schuster2021you} perform a backdoor attack against the code completion model, including GPT-2.
In 2022, Wan~\etal~\cite{wan2022you} perform the first backdoor attack for code search models, including an encoder-only Code LLM CodeBERT.
The attack utilizes two types of a piece of dead code as the backdoor trigger, including a piece of fixed logging code and a grammar trigger generated by the probabilistic context-free grammar.

However, the previously designed triggers (\ie dead code snippets) are very suspicious and can be easily identified by developers~\cite{wan2022you}.
Thus, focusing on more stealthy attack, in 2023, Sun~\etal~\cite{sun2023backdooring} propose BADCODE, a backdoor attack approach targeting neural code search models by altering function and variable names, 
BADCODE mutates function and/or variable names in the original code snippet by adding extensions to existing function/variable names, such as changing ``function()" to ``function\_aux()".
BADCODE utilizes LLMs CodeBERT and CodeT5, and fine-tunes them on the CodeSearchNet dataset, using both fixed triggers and grammar triggers (PCFG) as baselines.
Recently, AFRAIDOOR~\cite{yang2024stealthy} leverages adversarial perturbations to inject adaptive triggers into code LLMs. 
Unlike previous studies designed for specific tasks, Li~\etal~\cite{li2023poison} propose CodePoisoner, a general backdoor attack approach for three code-related tasks (\ie defect detection, clone detection, and code repair) and three models, including an LLM-based one CodeBERT.
In parallel to CodePoisoner, Li~\etal~\cite{li2023multi} propose a task-agnostic attack approach to train backdoored models during pre-training, so as to support the multi-target downstream tasks.
The attack is designed on two LLMs (\ie PLBART and CodeT5) and two code understanding tasks (\ie defect detection, clone detection) and three code generation tasks (\ie code translation, code refinement, and code generation) from CodeXGLUE.

\textbf{Imitation Attack for Code LLMs.}
An imitation attack refers to an attempt to create a local model (also known as an imitation model) that mimics the behavior of a target model without having access to the target's internal architecture or training data.
In 2023, Li~\etal~\cite{li2023feasibility} propose the first imitation attack work to explore the feasibility of extracting specialized code abilities from LLMs sing common medium-sized models.
The attack employs OpenAI’s text-davinci-003 as the target LLM, CodeBERT and CodeT5 as imitation LLMs for training, and considers three code-related tasks, \ie code synthesis, code translation and code summarization.
The attack pipeline is four-fold, including (1) generating queries from LLMs tailored to different code-related tasks and query schemes; (2)designing a rule-based filter to select high-quality responses suitable for training; and (3) fine-tuning medium-sized backbone models with these filtered responses to train the imitation model.

\textbf{Others}.
In addition to the studies detailed above, researchers also explore the security and reliability of LLMs from a mass of perspectives, including human study~\cite{wang2024rocks,nam2024using,serafini2024chatgpt,tanzil2024chatgpt}, glitch token analysis~\cite{li2024glitch}, testing~\cite{jiang2024costello,yang2024distillseq,hyun2024metal}, memorization detection~\cite{al2024traces}, auto-generated code quality assurance~\cite{liu2024refining,jesse2023large,liu2024no}, auto-generated code detection~\cite{idialu2024whodunit}, knowledge understanding~\cite{karmakar2023inspect}, model analysis~\cite{song2024luna}, and jailbreak vulnerability fuzzing~\cite{yao2024fuzzllm}.

\subsection{Domain Tuning}
\label{sec:rq3.3_tuning}

As mentioned in Section~\ref{sec:rq2_se}, the pre-training-and-fine-tuning paradigm has been a crucial means of adapting LLMs to special domains.
Typically, LLMs are first pre-trained to learn the general purpose code representations on a large amount of data and then fine-tuned to targeted tasks.
While fine-tuning LLMs has proven to be effective, it comes with significant computational and energy costs due to the record-breaking parameter scale.
For example, it takes about 2 days to train CodeT5 with the parameter size of 220M on NVIDIA A100-40G GPUs for program repair~\cite{wang2023rap}, let alone more advanced LLMs with hundreds of millions or even billions of parameters.
Besides, when fine-tuning LLMs on new datasets, it is inevitable to suffer from the catastrophic forgetting problem~\cite{yuan2022circle}, \ie forgetting the knowledge learned from previous datasets.
In the field of SE, there has been some preliminary exploration of the optimizations on LLMs during the fine-tuning phase.
In the following, we summarize existing studies on fine-tuning LLMs for SE, which mainly fall into three categories according to previous issues.

\textbf{Efficient Parameter Fine-tuning.}
Such studies involve efficient training strategies to reduce the time and resource costs during fine-tuning~\cite{shi2023towards, wang2023one, weyssow2023exploring, liu2023empirical, zhuo2024astraios}.
For example, Liu~\etal~\cite{liu2023empirical} conduct an empirical study to explore the performance of parameter-efficient fine-tuning methods on two LLMs and four code-related tasks, including adapter tuning, prefix tuning and low-rank adaptation.
In 2023, considering that fine-tuning LLMs incurs a high computational cost, 
Shi~\etal~\cite{shi2023towards} attempt to reduce the number of parameters that need to be updated by selectively freezing layers that encode code basic properties well and only fine-tuning the more dynamically changing upper layers.
Wang~\etal~\cite{wang2023one} utilize adapter tuning to improve performance in code search and summarization tasks across multiple programming languages.

\textbf{Effective Continual Fine-tuning}
Such studies involve effective training strategies to address the catastrophic forgetting issue during fine-tuning~\cite{gao2023keeping,yuan2022circle,weyssow2023usage,weyssow2023usage}.
For example, As early as 2022, Yuan~\etal~\cite{yuan2022circle} propose a T5-based program repair CIRCLE equipped with continual learning ability across multiple programming languages.
The experimental results demonstrate that CIRCLE not only effectively repairs multiple programming languages in continual learning settings, but also achieves state-of-the-art performance on five benchmarks with a single repair model.
Since LLMs can easily forget knowledge learned from previous datasets when learning from the new dataset, 
in 2023, 
Gao~\etal~\cite{gao2023keeping} introduce REPEAT, 
a method to address forgetting issues in LLMs during continual learning.
REPEAT incorporates representative exemplars replay, where selected diverse and informative samples from previous datasets are used to retrain the model, preventing memory loss. 
Besides, Weyssow~\etal~\cite{weyssow2023usage} investigate how to adapt PLMs to the dynamic nature of software development with continual learning, including replay-based and regularization-based strategies.

\textbf{Others}.
In addition to the studies mentioned above, researchers utilize various tuning strategies to adapt LLMs for the SE domain, including noise-tolerant training~\cite{gao2024learning}, instruction tuning~\cite{du2024generalization,yang2024security}, reinforcement learning~\cite{jana2023cotran,wang2024rlcoder}, and prompt learning~\cite{sun2023prompt}.

\subsection{Compressing and Distillation}
\label{sec:rq3.4_compressing}

Once LLMs are well-trained and achieve impressive results in various SE tasks, they need to be further deployed in real-world scenarios, such as integrated development environments.
However, such LLMs consume hundreds of megabytes of memory and run slowly on personal devices, which results in an impediment to the wide and fluent adoption of these powerful models in the daily workflow of software developers.
To address these challenges, some optimization strategies have been utilized by SE researchers to enhance the usability and practicality of LLMs during deployment.
For example, Shi~\etal~\cite{shi2022compressing} propose Compressor to compress LLMs into extremely small models without compromising performance.
Compressor utilizes a genetic algorithm-based strategy to guide the process of model simplification, and adopts knowledge distillation to obtain a well-performing small model.
Compressor compresses two well-known LLMs (\ie CodeBERT and GraphCodeBERT) to a size of 3 MB on two important tasks (\ie vulnerability prediction and clone detection).
Besides, Su~\etal~\cite{su2024distilled} introduce a smaller, distilled model with outputs generated by GPT-3.5 to generate concise natural language descriptions of source code.

\finding{3}{
Overall, although a large amount of research effort has been devoted to how LLMs can be adapted to automate SE tasks more effectively, the literature has also seen some orthogonal works discussing the unique challenges encountered during the process of adaption.
First, benchmarks play a pivotal role in shaping the trajectory of research advancements. 
One typical trend is the construction of new benchmarks to address the data leakage issue, \eg HumanEval, and EvalGPTFix.
The other typical trend is the application of various SE tasks, \eg CodeXGLUE and CrossCodeBench.
Second, researchers have constructed a mass of empirical studies to explore the actual performance of LLMs from different aspects, including multiple tasks, human study and SE education.
Third, LLMs can be attacked to generate vulnerable code snippets or return the wrong classifications with different attack strategies, such as adversarial attack, backdoor attack, and imitation attack.
Fouth, considering the huge parameter scale of LLMs, it is crucial to design domain-tuning strategies to adapt such LLMs in SE tasks, such as parameter and continual fine-tuning.
Fifth, after LLMs are well-trained, deploying such LLMs within the development workflow necessitates further consideration of factors such as inference time and resource expenditure.
}

\section{Challenges and Opportunities}
\label{sec:challenges}
Our survey reveals that advances in LLMs for SE have a significant impact on both academia and industry.
Despite achieving promising progress, there are still numerous challenges that need to be addressed, providing abundant opportunities for further research and applications.
We discuss the following important practical guidelines for future LLM-based SE research.

\textbf{Trade-off between Effectiveness and Model Size}.
As discussed in Section~\ref{sec:rq1.1_llms}, the community tends to introduce the growing size of models, resulting in the recent emergence of LLMs with record-breaking parameters, \eg from 117M parameters of GPT-1 to 175B parameters of GPT-3.
This trend is reasonable, as existing studies have demonstrated that larger models usually yield better performance, \eg code generation~\cite{zan2023large} and program repair~\cite{xia2023automated}, highlighting the significance of the number of model parameters in performance enhancement.
However, models with such a large number of parameters may raise some concerns during training and deployment.
First, it is extremely time-consuming and resource-intensive to train such LLMs, especially since the GPU resources required are unaffordable for most researchers in academia and even in the corporate world.
For example, \eg it takes 12 days to train the medium-sized CodeT5-base model (220M parameters) with 16 NVIDIA A100 GPUs.
Second, it is difficult to deploy such LLMs in real-world scenarios, as they consume hundreds of megabytes of memory and disk space, \eg Code Llama-34B takes about 63GB of disk space\footnote{The model size is according to Code Llama's checkpoint implemented by HuggingFace in \url{https://huggingface.co/codellama/CodeLlama-34b-hf}}.
Thus, LLMs may run slowly on personal devices and cannot be deployed on resource-constrained or real-time terminal devices, such as mobile devices and autonomous driving.

We recommend that future work can be conducted in the following three directions.
First, it is promising to optimize the size of LLMs without significantly compromising their performance, such as model pruning, quantization, and knowledge distillation~\cite{shi2022compressing}.
Second, researchers can develop lightweight models tailored for specific applications or techniques for distributed computing, enabling parts of a model to run on different devices.

\textbf{Exploring Task-oriented Domain LLMs}.
As discussed in Section~\ref{sec:rq1.3_downstream}, although LLMs are increasingly being applied in the SE community, the majority of these models are designed with general-purpose training strategies to support multiple downstream tasks.
However, there are some concerns with the adoption of such general LLMs. 
First, LLMs need to learn general knowledge about natural language and different programming languages from extremely large datasets, leading to the model's vast size.
Second, LLMs contain a vast array of knowledge, much of which is irrelevant to specific tasks. 
Third, their pre-training tasks are universal, creating a certain gap with the downstream tasks.
For example, existing LLMs are usually trained with a given code snippet and the corresponding description, which can hardly be exploited to learn the code change patterns for some code-editing tasks that involve two code snippets.
Thus, employing existing LLMs for such editing tasks will inevitably lead to inconsistent inputs and objectives between pre-training and fine-tuning.

We recommend future work to explore domain LLMs for specific tasks.
For example, researchers can design LLMs specifically for unit testing scenarios (\eg test generation and update), focusing solely on learning domain knowledge relevant to unit testing with specific pre-training objectives.

\textbf{Clean Evaluation Datasets}.
LLMs have been gaining increasing attention and demonstrated promising performance across a variety of SE tasks, such as program repair and code generation.
However, there exists a potential risk of data leakage since such LLMS are usually trained with all possible public repositories in the wild.
As mentioned in Section~\ref{sec:rq3.1.1_benchmark}, researchers~\cite{zhang2023gamma} find that some code snippets in Defects4J, the widely-adopted benchmark in the program repair literature, are leaked in CodeSearchNet, which is the most popular dataset to train LLMs, \eg CodeT5, CodeBERT, and UniXcoder.
More importantly, the greater concern arises from the black-box LLMs developed by commercial companies, which often outperform open-source LLMs. 
It is difficult to ensure whether or not the evaluation dataset has been seen by such LLMs during training as these LLMs are usually closed-source with unknown specific training details, \eg pre-training datasets.
For example, ChatGPT, the latest black-box LLM, has been investigated by numerous recent research studies and has shown impressive performance in various code-related tasks.
However, researchers find that ChatGPT can directly provide complete descriptions and the corresponding solution by simply providing it with the number of a programming problem in LeetCode.
Considering the fact there exist a quite number of black-box LLMs for which no architecture or training data information has been released.
The data leaking on such LLMs is a significant concern when it comes to evaluating their performance in some code-related tasks in the SE community.

We recommend that future work can be carried out from two perspectives.
First, the construction of clean datasets is crucial to ensure they have not been contaminated by LLMs.
Three potential sources can be utilized for this purpose, 
(1) The first source comes from manually written programs.
Similar to HumanEval~\cite{chen2021evaluating}, researchers can create evaluation programs by hand so as to provide a unique and uncontaminated benchmark.
(2) The second source is the most recently released programs.
Similar to EvalGPTFix~\cite{zhang2023critical}, researchers can seek out the latest programs, such as those from recent competitions or coding challenge websites, as they often contain fresh and diverse problems that are less likely to have been included in the training sets of current LLMs.
(3) The third source is closed-source projects.
Researchers can evaluate LLMs with some internal projects within the company, which are not previously exposed to public repositories, thereby providing a more authentic evaluation of the models' capabilities in real-world scenarios.
In addition to clean dataset construction, it is essential to design some techniques to verify whether LLMs exhibit any form of data memorization for a given testing sample, such as detecting if a model is simply recalling information from its training data rather than genuinely understanding or solving a problem.

\textbf{Application on More SE Tasks}.
As discussed in Section~\ref{sec:rq2_se}, we observe a pronounced emphasis on the application of LLMs in software development, testing, and maintenance.
These areas have undoubtedly benefited from the capabilities of LLMs, leading to impressive performance in code completion~\cite{ciniselli2021bert}, fault detection~\cite{fu2022linevul}, program repair~\cite{zhang2023gamma} and so on.
Despite its success in these tasks, there are other tasks that have been popularly studied with traditional techniques or machine/deep learning techniques.
For example, the current application of LLMs in requirements engineering, software design, and software management remains relatively sparse. 

We suggest that future work should concentrate on two aspects to broaden the scope of LLM applications for more SE tasks.
First, for complex tasks, integrating LLMs into existing research workflows as a component rather than developing end-to-end solutions appears more pragmatic. 
For example, existing regression test case prioritization approaches tend to calculate the similarity of selected test cases and candidates based on code coverage and may fail to consider the semantic similarity between different test cases.
Researchers can boost existing similarity-based prioritization techniques via LLMs, which contain generic knowledge pre-trained with millions of code snippets from open-source projects, and provide accurate semantic information for test code.
This integration strategy leverages the strengths of LLMs in augmenting and enhancing current approaches, particularly in areas where conventional approaches have reached a plateau in terms of performance. 
By combining LLMs with established techniques, we can achieve more robust and efficient outcomes in complex scenarios.
Second, for rare SE tasks where LLMs may lack rich knowledge, it is promising to design domain-specific LLMs tailored to these underrepresented areas. 
For example, a variety of medium-sized LLMs are trained with CodeSearchNet without test cases, thus failing to benefit tasks such as unit test generation.

\textbf{Multi-task and Multi-dimensional Benchmarks}.
As the development of LLMs progresses, it is crucial to acquire and prepare benchmarks that are more diverse, comprehensive, and realistic to reflect capabilities in real-world scenarios.
However, existing datasets may face issues related to data bias, deficiency, quality, and credibility.
First, most well-constructed benchmarks are concentrated on some widely-investigated SE tasks (\eg HumanEval and CoderEval in code generation), while lacking in other less-explored tasks like unit test generation.
Second, the majority of existing evaluation dimensions focus primarily on performance metrics (\eg Pass-1 in code generation), paying little attention to other critical attributes of LLMs, such as time efficiency and robustness.

We recommend that future work can be directed in the following two areas.
First, to address the limitation in task scope, researchers can build diversified benchmarks to evaluate LLMs in emerging fields, such as unit test generation. 
Second, to address the limitations in evaluation dimensions, new metrics, and specialized benchmarks should be introduced to assess some crucial aspects, such as robustness and efficiency.

\textbf{Beyond Text-based LLMs for Vision-based SE}.
In the realm of SE, a predominant focus has been observed on LLMs that process text-based inputs, \ie natural language and source code, significantly benefiting a multitude of code-related tasks. 
However, alternative forms of input play an equally crucial role in SE tasks, notably images in mobile applications. 
For example, the graphical user interface has emerged as a crucial component of mobile applications, attracting substantial research attention in the area of GUI testing.
Recently, with advancements in computer vision technologies, vision-based GUI testing approaches have been developed and have shown promising progress.

We suggest that future works focus on the utilization of multi-modal LLMs in version-based SE tasks.
For example, it is potential to combine both text and image understanding capabilities of multi-modal LLMs to better capture the syntax and semantic information about source code, test scripts, and test reports in GUI testing, so as to benefit several tasks, \eg GUI test generation, GUI test record, and replay.

\textbf{Explainable LLM-based Research}.
Existing LLMs usually address SE tasks in a black-box manner due to the inherent limitations of DL and the vast parameters of LLMs.
The developers are unaware of why LLMs generate the predictions, thus unsure about the reliability of these results, hindering the adoption of LLM in practice.
In the literature, a majority of studies focus on improving the performance of pre-defined metrics (\eg accuracy and precision), while minor focus on improving the explainability of such LLMs.
Traditional rule-based SE approaches rely on pre-defined rules and logic, which makes them more interpretable and offers more transparency.

In the future, advanced explainable techniques can be considered to make the predictions of LLMs more practical, explainable, and actionable.
We suggest that future work should concentrate on two aspects to support the understanding of LLMs for SE research.
First, it is possible to incorporate XAI techniques to elucidate the decision-making process of LLMs, such as designing strategies to trace back the decision process to specific data points or model components.
Second, developing hybrid frameworks that combine the interpretability of traditional rule-based approaches with the predictive power of LLMs could help in bridging the gap between traditional SE approaches and advanced LLMs, providing a balance between transparency and performance.

\section{Conclusion}
\label{sec:conclusion}

Large language models (LLMs) are bringing significant changes to the software engineering (SE) field, with their ability to handle complex code-related tasks poised to fundamentally reshape numerous SE practices and approaches.
In this paper, we provide a comprehensive survey of existing LLM-based SE studies from both the LLM and SE perspectives.
We summarize \numllm{} representative \clm{} and discuss their distinct architectures, pre-training objectives, downstream tasks, and open science.
We illustrate the wide range of SE tasks where LLMs have been applied, involving \numse{} relevant studies for \numcode{} code-related tasks across five crucial SE phases.
We highlight several crucial aspects of the optimization and application for the LLM-based  SE research, including empirical evaluation, security and reliability, domain tuning, and compressing and distillation.
Finally, we point out several challenges (such as the data leakage issue) and provide possible directions for future study.
Overall, our work serves as a roadmap for promising future research and is valuable to both researchers and practitioners, assisting them in leveraging LLMs to improve existing SE practices.

\bibliographystyle{IEEEtran}
\bibliography{reference}

\clearpage
\newpage
\setcounter{page}{1}
\setcounter{section}{0}

\section*{Appendix A: Summarization of Existing LLMs of Code}

In the following, we summarize some representative LLMs of code.

\subsection{Encoder-only LLMs}
\label{sec:rq1.1.1}

\textbf{CuBERT.}
CuBERT ~\cite{kanade2020learning} is the first attempt to apply BERT to source code by replicating the training procedure of BERT on a code corpus.
In particular, Kanade~\etal~\cite{kanade2020learning} construct a massive corpus of 7.4M Python files from GitHub and pre-train CuBERT with masked language modeling and next sentence prediction as the objectives.
CuBERT is fine-tuned on six downstream tasks, including five classification tasks and one program repair task, demonstrating its superior performance over LSTM and vanilla Transformer models.

\textbf{CodeBERT.}
CodeBERT~\cite{feng2020codebert} represents a successful adaption of BERT from NLP to the source code domain.
CodeBERT follows the BERT architecture (\ie a multi-layer bidirectional Transformer model), but unlike BERT, which only considers natural language (NL), CodeBERT takes into account both NL and programming language (PL).
Regarding input representation, CodeBERT's input is divided into two parts: NL and PL, forming the format $[CLS], w_1, w_2, ..., w_n, [SEP], c_1, c_2, ..., c_m, [EOS]$, where the special marker $[CLS]$ is positioned before these two segments.
CodeBERT's output comprises contextual representations for each token and the representation of $[CLS]$.
In the pre-training phase, CodeBERT employs two training objectives: masked language modeling and replaced token detection.
The first objective aims to predict the original tokens that are masked, a goal established by Devlin~\etal~\cite{devlin2018bert}, where  only bimodal data (NL-PL pairs) are utilized for training. 
The second objective is optimized to train on both unimodal and multimodal data, implying that the generator uses both NL and PL data.

Compared with CuBERT~\cite{kanade2020learning}, CodeBERT is more powerful due to several improvements during pre-training.
First, CuBERT is pre-trained with code snippets, while CodeBERT is pre-trained with both bimodal NL-PL data and unimodal PL/NL data.
Second, CuBERT is only pre-trained with Python, while CodeBERT is pre-trained with six programming languages.
Third, CuEBRT follows the objectives of BERT, while CodeBERT is trained with a new learning objective based on replaced token detection.

\textbf{GraphCodeBERT: Structure-aware Pre-training for Source Code.}
Although CodeBERT introduces code snippets during pre-training, its training paradigm is still derived from NLP by regarding a code snippet as a sequence of tokens while overlooking the inherent structure of source code.
In 2020, Guo~\etal~\cite{guo2020graphcodebert} introduce GraphCodeBERT, a graph-based LLM built upon the BERT architecture designed for code-related applications. 
GraphCodeBERT employs a representation approach rooted in data flow learning for code. 
It involves extracting ASTs through tree-sitter and capturing variables from the ASTs to form a sequence of variables. 
The relationships between extracted variables, such as data source connections, are used to construct a data flow graph. During the model's pre-training phase, GraphCodeBERT introduces two innovative training tasks alongside the inherited MLM task from CodeBERT, \ie edge prediction and node alignment.
The edge prediction task aims to learn code structural information by predicting edges within the data flow graph, while the node alignment task aims to learn which specific node in the data flow graph corresponds to which code token in the input code.
Besides, to accommodate the structure of AST graphs, GraphCodeBERT employs graph-guided masked attention.

\subsection{Encoder-decoder LLMs}
\label{sec:rq1.1.2}

\textbf{PYMT5: First Attempt of Encoder-decoder LLM.}
Similar to CuBERT~\cite{kanade2020learning} in the encoder-only LLM domain, as early as 2020, PYMT5~\cite{clement2020pymt5} is the first attempt to apply encoder-decoder LLMs to source code by replicating the pre-training process of T5 on a code corpus.
PYMT5 is pre-trained with a similar span masking objective from T5 on 26 million Python code snippets and built on an encode-decoder Transformer with 374 million parameters.
PYMT5 is fine-tuned with two tasks, \ie method and comment generation, demonstrating superior performance against GPT-2.

\textbf{T5-Learning: Adaption of T5 for Source Code.}
In parallel with PYMT5~\cite{clement2020pymt5}, Mastropaolo~\etal~\cite{mastropaolo2021studying} propose T5-learning, to empirically investigate how the T5 model performs when pre-trained and fine-tuned to support code-related tasks.
T5-learning is first pre-trained in a self-supervised way from T5 on CodeSearchNet with both natural
language text and programming language code, \ie masking tokens in code and asking the model to guess the masked tokens.
T5-learning is then fine-tuned to support four downstream tasks, \ie program repair, mutant injection, assertion generation, and code summarization.
The results demonstrate that T5-learning outperforms previous baselines, showcasing the potential of T5 in code-related tasks.

\textbf{PLBART: BART-based LLM for Code.}
Unlike PYMT5~\cite{clement2020pymt5} only focusing on Python code generation, in 2021, 
Ahmad~\etal~\cite{ahmad2021unified} propose
PLBART, an encoder-decoder LLM capable of performing a broad spectrum of code understanding and generation tasks.
PLABRT is pre-trained with the denoising objective and built on the BART architecture.
During the pre-training, PLABRT learns to reconstruct an original text that is corrupted using an arbitrary noise function, including three noise strategies in this work, \ie token masking, token deletion, and token infilling.
PLBART is fine-tuned for two categories of four downstream tasks (\ie code generation, translation, summarization, and classification) across seven programming languages.
The experimental results demonstrate that PLBART outperforms previous LLMs, such as CodeBERT and GraphCodeBERT, demonstrating its promise in both code understanding and generation.

\textbf{CodeT5: Code-aware T5-based LLM.}
Despite introducing source code, PLBART simply processes code snippets as natural language and ignores the code-specific characteristics.
CodeT5~\cite{feng2020codebert} represents a successful adaption of encoder-decoder LLMs from NLP to the source code domain and has been widely used in SE research.
In 2021, Wang~\etal~\cite{wang2021codet5} introduce CodeT5, a unified encoder-decoder LLM based on the T5 architecture by leveraging the code semantics from the developer-assigned identifiers.
CodeT5 considers two types of input representations based on whether a code snippet has a corresponding NL description: unimodal (\ie PL) and bimodal (\ie PL-NL pairs) data.
To encode the input data, CodeT5 concatenates PL and NL into a whole sequence $X$ with a delimiter token $[SEP]$, \ie $X=(w_1, \cdots, w_n, [SEP], c_1, \cdots, c_m, [SEP]])$, where $n$ and $m$ denote the number of NL word tokens and PL code tokens, respectively.
CodeT5 employs three identifier-aware pre-training tasks (\ie masked span prediction, masked identifier prediction, and identifier tagging) to consider the crucial token type information and a bimodal dual generation pre-training task to learn a better NL-PL alignment between the code and its accompanying comment.
CodeT5 is then fine-tuned with the CodeXGLUE benchmark to perform both code generation and understanding tasks, \ie code summarization, code generation, code translation, code refinement, defect detection, and clone detection.
The results demonstrate that CodeT5 significantly outperforms previous LLMs in most downstream tasks, such as RoBERTa, CodeBERT, GraphCodeBERT, GPT2, CodeGPT, and PLBART.

\textbf{SPT-Code.}
However, previous LLMs simply reuse the pre-training tasks designed for NL, while failing to learn the the connection between a piece of code and the associated NL for code-related tasks.
In May 2022, Niu~\etal~\cite{niu2022spt} introduce SPT-Code, which is a sequence-to-sequence LLM designed for source code. 
When given a complete method, SPT-Code aims to acquire general knowledge from the method's source code, its underlying code structure, and the corresponding natural language description.
The input is represented as $\{c_1, \cdots, c_l, [SEP], a_1, \cdots, a_m, [SEP], n_1, \cdots, n_p\}$, where $l$ represents the number of code tokens, $m$ denotes the length of the linearized Abstract Syntax Tree (AST) sequence, and $p$ signifies the number of tokens in the natural language description.
SPT-Code introduces three specialized code-specific pre-training tasks, \ie Code-AST Prediction (CAP), Masked Sequence to Sequence (MASS), and Method Name Generation (MNG). 
Each of these tasks enables SPT-Code to capture a distinct aspect of the data instance. 
Specifically, CAP focuses on understanding the source code by masking a random fragment of the code tokens. 
MNG aims to predict whether a given AST accurately represents a particular code fragment, thereby gaining insights into the syntactic structure. 
Finally, MNG's objective is to generate subtokens corresponding to the method name, a concise natural language description of the method. 
These three pre-training tasks are meticulously designed to enable SPT-Code to learn about source code, its underlying structure, and the natural language descriptions associated with it.
Importantly, SPT-Code does not rely on any bilingual corpora. 
This knowledge is leveraged when SPT-Code is applied to downstream tasks, making use of these three informational sources.

\textbf{CodeRL: CodeT5-derived LLM for program synthesis.}
Unlike previous general-propose LLMs, in 2022, Le~\etal \cite{le2022coderl} propose CodeRL, a successor of CodeT5, for the program synthesis task based on deep reinforcement learning.
CodeRL is built on top of CodeT5-large architecture with (1) an enlarged pre-training dataset, which has 10.5B tokens and is 10x larger than the CodeSearchNet corpus used in the original CodeT5; and (2) enhanced learning objectives, \ie masked span prediction and next-token prediction.
In particular, CodeRL considers program synthesis as a reinforcement learning problem and applies the actor-critic reinforcement learning method, enhancing CodeT5's performance by leveraging unit test signals during model optimization and generation.

\textbf{CoditT5: CodeT5-derived LLM for Code Editing.}
Despite achieving impressive performance in numerous code-related generation tasks, previous LLMs are not well-suited for editing tasks. 
In 2022, Zhang~\etal~\cite{zhang2022coditt5} propose CoditT5, an encoder-decoder LLM for code-related editing tasks based on CodeT5.
Initialized from the CodeT5-base model, CoditT5 is pre-trained with an edit-aware pre-training objective on the CodeSearchNet dataset, \ie generating the edit-based output sequence given the corrupted input sequence.
CoditT5 is fine-tuned on three downstream tasks, including comment updating, bug fixing, and automated code review, demonstrating superior performance against previous generation-based LLMs (\eg PLBART and CodeT5) in tackling code editing tasks, such as program repair.

\textbf{AlphaCode: Competition-level Code Generation LLM.}
Despite demonstrating remarkable abilities in code generation, previous LLMs have shown limited success when confronted with competition-level programming problems that require problem-solving skills beyond simply translating instructions into code.
In 2022, Li~\etal~\cite{li2022competition} from DeepMind propose AlphaCode, an encoder-decoder LLM specifically designed to generate solutions for competitive programming solutions problems that require deep reasoning.
AlphaCode is built on top of an encoder-decoder transformer-based architecture and is pre-trained with 86.31 million files across 13 programming languages from public GitHub repositories.
The encoder and decoder are pre-trained with masked language modeling and next-token prediction objectives, respectively.
AlphaCode takes the problem description as input to the encoder and generates a code autoregressively from the decoder one token at a time until an end-of-code token is produced.
AlphaCode is then fine-tuned with the CodeContests dataset and the results show that AlphaCode performs roughly at the level of the median competitor, \ie achieving on average a ranking of top 54.3\% in competitions with more than 5,000 participants.

\textbf{CodeT5+: successor LLM of CodeT5.}
Although existing LLMs are adept at learning rich contextual representations applicable to a variety of code-related tasks, they often rely on a limited set of pre-training objectives.
Such objectives might result in substantial performance degradation in certain downstream tasks due to the discrepancy between the pre-training and fine-tuning stages.
In 2023, Wang~\etal ~\cite{wang2023codet5+} present CodeT5+, a successor of CodeT5 where component modules can be flexibly combined to accommodate a wide range of downstream code tasks.
CodeT5+ is pre-trained with two objectives (\ie span denoising and causal language modeling) on unimodal code corpora and two objectives(\ie text-code contrastive learning and text-code matching) on bimodal text-code corpora.
Codet5+ is built on top of the encoder-decoder Transformer architecture and is classified into two groups according to mode size.
CodeT5+ 220M and 770M are trained from scratch following T5’s architecture and CodeT5+ 2B, 6B, 16B are initialized from off-the-shelf CodeGen checkpoints~\cite{nijkamp2022codegen}. 
The evaluation experiments are conducted on 20 code-related benchmarks under different settings, including zero-shot, fine-tuning, and instruction-tuning.
The experimental results demonstrate that CodeT5+ achieves substantial performance on various code-related tasks, such as code generation and completion, math programming, and text-to-code retrieval tasks

\textbf{JuPyT5: PyMT5-derived LLM for Jupyter Notebook.}
Unlike existing LLMs generating code from descriptions, in 2022, Chandel~\etal~\cite{chandel2022training} propose JuPyT5, an encoder-decoder LLM designed as a data science assistant for the Jupyter Notebook.
JuPyT5 is built on the BART architecture and initialized from a pre-trained PyMT5 checkpoint with the same training hyperparameters.
JuPyT5 is then pre-trained with a cell-infilling objective on a Data Science Problems (DSP) dataset, which is constructed from almost all publicly available Jupyter Notebook GitHub repositories.
DSP consists of 1119 problems curated from 306 pedagogical notebooks with 92 dataset dependencies, natural language and Markdown problem descriptions, and assert-based unit tests. 
These problems are designed to assess university students' mastery of various Python implementations in Math and Data Science.
The experimental results demonstrate that JuPyT5 achieves a 77.5\% success rate in solving DSP problems based on 100 sampling attempts, proving the potential of using LLMs as data science assistants.

\textbf{ERNIE-Code: Multilingual-NL-and-PL LLM.}
Despite achieving impressive performance in various SE tasks, existing LLMs has been essentially connecting English texts (\eg comments or docstring) and multilingual code snippets (\eg Python and Java).
Such an English-centricity issue dramatically limits the application of such LLMs in practice, given that 95\% of the world's population are non-native English speakers.
In 2023, Chai~\etal~\cite{chai2022ernie} from Baidu propose ERNIE-Code, which is a unified LLM designed to bridge the gap between multilingual natural languages (NLs) and multilingual programming languages (PLs). 
The cross-lingual NL-PL ability of ERNIE-Code is learned from two pre-training tasks, \ie span-corruption language modeling and Pivot-based translation language modeling.
The former learns intra-modal patterns from PL or NL only, while the latter learns cross-modal alignment from many NLs and PLs.
ERNIE-Code is built on the T5 encoder-decoder architecture and trained on PL corpus (\ie CodeSearchNe with six PLs), monolingual NL corpus (\ie CC100 with monolingual NLs), and parallel NL corpus (\ie OPUS with 105 bilingual pairs).
ERNIE-Code is capable of understanding and generating code and text in 116 different NLs and 6 PLs, and outperforms previous LLMs such as PLBART and CodeT5 in various code tasks, such as code-to-text, text-to-code, code-to-code, and text-to-text generation.
Importantly, ERNIE-Code demonstrates superior performance in zero-shot prompting for multilingual code summarization and text-to-text translation.

\subsection{Decoder-only LLMs}
\label{sec:rq1.1.3}

\textbf{GPT-C: first attempt LLM for Code Generation.} 
As early as 2020, Svyatkovskiy~\etal~\cite{svyatkovskiy2020intellicode} from Microsoft propose GPT-C, a variant of the GPT-2 trained from scratch on a large unsupervised multilingual source code dataset.
GPT-C is a multi-layer generative transformer model trained to predict sequences of code tokens of arbitrary types, generating up to entire lines of syntactically correct code.
The pre-training dataset contains 1.2 billion lines of code in Python, C\#, JavaScript, and TypeScript.
The experimental results demonstrate that GPT-C achieves an average edit similarity of 86.7\% on code completion tasks for Python programming language.
Importantly, GPT-C is implemented as a cloud-based web service, offering real-time code completion suggestions in Visual Studio Code IDE and Azure Notebook environments.

\textbf{CodeGPT: a variant of GPT-2 for Source Code}
In 2021, similar to GPT-C, Lu~\etal~\cite{lu2021codexglue} from Microsoft propose CodeGPT, a decoder-only Transformer-based LLM, following the model architecture and training objectives of GPT-2.
As one of the baseline LLMs in CodeXGLUE, CodeGPT is designed to support code completion and text-to-code generation tasks.
CodeGPT undergoes pre-training on the CodeSearchNet dataset, particularly focusing on the Python and Java corpora.
There exist two versions of CodeGPT, \ie the original CodeGPT, which is pre-trained from scratch with randomly initialized parameters; and CodeGPT-adapted, which is re-trained from the checkpoint of GPT-2 on the code corpus.
The experimental results show that in code completion tasks on the PY150 and Github Java Corpus datasets, CodeGPT achieves a performance of 70.65\%, while its enhanced version, CodeGPT-adapted, reaches 71.28\%. 
In the text-to-code generation task on the CONCODE dataset, CodeGPT attains a CodeBLEU performance of 32.71\%, and CodeGPT-adapted achieves 35.98\%.

\textbf{Codex: A descendant of GPT-3 for Code Tasks} 
Inspired by the considerable success of LLMs (such as GPT-3) in NLP and the abundance of publicly available code, Chen~\etal~\cite{chen2021evaluating} from OpenAI propose Codex, a descendant of GPT-3 model fine-tuned with publicly available code corpus from GitHub.
Codex is primarily trained for the task of generating independent Python functions from docstrings.
The HumanEval benchmark is constructed to evaluate the functional correctness of generated code with 164 handwritten programming problems, each accompanied by a function signature, docstring, body, and several unit tests.
The experimental results demonstrate that Codex exhibits remarkable performance, with its model solving more problems on the HumanEval dataset than GPT-3 and GPT-J, achieving a success rate of 28.8\%.
Furthermore, Codex can solve 70.2\% of the questions using a repeated sampling strategy, with 100 samples per question. This suggests that generating multiple samples from the model and selecting the optimal solution is a highly effective approach for challenging prompts.
Importantly, Codex and descendants are deployed in GitHub Copilot, indicating the power of LLMs in transforming the landscape of code-related tasks.

\textbf{PolyCoder: open-sourced LLM comparable to Codex.} 
Despite the impressive success of LLMs of code, some powerful LLMs (such as Codex) are not publicly available, preventing the research community from studying and improving such LLMs.
In 2022, Xu~\etal~\cite{xu2022systematic} propose PolyCoder, a decoder-only LLM based on GPT-2 architecture. 
PolyCoder is trained with 249GB of code from 12 programming languages and contains three sizes,  160M, 400M, and 2.7B parameters.
The results demonstrate that despite Codex's primary focus on Python, it still performs well on other programming languages, even outperforming GPT-J and GPT-NeoX.
However, for the C programming language, PolyCoder outperforms all LLMs including Codex.
Importantly, unlike Codex, three PolyCoder models of different sizes are made available for the research community.

\textbf{CodeGen: LLM for program synthesis.}
To investigate program synthesis with LLMs, in 2023, Nijkamp~\etal~\cite{nijkamp2022codegen} from Salesforce introduce CodeGen, which employs a self-regressive Transformer architecture and is trained sequentially on natural language and programming language datasets (THEPILE, BIGQUERY, and BIGPYTHON). 
It is designed for multi-round program synthesis. CodeGen undergoes evaluation for both single-round and multi-round program synthesis. In single-round evaluation, the synthetic benchmark HumanEval is utilized, and it is observed that CodeGen performance improves with data sizes. Experimental results demonstrate that the performance of the CodeGen NL model either surpasses or is comparable to that of GPT-NEO and GPT-J. CodeGen-Multi demonstrates a significant performance advantage over GPT-NEO, GPT-J, and CodeGen-NL. 
Furthermore, CodeGen-Mono, fine-tuned on a pure Python dataset, exhibits remarkable enhancements in program synthesis.

\textbf{InCoder: LLM for Code Infilling and Synthesis.}
Existing LLMs generate code in a left-to-right manner, which may be unsuitable to many many ubiquitous code editing tasks, such as bug fixing.
In 2022, Fried~\etal~\cite{fried2022incoder} from Facebook propose InCoder, a decoder-only LLM designed for program synthesis and editing. 
InCoder is pre-trained by a causal masking objective, \ie learning through the random replacement of code segments with sentinel tokens, moving them to the end of the sequence.
InCoder's training data consists solely of open-licensed code (Apache 2.0, MIT, BSD-2, and BSD-3 licenses) from online sources such as GitHub, GitLab, and StackOverflow. It primarily focuses on Python and JavaScript but encompasses a total of 28 languages, amounting to approximately 200GB of data in total. There are two versions of the publicly released pre-trained models: one with 6.7 billion parameters and another with 1.3 billion parameters.
The experimental results demonstrate that InCoder is able to infill arbitrary regions of code under a zero-shot setting for several tasks, such as type inference and comment generation ,InCoder achieves performance roughly equivalent to CodeGen-Multi on the HumanEval benchmark.

\textbf{PyCodeGPT: LLM for library-oriented code generation.}
Previous state-of-the-art LLMs are not publicly available, hindering the progress of related research topics and applications.
Similar to PolyCoder~\cite{xu2022systematic}, in 2022, Zan~\etal~\cite{zan2022cert} propose PyCodeGPT, a publicly available LLM particular designed for Python. 
PyCodeGPT is derived from GPT-Neo 125M with a vocabulary size of 32K and incorporates a novel byte-level BPE tokenizer tailored for Python source code.
The training dataset consists of 13 million Python files with 96GB crawled from GitHub.
The experimental results demonstrate that PyCodeGPT achieves a pass@1 score of 8.33\% and pass@10 of 13.53\% on the HumanEval benchmark, surpassing other LLMs with similar parameters, such as AlphaCode, CodeClippy, and CodeParrot.

\textbf{SantaCoder.}
Regarding the removal of personally identifiable information, the BigCode community~\cite{allal2023santacoder} propose SantaCoder, a decoder-only LLM with 1.1 billion parameters. 
SantaCoder's architecture is based on GPT-2 with multi-query attention and Fill-in-the-Middle objective.
Its training dataset consists of Python, Java, and JavaScript files from The Stack v1.1. The dataset has undergone several preprocessing steps, including partial data removal, near-duplication removal, de-identification of personally identifiable information, and filtering based on line length and the percentage of alphanumeric characters. Files containing test samples from benchmarks such as HumanEval, APPS, MBPP, and MultiPL-E have also been excluded.
The experimental results on the MultiPL-E benchmark demonstrate that SantaCoder outperforms InCoder-6.7B and CodeGen-2.7B in code generation and filling tasks.

\textbf{StarCoder.}
Committed to developing responsible LLMs, Li~\etal~\cite{li2023starcoder} from Hugging Face present StarCoder and StarCoderBase, which are LLMs for code. StarCoder is trained on Stack v1.2, and to ensure the secure release of open-source LLMs, it has improved the personally identifiable information editing pipeline and introduced innovative attribution tracking tools. 
StarCoder undergoes evaluations on HumanEval and MBPP, the experimental results show that StarCoder outperforms PaLM, LaMDA, LLaMA, CodeGen-16B-Mono, and OpenAI's code-cushman-001 (12B) on HumanEval.

\textbf{PanGu-Coder: LLM for text-to-code generation.} 
To address the specific task of text-to-code generation, adapt to more specific language domains, and handle signals beyond natural language,
In 2022, Christopoulou1~\etal~\cite{christopoulou2022pangu} from Huawei propose PanGu-Coder, a decoder-only LLM for text-to-code generation, \ie generating stand-alone Python functions from docstrings and evaluating the correctness of code examples through unit tests.
PanGu-Coder is built on top of the PanGu-Alpha architecture, a uni-directional decoder-only transformer with an extra query layer stacked on top.
PanGu-Coder is trained with two objectives, \ie a causal language modeling on raw programming language data, and a combination of causal language modeling and masked language modeling for the downstream task of text-to-code generation.
The results under a zero-shot manner show that PanGu-Coder outperforms industry LLMs such as Codex and AlphaCode on the HumanEval and MBPP datasets.

\textbf{PanGu-Coder2: LLM with Reinforcement Learning.}
Inspired by the success of Reinforcement Learning from Human Feedback in LLMs, 
in 2023, Shen~\etal~\cite{shen2023pangu} propose PanGu-Coder2, a successor of PanGu-Coder with more powerful code generation capability.
PanGu-Coder2 is trained with a new training paradigm, \ie Rank Responses to align Test\&Teacher Feedback and built on top of the advanced StarCoder 15B model. 
The experimental results demonstrate that PanGu-Coder2 is able to outperform previous LLMs, such as StarCoder, CodeT5+, and AlphaCode on HumanEval, CodeEval, and LeetCode
benchmarks.

\textbf{PaLM-Coder: a variant of PaLM for source code.} 
To investigate the captivity of PaLM for source code, in 2022, Chowdhery~\etal~\cite{chowdhery2022palm} from Google propose PaLM-Coder, a variant of PaLM by code-specific fine-tuning.
The based model PaLM is pre-trained with a high-quality corpus of 780 billion tokens, including 196GB of source code from open-source repositories on GitHub.
PaLM-Coder is further derived from PaLM with a two-stage fine-tuning process, \ie (1) an initial fine-tuning over 6.5 billion tokens, consisting of a blend with 60\% Python code, 30\% multi-language code and 10\% natural language; and (2) an extended fine-tuning with 1.9 billion Python code tokens.
The experimental results show that PaLM-Coder is able to achieve 88.4\% pass@100 on HumanEval and 80.8\% pass@80 on MBPP.
Besides, 
PaLM-Coder demonstrates impressive performance on the
DeepFix code repair task with a compile rate of 82.1\%, opening up opportunities for fixing complex errors that arise during software development.

\textbf{CodeGeeX: LLM for Multilingual Code Generation.} 
Despite demonstrating impressive performance, previous LLMs (such as Codex) mainly focus on code generation and are closed-source.
In 2023, Zheng~\etal~\cite{zheng2023codegeex} introduce CodeGeeX, a multilingual decoder-only open-sourced  LLM with 13 billion parameters for both code generation and translation tasks.
CodeGeeX is implemented with the Huawei MindSpore framework and pre-trained on 850 billion tokens from 23 programming languages, including C++, Java, JavaScript, and Go.
Besides, on top of the well-known HumanEval benchmark, a multilingual code generation benchmark  HumanEval-X is constructed to evaluate CodeGeeX by hand-writing the solutions in C++, Java, JavaScript, and Go.
The experimental results demonstrate that CodeGeeX performs exceptionally well in code generation and translation tasks on the HumanEval-X benchmark. 
Importantly, CodeGeeX has been integrated with Visual Studio Code, JetBrains, and Cloud Studio. It generates 4.7 billion tokens weekly for tens of thousands of active developers, enhancing the coding efficiency of 83.4\% of its users.
Besides, CodeGeeX is open-sourced, as well as its code, model weights, API, and HumanEval-X, facilitating the understanding and advances in the community.

\textbf{CodeGen2: A Successor of CodeGen.} 
Considering the high computational cost of training LLMs, in 2023, Nijkamp~\etal~\cite{nijkamp2023codegen2} propose CodeGen2, an successor of CodeGen, 
aimed at addressing the challenge of more efficiently training LLMs for program synthesis and understanding tasks. 
CodeGen2 provide a training framework along with open-source CodeGen2 models in four variations, including 1B, 3.7B, 7B, and 16B parameters in size.
CodeGen2 is trained on the BigPython dataset and evaluated on the Stack dataset to assess its learning performance in HumanEval and HumanEval-Infill tasks.
The experimental results demonstrate that CodeGen2 performs well across various model sizes and program synthesis and understanding tasks, outperforming InCoder in the evaluation on HumanEval.

\textbf{Code Llama: Llama-based LLM for Source Code.}
On top of the powerful Llama 2 model in NLP, in 2023, Roziere~\etal~\cite{roziere2023code} from Meta AI propose Code Llama, a series of LLMs specialized in handling code-related tasks. 
Code Llama exhibits capabilities such as infilling, support for large input contexts, and zero-shot instruction-following abilities.
The dataset of Code Llama primarily comprises an extensive collection of programming language content, with a special emphasis on Python language, trained through a code-heavy dataset containing 500B tokens and an additional Python-intensive data mix of 100B tokens.
It comprises multiple versions, covering Code Llama, Code Llama - Python, and Code Llama - Instruct with 7B, 13B and 34B parameters each.
Code Llama undergoes evaluation on HumanEval and MBPP, the experimental results indicate that Code Llama outperforms LLama and Llama 2 in terms of performance.

\clearpage
\newpage

\section*{Appendix B: Detailed Summarization of Existing LLM-based SE studies}
\label{appendix:rq2}

\definecolor{line-color}{RGB}{0, 0, 0}
\definecolor{fill-color}{RGB}{114, 200, 222}
\tikzstyle{category}=[
    rectangle,
    draw=line-color,
    rounded corners,
    text opacity=1,
    minimum height=1.5em,
    minimum width=5em,
    inner sep=2pt,
    align=center,
    fill opacity=.5,
]

\tikzstyle{leaf}=[category,
minimum height=1em,
fill=mylightpurple!60, 
text width=28em,  
text=black,
align=left,
font=\tiny,
inner xsep=2pt,
inner ysep=1pt,
]

\begin{figure*}[htbp]
  \centering
\begin{forest}
  forked edges,
  for tree={
  grow=east,
  reversed=true,%
  anchor=base west,
  parent anchor=east,
  child anchor=west,
  base=left,
  font=\small,
  rectangle,
  draw=line-color,
  rounded corners,align=left,
  minimum width=2.5em,
s sep=3pt,
inner xsep=2pt,
inner ysep=1pt,
ver/.style={rotate=90, child anchor=north, parent anchor=south, anchor=center},
  },
  where level=1{text width=12em,font=\scriptsize,}{},
  where level=2{text width=8em,font=\tiny}{},
  where level=3{text width=4.0em,font=\tiny}{},%
  where level=4{text width=4.2em,font=\tiny}{},%
  [RQ2, ver
                [Software Requirements \& Design, 
                    ver, 
                    [Ambiguity detection
                        [\cite{ezzini2022automated} \cite{moharil2022identification} \cite{moharil2023tabasco} \cite{gartner2024automated} \cite{sridhara2023chatgpt}, leaf]
                    ]
                    [GUI Layouts
                        [\cite{kolthoff2023data} \cite{brie2023evaluating}, leaf]
                    ]
                    [Requirement Classification
                        [\cite{subahi2023bert} \cite{han2023improving} \cite{khan2023non} \cite{9218141} \cite{luo2022prcbert} \cite{rahman2023pre} \cite{el2023ai}, leaf]
                    ]
                    [Requirement Completeness\\ Detection
                        [\cite{luitel2024improving}, leaf]
                    ]
                    [Requirement Elicitation
                        [\cite{ren2024combining}, leaf]
                    ]
                    [Requirement Engineering
                        [\cite{arora2024advancing} \cite{hassani2024enhancing} \cite{vogelsang2024specifications} \cite{fazelnia2024lessons} \cite{jin2024mare} \cite{pilone2024multilingual} \cite{ronanki2024requirements}, leaf]
                    ]
                    [Use Case Generation
                        [\cite{zhang2024experimenting}, leaf]
                    ]
                    [Requirement Prioritization
                        [\cite{sami2024prioritizing}, leaf]
                    ]
                    [Requirement \\Summarization
                        [\cite{jain2023transformer}, leaf]
                    ]
                    [Requirement Traceability
                        [\cite{guo2024natural} \cite{lin2021traceability}, leaf]
                    ]
                    [Requirements Quality \\ Assurance
                        [\cite{ronanki2022chatgpt} \cite{lubos2024leveraging} \cite{poudel2023leveraging} \cite{preda2024supporting}, leaf]
                    ]
                    [Software Modeling
                        [\cite{wang2024llms} \cite{tinnes2024leveraging} \cite{ferrari2024model} \cite{wei2023natural} \cite{chaaben2023towards}, leaf]
                    ]
                    [Specification Generation
                        [\cite{xie2023impact} \cite{mandal2023large} \cite{ma2024specgen}, leaf]
                    ]
                    [Specifications Repair
                        [\cite{hasan2023automated}, leaf]
                    ]
                    [Class Diagram \\Derivation
                        [\cite{li2024llm2} \cite{sanyal2024hybrid}, leaf]
                    ]
                ]
            ]
\end{forest}
\caption{Taxonomy of the application of LLMs in different domains within software engineering}
\end{figure*}
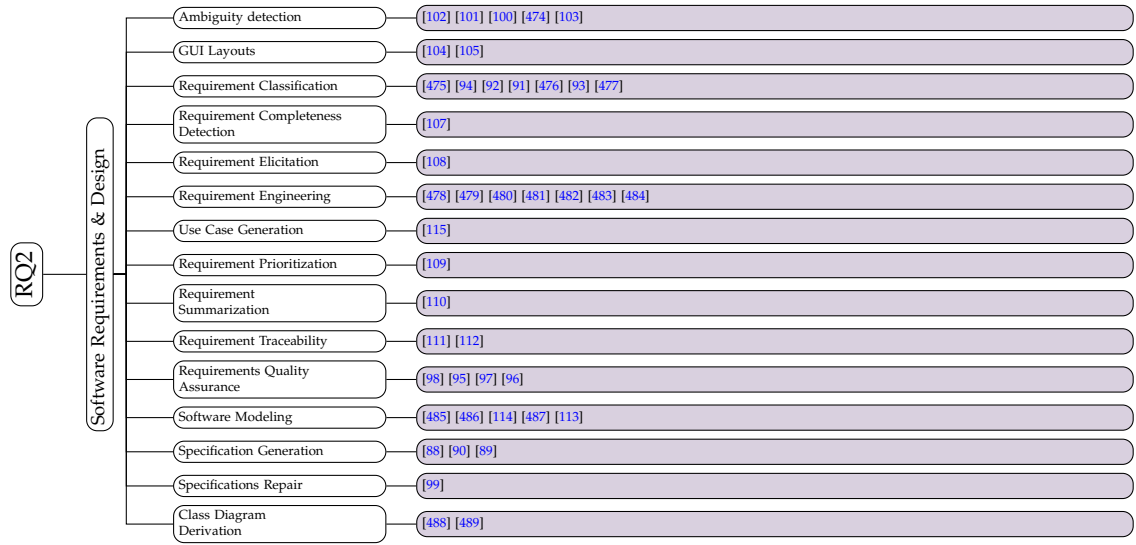

\begin{figure*}[htbp]
\ContinuedFloat
  \centering
\begin{forest}
  forked edges,
  for tree={
  grow=east,
  reversed=true,%
  anchor=base west,
  parent anchor=east,
  child anchor=west,
  base=left,
  font=\small,
  rectangle,
  draw=line-color,
  rounded corners,align=left,
  minimum width=2.5em,
s sep=3pt,
inner xsep=2pt,
inner ysep=1pt,
ver/.style={rotate=90, child anchor=north, parent anchor=south, anchor=center},
  },
  where level=1{text width=8em,font=\scriptsize}{},
  where level=2{text width=8em,font=\tiny}{},
  where level=3{text width=4.0em,font=\tiny}{},%
  where level=4{text width=4.2em,font=\tiny}{},%
  [RQ2, ver
                [Software Development, 
                    ver,
                    [API Documentation\\ Smells
                        [\cite{khan2021automatic}, leaf]
                    ]
                    [API Inference
                        [\cite{huang2023adaptive}
                        \cite{patil2023gorilla}
                        \cite{wang2023measuring}
                        \cite{zhuo2023pop}
                        , leaf]
                    ]
                    [API recommendation
                        [\cite{chen2024apigen}
                        \cite{wu2024automatic}
                        \cite{wei2022clear}
                        \cite{huang2023let}
                        \cite{li2024ptm}
                        \cite{zhang2023toolcoder}
                        , leaf]
                    ]
                    [Code Comment \\Completion
                        [\cite{mastropaolo2021empirical}
                        , leaf]
                    ]
                    [Code Completion 
                        [\cite{shi2023towards}
                        \cite{ding2023static}
                        \cite{xu2022systematic}
                        \cite{ciniselli2021bert}
                        \cite{ciniselli2021transformer}
                        \cite{khan2022automatic}
                        \cite{li2023cctest}
                        \cite{izadi2022codefill}
                        \cite{nashid2024contextual}
                        \cite{ding2024crosscodeeval}
                        \cite{cheng2024dataflow}
                        \cite{eghbali2024hallucinator}
                        \cite{tang2023domain}
                        \cite{li2024enhancing}
                        \cite{van2023enriching}
                        \cite{ochs2023evaluating}
                        \cite{gong2024evaluation}\\
                        \cite{pudari2023copilot}
                        \cite{liu2024graphcoder}
                        \cite{zhang2024hierarchical}
                        \cite{dinh2024large}
                        \cite{nie2023learning}
                        \cite{zhang2024llm}
                        \cite{prenner2021making}
                        \cite{doderlein2022piloting}
                        \cite{tan2024prompt}
                        \cite{deng2024r2c2}
                        \cite{liu2023repobench}
                        \cite{wu2024repoformer}
                        \cite{phan2024repohyper}
                        \cite{wang2024rlcoder}
                        \cite{liu2024stall+}
                        \cite{li2021toward}
                        \cite{sun2024neural}\\
                        \cite{kabir2024zs4c}
                        , leaf]
                    ]
                    [Code Compression
                        [\cite{von2022validity}
                        \cite{gilbert2023semantic}
                        , leaf]
                    ]
                    [Code Editing
                        [\cite{liu2024automated}
                        \cite{li2023codeeditor}
                        \cite{moon2023coffee}
                        \cite{gupta2023grace}
                        \cite{dilhara2024unprecedented}
                        , leaf]
                    ]
                    [Code Generation
                        [\cite{wang2023two}
                        \cite{zeng2022extensive}
                        \cite{yan2023closer}
                        \cite{buscemi2023comparative}
                        \cite{zhang2024lightweight}
                        \cite{siddiq2023lightweight}
                        \cite{li2023novel}
                        \cite{coignion2024performance}
                        \cite{yang2023syntax}
                        \cite{laskar2023systematic}
                        \cite{li2024acecoder}
                        \cite{huang2023agentcoder}
                        \cite{rao2024ai}
                        \cite{dibia2022aligning}
                        \cite{huang2023anpl}
                        \cite{han2024archcode}
                        \cite{okuda2024askit}\\
                        \cite{antal2024assessing}
                        \cite{tambon2024assessing}
                        \cite{khan2023assessing}
                        \cite{alshahwan2024assured}
                        \cite{ji2023benchmarking}
                        \cite{ghosh2024benchmarks}
                        \cite{huang2023bias}
                        \cite{singh2024brevity}
                        \cite{li2024bridging}
                        \cite{zhong2023can}
                        \cite{jin2024can}
                        \cite{nichols2024can}
                        \cite{zhong2023study}
                        \cite{jones2022capturing}
                        \cite{wang2023chatcoder}
                        \cite{mu2023clarifygpt}
                        \cite{sun2024clover}\\
                        \cite{jain2023coarse}
                        \cite{bareiss2022code}
                        \cite{ridnik2024code}
                        \cite{tsai2024code}
                        \cite{chen2024code}
                        \cite{zhang2024codeagent}
                        \cite{zhou2023codebertscore}
                        \cite{le2023codechain}
                        \cite{murali2023codecompose}
                        \cite{huang2023codecot}
                        \cite{liu2023codegen4libs}
                        \cite{du2024codegrag}
                        \cite{li2023codeie}
                        \cite{bairi2024codeplan}
                        \cite{zhang2023coder} 
                        \cite{yu2024codereval}
                        \cite{dong2023codescore}\\
                        \cite{yen2023coladder}
                        \cite{chen2024comments}
                        \cite{nascimento2023comparing}
                        \cite{ma2024compositional}
                        \cite{tan2023copilot}
                        \cite{wu2023deceptprompt}
                        \cite{olausson2023demystifying}
                        \cite{li2024deveval}
                        \cite{li2023discriminating}
                        \cite{li2024distilling}
                        \cite{kou2024large}
                        \cite{kou2023model}
                        \cite{lai2023ds}
                        \cite{li2023enabling}
                        \cite{yang2023enhancing}
                        \cite{huang2023enhancing}
                        \cite{shapkin2023entity}\\
                        \cite{patel2023evaluating}
                        \cite{chen2021evaluating}
                        \cite{assogba2024evaluating}
                        \cite{yeticstiren2023evaluating}
                        \cite{wang2024fine}
                        \cite{ren2023misuse}
                        \cite{to2024functional}
                        \cite{fu2022gpt2sp}
                        \cite{liu2024guiding}
                        \cite{dong2023abilities}
                        \cite{kazemitabaar2023novices}
                        \cite{liu2023improving}
                        \cite{malkadi2023improving}
                        \cite{chen2023improving}
                        \cite{ugare2024improving}
                        \cite{lahiri2022interactive}
                        \cite{feng2023investigating}\\
                        \cite{li2024programming}
                        \cite{nguyen2023snippet}
                        \cite{koziolek2024llm}
                        \cite{chen2024jumpcoder}
                        \cite{huang2024knowledge}
                        \cite{ni2023l2ceval}
                        \cite{xu2024large}
                        \cite{wang2023leti}
                        \cite{ni2023lever}
                        \cite{hu2024leveraging}
                        \cite{ouyang2023llm}
                        \cite{jain2023llm}
                        \cite{koziolek2024llm}
                        \cite{gu2023llm}
                        \cite{hendrycks2021measuring}
                        \cite{hong2023metagpt}
                        \cite{chen2024model}\\
                        \cite{li2023motcoder}
                        \cite{dai2024mpcoder}
                        \cite{chen2023effectiveness}
                        \cite{mastropaolo2023robustness}
                        \cite{wang2024oop}
                        \cite{arora2024optimizing}
                        \cite{fan2018oracle}
                        \cite{zheng2023outline}
                        \cite{nichols2024performance}
                        \cite{zhang2023planning}
                        \cite{zan2023private}
                        \cite{pegolotti2023qigen}
                        \cite{siddiq2024quality}
                        \cite{riddell2024quantifying}
                        \cite{wang2022recode}
                        \cite{kang2024revisiting}
                        \cite{wurustgen}\\
                        \cite{su2022selective}
                        \cite{tian2024selective}
                        \cite{dong2023self}
                        \cite{zhang2023self}
                        \cite{zheng2023self}
                        \cite{jiang2023self}
                        \cite{zelikman2023self}
                        \cite{jiang2023selfevolve}
                        \cite{li2023skcoder}
                        \cite{zhu2024sketch}
                        \cite{li2023structured}
                        \cite{jimenez2023swe}
                        \cite{sarker2024syntactic}
                        \cite{wang2024teaching}
                        \cite{chen2023teaching}
                        \cite{tian2023test}
                        \cite{hassid2024larger}\\
                        \cite{tarassow2023potential}
                        \cite{li2023think}
                        \cite{lyu2024top}
                        \cite{fakhoury2023towards}
                        \cite{busch2023towards}
                        \cite{wu2024uicoder}
                        \cite{allamanis2024unsupervised}
                        \cite{thakur2024verigen}
                        \cite{gong2023intended}
                        \cite{zan2022language}
                        \cite{lin2024llm}
                        \cite{guo2024stop}
                        \cite{liu2023better}
                        \cite{khan2023xcodeeval}
                        \cite{yang2023exploitgen}
                        \cite{zhu2024hot}
                        \cite{zan2022cert}
                        , leaf]
                    ]
                    [Code Representation
                        [\cite{cui2024api2vec++}
                        \cite{liu2023contrabert}
                        \cite{saberi2023model}
                        \cite{he2024representation}
                        \cite{agarwal2024structured}
                        \cite{lin2024vargan}
                        , leaf]
                    ]
                    [Code Search
                        [\cite{chi2024empirical}
                        \cite{shi2023cocosoda}
                        \cite{varkey2024codecse}
                        \cite{li2022coderetriever}
                        \cite{shi2022cross}
                        \cite{li2023pre}
                        \cite{li2022generation}
                        \cite{shi2023improving}
                        \cite{jain2024llm}
                        \cite{li2023mcodesearcher}
                        \cite{saieva2023contrastive}
                        \cite{salza2022effectiveness}
                        \cite{fan2024rapid}
                        \cite{li2024rewriting}
                        \cite{mao2023self}
                        \cite{wang2023you}
                        \cite{wang2023natural}
                        , leaf]
                    ]
                    [Code Summarization
                        [\cite{su2024distilled}
                        \cite{dvivedi2023comparative}
                        \cite{sun2023prompt}
                        \cite{rukmono2023achieving}
                        \cite{sadik2023analysis}
                        \cite{haldar2024analyzing}
                        \cite{gu2022assemble}
                        \cite{sun2023automatic}
                        \cite{sun2023automatic}
                        \cite{ahmed2024automatic}
                        \cite{zhao2024automatic}
                        \cite{jin2023binary}
                        \cite{kumar2024code}
                        \cite{gao2023constructing}
                        \cite{su2024context}
                        \cite{li2024cross}
                        \cite{oh2024csa}\\
                        \cite{li2024machines}
                        \cite{poudel2024documint}
                        \cite{virk2024enhancing}
                        \cite{fang2024esale}
                        \cite{arakelyan2023exploring}
                        \cite{pordanesh2024exploring}
                        \cite{al2023extending}
                        \cite{wang2023generating}
                        \cite{kang2024identifying}
                        \cite{geng2024large}
                        \cite{szalontai2024largelanguagemodelscode}
                        \cite{lu2024malsight}
                        \cite{saberi2023multilingual}
                        \cite{wang2024natural}
                        \cite{shi2024natural}
                        \cite{chen2022transferability}
                        \cite{su2023semantic}\\
                        \cite{jin2024simllm}
                        \cite{shi2023sotana}
                        \cite{sun2024source}
                        \cite{wang2024sparsecoder}
                        \cite{fan2023dialog}
                        \cite{shen2024bash}
                        \cite{jin2023binary}
                        \cite{shin2023prompt}
                        , leaf]
                    ]
                    [Code Translation
                        [\cite{xueinterpretable}
                        \cite{yang2023assessing}
                        \cite{bui2023codetf}
                        \cite{yan2023codetransocean}
                        \cite{jana2023cotran}
                        \cite{bhattarai2024enhancing}
                        \cite{tang2023explain}
                        \cite{yang2024exploring}
                        \cite{macedo2024exploring}
                        \cite{dearing2024lassi}
                        \cite{baltaji2023learning}
                        \cite{pan2024lost}
                        \cite{huang2023program}
                        \cite{yin2024rectifier}
                        \cite{nitin2024spectra}
                        \cite{pan2023stelocoder}
                        \cite{qi2023sut}
                        \cite{eniser2024towards}\\
                        \cite{wang2023transmap}
                        \cite{pan2308understanding}
                        \cite{yang2024vert}
                        \cite{li2024few}
                        \cite{zhu2022multilingual}
                        \cite{jiao2023evaluation}
                        , leaf]
                    ]
                    [Code Understanding
                        [\cite{paranjape2023art}
                        \cite{shen2022benchmarking}
                        \cite{khakhar2023pac}
                        \cite{ding2024semcoder}
                        \cite{ma2023scope}
                        \cite{zhao2023understanding}
                        \cite{artuso2024binbert}
                        \cite{utpala2023language}
                        \cite{pei2023better}
                        \cite{shi2023shellgpt}
                        \cite{abdelaziz2022can}
                        \cite{nam2023ide}
                        , leaf]
                    ]
                    [Continuous Development
                        [\cite{baral2023optimizing}
                        , leaf]
                    ]
                    [Identifier normalization \\Derivation
                        [\cite{zhang2022beqain}
                        , leaf]
                    ]
                    [Microservice \\Recommendation
                        [\cite{alsayed2024microrec}
                        , leaf]
                    ]
                    [Neural Architecture\\ Search
                        [\cite{nasir2024llmatic}
                        , leaf]
                    ]
                    [Program Synthesis
                        [\cite{tao2024enhancing}
                        \cite{singla2023evaluating}
                        \cite{shirafuji2023exploring}
                        \cite{liventsev2023fully}
                        \cite{hajali2023function}
                        \cite{ye2023generating}
                        \cite{li2024guiding}
                        \cite{barke2024hysynth}
                        \cite{jain2022jigsaw}
                        \cite{kuznia2022less}
                        \cite{gandhi2023natural}
                        \cite{austin2021program}
                        \cite{vella2024synergistic}
                        , leaf]
                    ]
                    [SO post title\\ generation
                        [\cite{yang2024automatic}
                        \cite{le2024good}
                        , leaf]
                    ]
                    [Type inference
                        [\cite{jesse2022learning}
                        , leaf]
                    ]
                    [Unified Development
                        [\cite{qian2024chatdev}
                        , leaf]
                    ]
                    [Code Recommendation
                        [\cite{rahmani2023improving}
                        \cite{zong2024graphpyrec}
                        , leaf]
                    ]
                    [Control flow\\ graph generation
                        [\cite{huang2023ai}, leaf]
                    ]
                    [Data Analysis
                        [\cite{cheng2023gpt}, leaf]
                    ]
                    [Method Name Generation
                        [\cite{zhu2023automating}, leaf]
                    ]
                    [Project Planning
                        [\cite{schroder2023autoscrum}, leaf]
                    ]
                    [SO Question Answering
                        [\cite{kabir2024stack}
                        \cite{firouzi2024time}
                        , leaf]
                    ]
                    [Data Augmentation
                        [\cite{abdellatif2024transformer}
                        \cite{banday2024perfgen}
                        , leaf]
                    ]
                ]
            ]
\end{forest}
\caption{Taxonomy of the application of LLMs in different domains within software engineering (Continue)}
\end{figure*}

\begin{figure*}[htbp!]
\ContinuedFloat
  \centering
  \begin{forest}
  forked edges,
  for tree={
  grow=east,
  reversed=true,%
  anchor=base west,
  parent anchor=east,
  child anchor=west,
  base=left,
  font=\small,
  rectangle,
  draw=line-color,
  rounded corners,align=left,
  minimum width=2.5em,
s sep=3pt,
inner xsep=2pt,
inner ysep=1pt,
ver/.style={rotate=90, child anchor=north, parent anchor=south, anchor=center},
  },
  where level=1{text width=7em,font=\scriptsize,}{},
  where level=2{text width=8em,font=\tiny}{},
  where level=3{text width=4.0em,font=\tiny}{},%
  where level=4{text width=4.2em,font=\tiny}{},%
  [RQ2, ver
                [Software Testing, 
                    ver, 
                    [Invariant Prediction
                        [\cite{pei2023can}, leaf]
                    ]
                    [GUI Testing
                        [\cite{yoon2023autonomous}
                        \cite{liu2022fill}
                        \cite{yoon2024intent}
                        \cite{liu2023make}
                        \cite{liu2024vision}
                        , leaf]
                    ]
                    [Proof Generation
                        [\cite{zhang2024selene}
                        , leaf]
                    ]
                    [Indirect Call Analysis
                        [\cite{cheng2024semantic}, leaf]
                    ]
                    [Resource Leak Detection
                        [\cite{wang2023boosting}, leaf]
                    ]
                    [Mutation Testing
                        [\cite{wang2024exploratory}
                        \cite{ibrahimzada2023automated}
                        \cite{jain2023contextual}
                        \cite{khanfir2023efficient}
                        \cite{tian2024large}
                        \cite{richter2022learning}
                        \cite{hassan2024llm}
                        \cite{tip2024llmorpheus}
                        \cite{li2024mutation}
                        \cite{garg2024coupling}
                        \cite{nong2023vulgen}
                        \cite{degiovanni2022mubert}
                        , leaf]
                    ]
                    [Taint Analysis
                        [\cite{liu2023harnessing}, leaf]
                    ]
                    [NLP Testing
                        [\cite{yu2023automated}
                        \cite{liu2021dialtest}
                        \cite{sun2022improving}
                        \cite{gupta2020machine}
                        \cite{wang2023mttm}
                        \cite{liu2022qatest}
                        \cite{he2020structure}
                        , leaf]
                    ]
                    [Vulnerability Detection
                        [\cite{steenhoek2024comprehensive}
                        \cite{kholoosi2024qualitative}
                        \cite{ju2024study}
                        \cite{saad2024alpine}
                        \cite{steenhoek2023empirical}
                        \cite{pelofske2024automated}
                        \cite{gomes2023bert}
                        \cite{chen2024bridge}
                        \cite{wan2024bridging}
                        \cite{lee2024bug}
                        \cite{noever2023can}
                        \cite{ullah2023can}
                        \cite{nong2024chain}
                        \cite{wang2024code}
                        \cite{zhao2024coding}
                        \cite{zhou2024comparison}
                        \cite{tang2023csgvd}\\
                        \cite{steenhoek2024dataflow}
                        \cite{koide2023detecting}
                        \cite{sun2023dexbert}
                        \cite{chen2023diversevul}
                        \cite{yang2024dlap}
                        \cite{liueatvul}
                        \cite{yuan2023enhancing}
                        \cite{lucas2024evaluating}
                        \cite{daneshvar2024exploring}
                        \cite{shestov2024finetuning}
                        \cite{ahmad2023flag}
                        \cite{du2024generalization}
                        \cite{sun2024gptscan}
                        \cite{lu2024grace}
                        \cite{tamberg2024harnessing}
                        \cite{mahyari2024harnessing}
                        \cite{gao2023far}\\
                        \cite{tang2023just}
                        \cite{zhou2024large}
                        \cite{zhou2024large}
                        \cite{dozono2024large}
                        \cite{fu2022linevul}
                        \cite{mathews2024llbezpeky}
                        \cite{li2024llm}
                        \cite{ashiwal2024llm}
                        \cite{cheng2024llm}
                        \cite{sun2024llm4vuln}
                        \cite{alqarni2022low}
                        \cite{wang2024m2cvd}
                        \cite{mao2024multi}
                        \cite{yin2024multitask}
                        \cite{wong2023natural}
                        \cite{do2024optimizing}
                        \cite{zhang2023prompt}\\
                        \cite{yin2024pros}
                        \cite{gonccalves2024scope}
                        \cite{yu2024security}
                        \cite{yang2024security}
                        \cite{grishina2023earlybird}
                        \cite{mao2024towards}
                        \cite{thapa2022transformer}
                        \cite{chan2023transformer}
                        \cite{khare2023understanding}
                        \cite{taghavi2024using}
                        \cite{du2024vul}
                        \cite{hanif2022vulberta}
                        \cite{chukkol2024vulcatch}
                        \cite{liu2024vuldetectbench}
                        \cite{wen2024vuleval}
                        \cite{akuthota2023vulnerability}
                        \cite{ding2024vulnerability}\\
                        \cite{quan2023xgv}
                        \cite{wu2021peculiar}
                        \cite{jiang2024dfept}
                        , leaf]
                    ]
                    [Penetration Testing
                        [\cite{pratama2024cipher}
                        \cite{happe2023getting}
                        \cite{deng2023pentestgpt}
                        \cite{wu2024ptgroup}
                        , leaf]
                    ]
                    [Program Analysis
                        [\cite{su2024cfstra}
                        , leaf]
                    ]
                    [Adversarial Attack
                        [\cite{zhang2024codebert}
                        \cite{yan2024llm}
                        , leaf]
                    ]
                    [Program Reduction
                        [\cite{zhang2024lpr}
                        , leaf]
                    ]
                    [API Misuse Detection
                        [\cite{xia2024exploring}, leaf]
                    ]
                    [API Testing
                        [\cite{le2024kat}, leaf]
                    ]
                    [Assertion Generation
                        [\cite{he2024empirical}
                        \cite{pulavarthi2024assertionbench}
                        \cite{he2024beyond}
                        \cite{endres2024can}
                        \cite{wang2024chat}
                        \cite{mali2024chiraag}
                        \cite{tufano2022generating}
                        \cite{nashid2023retrieval}
                        \cite{dinella2022toga}
                        \cite{hossain2024togll}
                        , leaf]
                    ]
                    [Binary Code \\Similarity Detection
                        [\cite{feng2024crabs}
                        \cite{wang2022jtrans}
                        \cite{ahn2022practical}
                        \cite{yu2020order}
                        , leaf]
                    ]
                    [Code Execution
                        [\cite{xue2024selfpico}
                        , leaf]
                    ]
                    [Decompilation
                        [\cite{hu2024degpt}
                        \cite{shang2024far}
                        \cite{xu2023lmpa}
                        \cite{jiang2023nova}
                        \cite{wong2023refining}
                        \cite{armengol2024slade}
                        \cite{she2024wadec}
                        \cite{tan2024llm4decompile}
                        , leaf]
                    ]
                    [Failure-Inducing Testing
                        [\cite{li2023finding}
                        , leaf]
                    ]
                    [Fault Localization
                        [\cite{kang2024quantitative}
                        \cite{qin2024agentfl}
                        \cite{yan2024better}
                        \cite{widyasari2024demystifying}
                        \cite{widyasari2024demystifying}
                        \cite{mohsen2023enhancing}
                        \cite{shan2024face}
                        \cite{ciborowska2022fast}
                        \cite{ji2024impact}
                        \cite{yang2024large}
                        \cite{wu2023large}
                        \cite{bin2024llm}
                        \cite{du2023pre}
                        \cite{chandramohan2024supporting}
                        \cite{ciborowska2023too}
                        \cite{zhu2021trobo}
                        \cite{wu2024condefects}
                        , leaf]
                    ]
                    [Fuzzing
                        [\cite{hu2023augmenting}
                        \cite{eom2024covrl}
                        \cite{xia2024fuzz4all}
                        \cite{oliinyk2024fuzzing}
                        \cite{yang2023kernelgpt}
                        \cite{meng2024large}
                        \cite{deng2023large2}
                        \cite{deng2023large2}
                        \cite{deng2023large1}
                        \cite{pearce2023examining}
                        \cite{zhang2024llamafuzz}
                        \cite{shou2024llm4fuzz}
                        \cite{wang2024llmif}
                        \cite{dakhama2023searchgem5}
                        \cite{zhang2023understanding}
                        \cite{yang2023white}
                        , leaf]
                    ]
                    [Formal Verification
                        [\cite{charalambous2023new}
                        \cite{liu2024propertygpt}
                        \cite{tihanyi2023formai}
                        \cite{wen2024enchanting}
                        , leaf]
                    ]
                    [Static Analysis
                        [\cite{li2023assisting}
                        \cite{hao2023v}
                        \cite{li2024enhancing}
                        \cite{mohajer2023skipanalyzer}
                        \cite{chapman2024interleaving}
                        \cite{fang2024large}
                        , leaf]
                    ]
                    [Static Warning \\Validating
                        [\cite{wen2024automatically}
                        , leaf]
                    ]
                    [Test Generation
                        [\cite{lops2024system}
                        \cite{yuan2024s}
                        \cite{zhang2023algo}
                        \cite{schafer2023empirical}
                        \cite{yang2024empirical}
                        \cite{guilherme2023initial}
                        \cite{zhou2024llm}
                        \cite{plein2024automatic}
                        \cite{ni2024casmodatest}
                        \cite{rao2023cat}
                        \cite{kirinuki2024chatgpt}
                        \cite{azaria2024chatgpt}
                        \cite{tang2023chatgpt}
                        \cite{xie2023chatunitest}
                        \cite{lemieux2023codamosa}
                        \cite{mundler2024code}
                        \cite{ryan2024code}\\
                        \cite{pizzorno2024coverup}
                        \cite{li2024dllens}
                        \cite{shin2023domain}
                        \cite{dakhel2023effective}
                        \cite{yang2024enhancing}
                        \cite{yuan2024evaluating}
                        \cite{he2024exploring}
                        \cite{siddiq2023exploring}
                        \cite{arora2024generating}
                        \cite{karanjai2024harnessing}
                        \cite{wang2024hits}
                        \cite{zhang2023well}
                        \cite{li2024large}
                        \cite{cui2024large}
                        \cite{ouedraogo2024large}
                        \cite{li2024leveraging}
                        \cite{deljouyi2024leveraging}\\
                        \cite{liu2024llm}
                        \cite{etemadi2024mokav}
                        \cite{karmarkar2024navigating}
                        \cite{yuan2023no}
                        \cite{xiao2024optimizing}
                        \cite{steenhoek2023reinforcement}
                        \cite{nabeel2024test}
                        \cite{gu2024testart}
                        \cite{xiong2023program}
                        \cite{tufano2020unit}
                        , leaf]
                    ]
                    [Test Suite Minimization
                        [\cite{pan2023ltm}
                        , leaf]
                    ]
                    [Vulnerable Dependency \\Alert Detection
                        [\cite{sun2023silent}
                        , leaf]
                    ]
                    [Theorem Proving
                        [\cite{liu2024llm}, leaf]
                    ]
                    [Property-based Testing
                        [\cite{vikram2023can}
                        , leaf]
                    ]
                ]
            ]
\end{forest}
\caption{Taxonomy of the application of LLMs in different domains within software engineering  (Continue)}
\end{figure*}

\begin{figure*}[htbp]
\ContinuedFloat
  \centering
  \begin{forest}
  forked edges,
  for tree={
  grow=east,
  reversed=true,%
  anchor=base west,
  parent anchor=east,
  child anchor=west,
  base=left,
  font=\small,
  rectangle,
  draw=line-color,
  rounded corners,align=left,
  minimum width=2.5em,
s sep=3pt,
inner xsep=2pt,
inner ysep=1pt,
ver/.style={rotate=90, child anchor=north, parent anchor=south, anchor=center},
  },
  where level=1{text width=8em,font=\scriptsize,}{},
  where level=2{text width=8em,font=\tiny}{},
  where level=3{text width=4.0em,font=\tiny}{},%
  where level=4{text width=4.2em,font=\tiny}{},%
  [RQ2, ver
                [Software Maintenance, 
                    ver, 
                    [Android Permissions
                        [\cite{oishwee2024large}, leaf]
                    ]
                    [APP Review Analysis
                        [\cite{motger2024t}
                        \cite{wang2022your}
                        , leaf]
                    ]
                    [Bug Report Detection
                        [\cite{plein2023can}
                        \cite{zhang2023cupid}
                        \cite{isotani2021duplicate}
                        \cite{wu2024refining}
                        \cite{helmeczi2023few}
                        , leaf]
                    ]
                    [Code Clone Detection
                        [\cite{sharma2022exploratory}
                        \cite{zhang2024assessing}
                        \cite{alam2023gptclonebench}
                        \cite{islam2024llm}
                        \cite{moumoula2024large}
                        \cite{dou2023towards}
                        \cite{chochlov2022using}
                        \cite{saberi2024utilization}
                        \cite{du2024adaccd}
                        , leaf]
                    ]
                    [Bug Reproduction
                        [\cite{huang2024crashtranslator}
                        \cite{kang2023evaluating}
                        \cite{kang2023large}
                        \cite{feng2023prompting}
                        , leaf]
                    ]
                    [Code Coverage \\Prediction
                        [\cite{tufano2023predicting}
                        , leaf]
                    ]
                    [Bug Triaging
                        [\cite{dipongkor2023comparative}
                        \cite{lee2022light}
                        \cite{colavito2024impact}
                        , leaf]
                    ]
                    [Code Evolution
                        [\cite{zhang2023multilingual}
                        , leaf]
                    ]
                    [Code Review
                        [\cite{lu2023improving}
                        \cite{dong2024gpt}
                        \cite{sghaier2023multi}
                        \cite{vijayvergiya2024ai}
                        \cite{rasheed2024ai}
                        \cite{yang2022aspect}
                        \cite{li2022auger}
                        \cite{ghadhab2021augmenting}
                        \cite{kou2023automated}
                        \cite{yin2023automatic}
                        \cite{li2022automating}
                        \cite{ahmed2024can}
                        \cite{tufano2024code}
                        \cite{koutcheme2024evaluating}
                        \cite{widyasari2023explaining}
                        \cite{fan2024exploring}
                        \cite{guo2023exploring}\\
                        \cite{pornprasit2024fine}
                        \cite{ben2024improving}
                        \cite{ferreira2024incivility}
                        \cite{lu2023llama}
                        \cite{mcaleese2024llm}
                        \cite{zhao2023right}
                        \cite{tufano2022using}
                        , leaf]
                    ]
                    [Code Refactoring
                        [\cite{pomiannext}
                        \cite{shirafuji2023refactoring}
                        \cite{zhang2024refactoring}
                        \cite{liu2023refbert}
                        , leaf]
                    ]
                    [Code Smells
                        [\cite{ma2023pre}
                        , leaf]
                    ]
                    [Commit Message \\Generation
                        [\cite{xue2024automated}
                        \cite{zhang2024automatic}
                        \cite{lopes2024commit}
                        \cite{jung2021commitbert}
                        \cite{li2024only}
                        , leaf]
                    ]
                    [Compiler Optimization
                        [\cite{tu2023isolating}
                        \cite{ye2024iterative}
                        \cite{cummins2023large}
                        \cite{shypula2023learning}
                        \cite{cummins2024meta}
                        \cite{grubisic2024priority}
                        \cite{romero2024should}
                        \cite{gao2024vic}
                        , leaf]
                    ]
                    [Debugging
                        [\cite{kang2023explainable}
                        , leaf]
                    ]
                    [Exception Handling\\ Recommendation
                        [\cite{cai2024programming}
                        , leaf]
                    ]
                    [Flaky Test Prediction
                        [\cite{fatima2022flakify}
                        , leaf]
                    ]
                    [Incident Management
                        [\cite{ahmed2023recommending}
                        \cite{jiang2024xpert}
                        , leaf]
                    ]
                    [Log Analysis
                        [\cite{mudgal2023assessment}
                        \cite{li2023exploring}
                        \cite{huang2024gloss}
                        \cite{liu2024interpretable}
                        \cite{ma2024knowlog}
                        \cite{jiang2023lilac}
                        \cite{ma2024llmparser}
                        \cite{yu2023log}
                        \cite{le2023log}
                        \cite{wu2024log}
                        \cite{le2023log2}
                        \cite{liu2024logprompt}
                        \cite{tao2022logstamp}
                        \cite{xiao2024stronger}
                        \cite{mehrabieffectiveness}
                        \cite{huang2024ulog}
                        \cite{xu2024unilog}\\
                        \cite{mastropaolo2022using2}
                        \cite{mastropaolo2024log}
                        , leaf]
                    ]
                    [Issue Labeling
                        [\cite{colavito2024leveraging}
                        \cite{colavito2024impact}
                        , leaf]
                    ]
                    [Mobile App Crash \\ Detection
                        [\cite{liu2024testing}
                        , leaf]
                    ]
                    [Log Anomaly Detection
                        [\cite{he2023parameter}
                        , leaf]
                    ]
                    [Outage Understanding
                        [\cite{jin2023assess}
                        , leaf]
                    ]                       
                    [Sentiment Analysis
                        [\cite{biswas2020achieving}
                        \cite{shafikuzzaman2024empirical}
                        \cite{zhang2023revisiting}
                        \cite{zhang2020sentiment}
                        , leaf]
                    ]
                    [Patch Correctness\\ Assessment
                        [\cite{zhang2023boosting}
                        \cite{tian2020evaluating}
                        \cite{molina2024improving}
                        \cite{tian2022change}
                        \cite{zhou2023patchzero}
                        \cite{tian2023best}
                        , leaf]
                    ]
                    [Tag Recommendation
                        [\cite{he2022ptm4tag}
                        , leaf]
                    ]
                    [Privacy Policy
                        [\cite{morales2024large}
                        , leaf]
                    ]
                    [Technical Debt\\ Management
                        [\cite{mastropaolo2023towards}
                        , leaf]
                    ]
                    [Program Repair
                        [\cite{huang2023chain}
                        \cite{hossain2024deep}
                        \cite{ruiz2024novel}
                        \cite{white2023prompt}
                        \cite{cao2023study}
                        \cite{lee2024unified}
                        \cite{widjojo2023addressing}
                        \cite{xu2024aligning}
                        \cite{sobania2023analysis}
                        \cite{kim2022empirical}
                        \cite{huang2023empirical}
                        \cite{xia2023automated}
                        \cite{charalambous2024automated}
                        \cite{fan2022automated}
                        \cite{first2023baldur}
                        \cite{hidvegi2024cigar}
                        \cite{yuan2022circle}\\
                        \cite{islam2024code}
                        \cite{xia2023conversational}
                        \cite{wei2023copiloting}
                        \cite{wadhwa2024core}
                        \cite{yang2024cref}
                        \cite{jiang2021cure}
                        \cite{li2022dear}
                        \cite{xin2024detecting}
                        \cite{peng2024domain}
                        \cite{paul2023enhancing}
                        \cite{zhao2024enhancing}
                        \cite{deligiannis2023fixing}
                        \cite{wadhwa2023frustrated}
                        \cite{zhang2023gamma}
                        \cite{drain2021generating}
                        \cite{xu2023guiding}
                        \cite{xiang2024far}\\
                        \cite{li2024hybrid}
                        \cite{jiang2023impact}
                        \cite{jin2023inferfix}
                        \cite{le2023invalidator}
                        \cite{xia2023keep}
                        \cite{xia2022less}
                        \cite{sun2024llm}
                        \cite{dehghan2024mergerepair}
                        \cite{yang2024multi}
                        \cite{zhang2023neural}
                        \cite{liu2024reliability}
                        \cite{xia2022practical}
                        \cite{zhang2024pydex}
                        \cite{chow2024pyty}
                        \cite{wang2023rap}
                        \cite{joshi2023repair}
                        \cite{zhao2024repair}\\
                        \cite{bouzenia2024repairagent}
                        \cite{silva2023repairllama}
                        \cite{du2023resolving}
                        \cite{wang2024revisiting}
                        \cite{yang2024revisiting}
                        \cite{henkel2021shipwright}
                        \cite{zhang2023steam}
                        \cite{berabi2021tfix}
                        \cite{xia2023revisiting}
                        \cite{yin2024thinkrepair}
                        \cite{lajko2022towards}
                        \cite{xin2024towards}
                        \cite{zhang2022using}
                        \cite{gharibi2024t5apr}
                        , leaf]
                    ]
                    [Test Update
                        [\cite{liu2024augmenting}
                        \cite{yaraghi2024automated}
                        \cite{hu2023identify}
                        , leaf]
                    ]
                    [Report Severity\\ Prediction
                        [\cite{ali2024bert}
                        \cite{acharya2024graph}
                        \cite{mashhadi2023method}
                        , leaf]
                    ]
                    [Traceability Link \\Recovery
                        [\cite{zhu2022enhancing}
                        , leaf]
                    ]
                    [Vulnerability Repair
                        [\cite{kulsum2024case}
                        \cite{zhang2023evaluating}
                        \cite{pearce2023examining}
                        \cite{wu2023exploring}
                        \cite{wu2023effective}
                        \cite{islam2024llm}
                        \cite{wang2024navrepair}
                        \cite{zhang2023pre}
                        \cite{sagodi2024reality}
                        \cite{fu2024vision}
                        \cite{fu2022vulrepair}
                        \cite{tol2023zeroleak}
                        , leaf]
                    ]
                ]
                [Software Management, 
                    ver, 
                    [Developers' Behavior\\ Analysis
                        [\cite{imran2024uncovering}
                        , leaf]
                    ]
                    [Effort Estimation
                        [\cite{alhamed2022evaluation}
                        \cite{li2024fine}
                        , leaf]
                    ]
                    [Software Tool \\ Configuration
                        [\cite{kannan2023can}
                        , leaf]
                    ]
                    [Software Repository\\ Mining
                        [\cite{abedu2024llm}
                        , leaf]
                    ]
                ]
            ]
\end{forest}
\caption{Taxonomy of the application of LLMs in different domains within software engineering  (Continue)}
\end{figure*}

\end{document}